{

\documentstyle[cup6b,psfig]{cupbook}  

\tolerance=10000 \hyphenpenalty10000 \exhyphenpenalty10000

\def\be{\begin{equation}}
\def\ee{\end{equation}}

\def\se#1{\S~\ref{sec:#1}}

\def\fig#1{Fig.~\ref{fig:#1}} 
 
\def\Fig#1{Figure~\ref{fig:#1}}

\def\today{\ifcase\month\or January\or February\or March\or April\or May\or 
  June\or July\or August\or September\or October\or November\or December\fi
  \space\number\day, \number\year}

\def\no{\noindent}

\def\ltsima{$\; \buildrel < \over \sim \;$}
\def\lsim{\lower.5ex\hbox{\ltsima}}
\def\gtsima{$\; \buildrel > \over \sim \;$}
\def\gsim{\lower.5ex\hbox{\gtsima}}
\def\ga{\mathrel{\hbox{\rlap{\hbox{\lower4pt\hbox{$\sim$}}}\hbox{$>$}}}}
\def\la{\mathrel{\hbox{\rlap{\hbox{\lower4pt\hbox{$\sim$}}}\hbox{$<$}}}}

\def\ifm#1{\relax\ifmmode#1\else$\mathsurround=0pt #1$\fi}
\def\kms{\,{\rm km\,s\ifm{^{-1}}}}
\def\hmpc{\,h\ifm{^{-1}}{\rm Mpc}}
\def\kmsmpc{\,{\rm km\,s\ifm{^{-1}}Mpc^\ifm{{-1}}}}

\def\sin{{\rm sin}}

\def\del{\delta}

\def\lcdm{$\Lambda$CDM}

\def\pmb#1{\setbox0=\hbox{#1}%
 \kern-.025em\copy0\kern-\wd0
 \kern.05em\copy0\kern-\wd0
 \kern-.025em\raise.0433em\box0}

\def\vr{\pmb{$r$}}

\def\del0{\delta_0}
\def\del1{\delta_1}

\def\jo{\it}

\def\jo{\it}
\def\apj#1{{\jo Ap.J.}~{\bf #1}}

\def\pasp#1{{\jo Proc. Ast. Soc. Pac.}~{\bf #1}}

\def\prl#1{{\jo Phys. Rev. Lett.}~{\bf #1}}

\def\sc#1{{\jo Science}~{\bf #1}}
\begin{document}

\def\vr{{\vec r}}
\def\hMpc{\hmpc}
\def\xic{\xi_c}
\def\ApJ#1{\APJ#1}
\def\ApJIP{{\sl Ap. J.}, in print}

\author[J. R. Primack]{Joel R. PRIMACK\\University of California, Santa
Cruz}
\chapter{Dark Matter and Structure Formation} 
\label{chap-pr}

\begin{abstract}
This chapter aims
to present an introduction to current research on
the nature of the cosmological dark matter and the origin of galaxies
and large scale structure within the standard theoretical framework:
gravitational collapse of fluctuations as the origin of structure in
the expanding universe.  General relativistic cosmology is summarized,
and the data on the basic cosmological parameters ($t_o$ and $H_0
\equiv 100 h \kmsmpc$, $\Omega_0$, $\Omega_\Lambda$
and $\Omega_b$) are reviewed.
Various particle physics candidates for hot, warm, and cold dark matter are
briefly reviewed, together with current constraints and experiments
that could detect or eliminate them.  Also included is a very brief
summary of the theory of cosmic defects, and a somewhat more extended
exposition of the idea of cosmological inflation with a summary of
some current models of inflation. The remainder is a discussion of
observational constraints on cosmological model building, emphasizing
models in which most of the dark matter is cold and the primordial
fluctuations are the sort predicted by inflation.  It is argued that the
simplest models that have a hope of working are Cold Dark Matter with
a cosmological constant ($\Lambda$CDM) if the Hubble parameter is high
($h \gsim 0.7$), and Cold + Hot Dark Matter (CHDM) if the Hubble parameter
and age permit an $\Omega=1$ cosmology, as seems plausible in light of the
data from the Hipparcos astrometric satellite.  The most
attractive variants of these models and the critical tests for each
are discussed.
\end{abstract}

\noindent
To be published as Chapter 1 of {\it Formation of Structure in the
Universe}, Proceedings of the Jerusalem Winter School 1996,
edited by A. Dekel and J.P. Ostriker (Cambridge University Press).

 \cleardoublepage
 \setcounter{tocdepth}{3}  
 \tableofcontents
 \cleardoublepage

\pagestyle{myheadings}
\markboth{Dark Matter and Structure Formation}{Dark Matter and
Structure Format-on}

\section{Introduction}
\label{sec:pr_intro}

The standard theory of cosmology is the Hot Big Bang, according to
which the early universe was hot, dense, very nearly homogeneous, and
expanding adiabatically according to the laws of general relativity
(GR). This theory nicely accounts for the cosmic background radiation,
and is at least roughly consistent with the abundances of the lightest
nuclides.  It is probably even true, as far as it goes; at least, I
will assume so here. But as a fundamental theory of cosmology, the
standard theory is seriously incomplete.  One way of putting this is
to say that it describes the middle of the story, but leaves us
guessing about both the beginning and the end.

Galaxies and clusters of galaxies are the largest bound systems, and
the filamentary or wall-like superclusters and the voids between them
are the largest scale structures visible in the universe, but their
origins are not yet entirely understood.  Moreover, within the
framework of the standard theory of gravity, there is compelling
observational evidence that most of the mass detected gravitationally
in galaxies and clusters, and especially on larger scales, is ``dark''
--- that is, visible neither in absorption nor emission of any
frequency of electromagnetic radiation.  But we still do not know what
this dark matter is.

Explaining the rich variety and correlations of galaxy and cluster
morphology will require filling in much more of the history of the
universe:
\begin{itemize}
\item {\it Beginnings,} in order to understand the origin of
the fluctuations that eventually collapse gravitationally to form
galaxies and large scale structure.  This is a mystery in the standard
hot big bang universe, because the matter that comprises a typical
galaxy, for example, first came into causal contact about a year after
the Big Bang. It is hard to see how galaxy-size fluctuations
could have formed after that, but even harder to see how they could
have formed earlier.  The best solution to this problem yet
discovered, and the one emphasized here, is cosmic inflation.  The
main alternative, discussed in less detail here, is cosmic topological
defects.
\item {\it Denouement,} since even given appropriate initial
fluctuations, we are far from understanding the evolution of galaxies,
clusters, and large scale structure ---
or even the origins of stars and the stellar initial mass function.
\item
And the {\it dark matter} is probably the key to
unravelling the plot since it appears to be gravitationally dominant on
all scales larger than the cores of galaxies.  The dark matter is therefore
crucial for understanding the evolution and present structure of galaxies,
clusters, superclusters and voids.
\end{itemize}

The present chapter
(updating Primack 1987-88, 1993, 1995-97) concentrates on the
period {\it after} the first three minutes, during which the universe
expands by a factor of $\sim 10^8$ to its present size, and all the
observed structures form.  This is now an area undergoing intense
development in astrophysics, both observationally and theoretically.
It is likely that the present decade will see the construction at last
of a fundamental theory of cosmology, with perhaps profound
implications for particle physics --- and possibly even for broader
areas of modern culture.

The current controversy over the amount of matter in the universe
will be emphasized, discussing especially the two leading alternatives: a
critical-density universe, i.e. with $\Omega_0 \equiv {\bar
\rho_0}/\rho_c = 1$ (see Table 1.1), vs. a low-density universe having
$\Omega_0 \approx 0.3$ with a positive cosmological constant $\Lambda
> 0$ such that $\Omega_\Lambda \equiv \Lambda/(3H_0^2) = 1-\Omega_0$
supplying the additional density required for the flatness predicted
by the simplest inflationary models. (The significance of the
cosmological parameters $\Omega_0$, $H_0$, $t_0$, and $\Lambda$ is
discussed in
\se{pr_cos}.)  
$\Omega=1$ requires that the
expansion rate of the universe, the Hubble parameter $H_0 \equiv 100 h
\kmsmpc \equiv 50 h_{50} \kmsmpc$, be relatively low, $h\lsim 0.6$,
in order that the age of the universe $t_0$ be as large as the minimum
estimate of the age of the stars in the oldest globular clusters.  If
the expansion rate turns out to be larger than this, we will see that
GR then requires that $\Omega_0 < 1$, with a positive cosmological
constant giving a larger age for any value of $\Omega_0$.

Although this chapter will concentrate on the implications of
CDM and alternative theories of dark matter for the development of
galaxies and large scale structure in the relatively ``recent''
universe, we can hardly avoid recalling some of the earlier parts of
the story.  Inflation or cosmic defects will be important in this
context for the nearly constant curvature (near-``Zel'dovich'')
spectrum of primordial fluctuations and as plausible solutions to the
problem of generating these large scale fluctuations without violating
causality; and primordial nucleosynthesis will be important as a
source of information on the amount of ordinary (``baryonic'') matter
in the universe.  The fact that the observational lower bound on
$\Omega_0$ --- namely $0.3 \lsim \Omega_0$ --- exceeds the most
conservative upper limit on baryonic mass $\Omega_b \lsim 0.03 h^{-2}$
from Big Bang Nucleosynthesis (Copi, Schramm, \& Turner 1995; cf. Hata
et al. 1995) is the main evidence that there must be such nonbaryonic
dark matter particles.

Of special concern will be evidence and arguments bearing on
the astrophysical properties of the dark matter, which can also help
to constrain possible particle physics candidates. The most popular of
these are few-eV neutrinos (the ``hot'' dark matter candidate), heavy
stable particles such as $\sim 100$ GeV photinos (or whatever
neutralino is the lightest supersymmetric partner particle) or
$10^{-6}-10^{-3}$ eV ``invisible'' axions (these remain the favorite
``cold'' dark matter candidates), and various more exotic ideas such
as keV gravitinos (``warm'' dark matter) or primordial black holes (BH).

Here we are using
the usual astrophysical classification of the dark matter candidates
into {\it hot, warm,} or {\it cold,} depending on
their thermal velocity in the early universe.  Hot dark matter, such
as few-eV neutrinos, is still relativistic when galaxy-size masses
($\sim 10^{12} M_\odot$) are first encompassed within the horizon.
Warm dark matter is just becoming nonrelativistic then. Cold dark
matter, such as axions or massive photinos, is nonrelativistic when
even globular cluster masses ($\sim 10^6 M_\odot$) come within the
horizon.  As a consequence, fluctuations on galaxy scales are wiped
out by the ``free streaming'' of the hot dark matter particles which
are moving at nearly the speed of light.  But galaxy-size fluctuations
are preserved with warm dark matter, and all cosmologically relevant
fluctuations survive in a universe dominated by the sluggishly moving
cold dark matter.

The first possibility for nonbaryonic dark matter that was examined in
detail was massive neutrinos, assumed to have mass $\sim 25$ eV ---
both because that mass corresponds to closure density for $h\approx
0.5$, and because in the late 1970s the Moscow tritium $\beta$-decay
experiment provided evidence (subsequently contradicted by other
experiments) that the electron neutrino has that mass.  Although this
picture leads to superclusters and voids of roughly the size seen,
superclusters are the first structures to collapse in this theory
since smaller size fluctuations do not survive.  The theory foundered
on this point, however, since galaxies are almost certainly older than
superclusters.  The standard (adiabatic) form of this theory has
recently been ruled out by the COBE data: if the amplitude of the
fluctuation spectrum is small enough for consistency with the COBE
fluctuations, superclusters would just be beginning to form at the
present epoch, and hardly any smaller-scale structures, including
galaxies, could have formed by the present epoch.

A currently popular possibility is that the dark matter is cold. After
Peebles (1982), we were among those who first proposed and worked out the
consequences of the Cold Dark Matter (CDM) model (Primack \&
Blumenthal 1983, 1984; Blumenthal et al. 1984). Its virtues include an
account of galaxy and cluster formation that at first sight appeared
to be very attractive. Its defects took longer to uncover, partly
because uncertainty about how to normalize the CDM fluctuation
amplitude allowed for a certain amount of fudging, at least until COBE
measured the fluctuation amplitude. The most serious problem with CDM
is probably the mismatch between supercluster-scale and galaxy-scale
structures and velocities, which suggests that the CDM fluctuation
spectrum is not quite the right shape --- which can perhaps be
remedied if the dark matter content is a mixture of hot and cold, or
if there is less than a critical density of cold dark matter.

The basic theoretical framework for cosmology is reviewed first,
followed by a discussion of the current knowledge about the
fundamental cosmological parameters.

Table 1.1 lists the values of the most important physical constants
used in
this chapter 
(cf. Barnett et al. 1996).  The distance to distant
galaxies is deduced from their redshifts; consequently, the parameter
$h$ appears in many formulas where the distance matters.

\begin{table}
\caption{Physical Constants for Cosmology}
\label{ta:consts}

\centerline{\vbox{\halign{\ \ #\hfill \quad \qquad &$#$\hfill \ &$#$
&#\hfill \ \ \cr
\noalign{\hrule}
\noalign{\vskip .10in}
parsec           &{\rm pc}  &= &$3.09 \times 10^{18}$ cm =
                                3.26 light years (lyr) \cr
Newton's const.  &G         &= &$6.67 \times 10^{-8}$ dyne cm$^2$ g$^{-2}$ \cr
Hubble parameter &H_0         &= &$100 \,h$ km s$^{-1}$ Mpc$^{-1}$ \ ,
                                \ $1/2 \lsim h \lsim 1$ \cr
Hubble time      &H_0^{-1}    &= &$h^{-1} \ 9.78$ Gyr \cr
Hubble radius    &R_H       &= &$cH^{-1} = 3.00\,h^{-1}$ Gpc \cr
critical density &\rho_c    &= &$3H^2/8\pi G = 1.88 \times 10^{-29} h^2$
                                g cm$^{-3}$ \cr
{ }              &{ }       &= &$10.5\,h^2$ keV cm$^{-3}$ = $2.78 \times 10
                                ^{11} h^2 \, M_\odot \, $ Mpc$^{-3}$ \cr
speed of light   &c         &= &$3.00 \times 10^{10}$ cm s$^{-1}$ =
                                306 Mpc Gyr$^{-1}$ \cr
solar mass       &M_\odot   &= &$1.99 \times 10^{33}$ g \cr
solar luminosity &L_\odot   &= &$3.85 \times 10^{33}$ erg s$^{-1}$ \cr
Planck's const.  &\hbar     &= &$1.05 \times 10^{-27}$ erg s =
                                $6.58 \times 10^{-16}$ eV s \cr
Planck mass      &m_{P\ell} &= &$(\hbar c/G)^{1/2} = 2.18 \times
                                10^{-5}$ g = $1.22 \times 10^{19}$ GeV\cr
\noalign{\vskip .10in}
\noalign{\hrule}
}}}
\end{table}

\section{Cosmology Basics}
\label{sec:pr_cos}

It is assumed here that Einstein's general theory of relativity
(GR) accurately describes gravity.  Although it is important to
appreciate that there is no observational confirmation of this on
scales larger than about 1 Mpc, the tests of GR on smaller scales are
becoming increasingly precise, especially with pulsars in binary star
systems (Will 1981, 1986, 1990; Taylor 1994).  On galaxy and cluster
scales, the general agreement between the mass estimated by velocity
measurements and by gravitational lensing provides evidence supporting
standard gravity.  There are two other reasons most cosmologists
believe in GR: it is conceptually so beautifully simple that it is
hard to believe it could be wrong, and anyway it has no serious
theoretical competition. Nevertheless, since a straightforward
interpretation of the available data in the context of the standard
theory of gravity leads to the disquieting conclusion that most of the
matter in the universe is dark, there have been suggestions that
perhaps our theory of gravity is inadequate on large scales.
They are mentioned briefly at the end of this section.

The ``Copernican'' or ``cosmological'' principle is logically
independent of our theory of gravity, so it is appropriate to state it
before discussing GR further.  First, some definitions are necessary:
\begin{itemize}
\item A {\it co-moving observer} is at rest and
unaccelerated with respect to nearby material (in practice, with
respect to the center of mass of galaxies within, say, 100 $h^{-1}$
Mpc).
\item The universe is {\it homogeneous} if all co-moving
observers see identical properties.
\item The universe is {\it isotropic} if all co-moving
observers see no preferred direction.
\end{itemize}

The {\it cosmological principle} asserts that the universe is
homogeneous and isotropic on large scales. (Isotropy about at least
three points actually implies homogeneity, but the counterexample of a
cylinder shows that the reverse is not true.) In reality, the matter
distribution in the universe is exceedingly inhomogeneous on small
scales, and increasingly homogeneous on scales approaching the entire
horizon.  The cosmological principle is in practice the assumption
that for cosmological purposes we can neglect this inhomogeneity, or
treat it perturbatively.  This has now been put on an improved basis,
based on the observed isotropy of the cosmic background radiation and
the (partially testable) Copernican assumption that other observers
also see a nearly homogeneous CBR.  The ``COBE-Copernicus'' theorem
(Stoeger, Maartens, \& Ellis 1995; Maartens, Ellis, \& Stoeger 1995;
reviewed by Ellis 1996) asserts that if all comoving observers measure
the cosmic microwave background radiation to be almost isotropic in a
region of the expanding universe, then the universe is locally almost
spatially homogeneous and isotropic in that region.

The great advantage of assuming homogeneity is that our own cosmic
neighborhood becomes representative of the whole universe, and the
range of cosmological models to be considered is also enormously
reduced. The cosmological principle also implies the existence of a
universal cosmic time, since all observers see the same sequence of
cosmic events with which to synchronize their clocks. (This assumption
is sometimes explicitly included in the statement of the cosmological
principle; e.g., Rindler (1977), p.~203.) In particular, they can all
start their clocks with the Big Bang.

Astronomers observe that the redshift
$z \equiv (\lambda-\lambda_0)/\lambda_0$
of distant galaxies is proportional to their distance.  We
assume, for lack of any viable alternative explanation, that
this redshift is due to the expansion of the universe.
Recent evidence for this includes higher CBR temperature at higher
redshift (Songaila et al. 1994b) and time dilation of high-redshift
Type~Ia supernovae (Goldhaber et al. 1996).
The cosmological principle then implies (see, for example,
Rowan-Robinson 1981, \S 4.3) that the expansion is homogeneous:
$r = a(t) r_0$, which immediately implies Hubble's law:
$v = \dot r = {\dot a} a^{-1} r = H_0 r$.
Here $r_0$ is the present distance of some distant galaxy
(the subscript ``$0$'' in cosmology denotes the present era),
$r$ is its distance as a function of time and $v$ is its
velocity, and $a(t)$ is the scale factor of the expansion
(scaled to be unity at the present: $a(t_0)=1$).
The scale factor is related to the redshift by $a=(1+z)^{-1}$.
Hubble's ``constant'' $H(t)$ (constant in space, but a function of
time except in an empty universe) is
$ H(t) = {\dot a} a^{-1}$.

Finally, it can be shown (see, e.g., Weinberg 1972, Rindler 1977) that
the most general metric satisfying
the cosmological principle is the Robertson-Walker metric
\begin{equation}
ds^2 = c^2 dt^2 - a(t)^2 \left[{{dr^2}\over{1-k r^2}} +
     r^2 (\sin^2 \theta d\phi^2 + d\theta^2) \right] ,
\end{equation}
where the curvature constant $k$, by a suitable choice of
units for $r$, has the value 1,0, or -1, depending on
whether the universe is closed, flat, or open, respectively.
For $k=1$ the spatial universe can be regarded as the
surface of a sphere of radius $a(t)$ in four-dimensional
Euclidean space; and although for $k=0$ or $-1$ no such
simple geometric interpretation is possible, $a(t)$ still
sets the scale of the geometry of space.

\def\bigint{{\displaystyle \int}}

\begin{table*}[!]
\caption{Theoretical Framework: GR Cosmology}
\label{GReqns}

\makeatletter
\def\setboxz@h{\setbox\z@\hbox}
\def\boxed#1{\setboxz@h{$\m@th\displaystyle{#1}$}\dimen@.4\ex@
 \advance\dimen@3\ex@\advance\dimen@\dp\z@
 \hbox{\lower\dimen@\hbox{%
 \vbox{\hrule height.4\ex@
 \hbox{\vrule width.4\ex@\hskip3\ex@\vbox{\vskip3\ex@\boxz@\vskip3\ex@}%
 \hskip3\ex@\vrule width.4\ex@}\hrule height.4\ex@}%
 }}}
\makeatother
\def\nicefrac#1#2{{\textstyle{#1\over #2}}}
\def\bigfrac#1#2{{\displaystyle{#1\over #2}}}
\parindent=0pt
\def\9{\hphantom 0}
\font\caps=cmcsc10

\vskip1pc

\hrule
\vskip .10in

{\caps
\renewcommand{\tabcolsep}{1pc}
\begin{tabular}{@{}ll@{\hspace{5pc}}l}
GR:& Matter tells space\qquad&  Curved space tells    \\
   & how to curve,           &  matter how to move.

\end{tabular}
}

\vskip1.5pc
 $(E)\quad R^{\mu\nu} -{1\over 2} Rg^{\mu\nu} =
           -8\pi G\,T^{\mu\nu}-\Lambda g^{\mu\nu}$
\vskip2pc
COBE - Copernicus Th:  If all observers measure nearly isotropic
     CBR, then universe is locally nearly homogeneous and isotropic --
   i.e., nearly FRW.
$$
\setlength{\arraycolsep}{0pt}
\begin{array}{ll}
{\rm FRW\ E}(00) \quad\bigfrac{\dot a^2}{a^2} = \bigfrac{8\pi}{3} G\rho
       - \bigfrac{k}{a^2} + \bigfrac\Lambda3&  \\[2pc]
           & \hspace*{2pc} H_0 \equiv 100 h\,{\rm km\,s}^{-1} {\rm Mpc}^{-2}\\[-2.5pc]
{\rm FRW\ E}(ii) \quad\bigfrac{2\ddot a}{a} + \bigfrac{\dot a^2}{a^2}
      = -8\pi Gp - \bigfrac{k}{a^2} +\Lambda & \\[10pt]
        & \hspace*{2pc}\phantom{H_0} \equiv \950 h_{50}\,{\rm km\,s}^{-1} {\rm Mpc}^{-2}
\end{array}
$$
\vspace*{-1pc}
$$
\setlength{\arraycolsep}{0pt}
\begin{array}{@{}lll}
 \bigfrac{E(00)}{H^2_0} \Rightarrow \;
         1=\Omega_0 - \bigfrac{k}{H^2_0} + \Omega_\Lambda \;
{\rm with}\; & H_0  \equiv {\dot a_0\over a_0},\ a_0\equiv 1,\
                     \Omega_0\equiv {\rho_0\over\rho_c},
    \Omega_\Lambda  \equiv {\Lambda\over3H^2_0},\\
    &\rho_c  \equiv {3H^2_0\over8\pi G} = 0.70\times 10^{11} h^2_{50}
                              M_\odot {\rm Mpc}^{-3}\\[3pt]
\end{array}
$$
$E(ii)-E(00) \Rightarrow \bigfrac{2\ddot a}{a} = -\bigfrac{8\pi}{3}G\rho
                 -8\pi Gp + \bigfrac23\Lambda$
\vskip6pt
Divide by $2E(00) \Rightarrow \quad  q_0 \equiv -\left(\bigfrac{\ddot a}a\
       \bigfrac{a^2}{\dot a^2}\right)_0 = \bigfrac{\Omega_0}2
            -\Omega_\Lambda$
$$
\setlength{\arraycolsep}{0pt}
\begin{array}{lll}
E(00) \Rightarrow &t_0 = \bigint^1_0 \bigfrac{\delta a}a
                     \left[\bigfrac{8\pi}3 G\rho
       - \bigfrac k{a^2} + \bigfrac{\Lambda}3 \right]^{1\over2}\!\! =
 H^{-1}_0 \bigint^1_0\!\bigfrac{\delta a}a
                        \!\left[\bigfrac{\Omega_0}{a^3}\right.
-&\bigfrac k{H^2_0a^2} + \Omega_\Lambda\biggl.\biggr]^{-{1\over2}} \\[1pc]
  &t_0  = H^{-1}_0  f(\Omega_0,\Omega_\Lambda)  \hskip1cm
                  H^{-1}_0 = 9.78 h^{-1} {\rm G\,yr}
 &    f(1,0)  = \nicefrac23        \\
&&    f(0,0)  = 1              \\
&&    f(0,1)  = \infty \qquad
\end{array}
$$

$[E(00)a^3]'$ vs.~$E(ii) \Rightarrow \quad\bigfrac\partial{\partial a}
       (\rho a^3) = -3p a^2  $ (``continuity'')

\vspace{1pc}
Given eq. of state $p = p(\rho)$,\quad integrate to determine $\rho(a)$,\\
\hspace*{4.9cm}                   integrate $E(00)$ to determine $a(t)$

\vspace{1pc}
Examples:
\vspace*{-1.8pc}
\begin{eqnarray*}
p &=& 0 \Rightarrow \rho = \rho_0 a^{-3}\;  {\rm (assumed\ above\ in}\
       q_0,\ t_0\ {\rm eqs.})  \\
p &=& {\rho\over 3},\ k=0 \Rightarrow \rho \propto a^{-4}
\end{eqnarray*}

\vskip .10in
\hrule

\end{table*}

Formally, GR consists of the assumption of the Equivalence Principle
(or the Principle of General Covariance) together with Einstein's
field equations, labeled (E) in Table 1.2, where
the key equations have been collected.
The Equivalence Principle implies that spacetime is
locally Minkowskian and globally (pseudo-)Riemannian, and the field
equations specify precisely how spacetime responds to its contents.
The essential physical idea underlying GR is that spacetime is not
just an arena, but rather an active participant in the dynamics, as
summarized by John Wheeler: ``Matter tells space how to curve, curved
space tells matter how to move.''

{\it Comoving coordinates} are coordinates with respect to which
comoving observers are at rest.  A comoving coordinate system expands
with the Hubble expansion.  It is convenient to specify linear
dimensions in comoving coordinates scaled to the present; for example,
if we say that two objects were 1 Mpc apart in comoving coordinates at
a redshift of $z=9$, their actual distance then was 0.1 Mpc.  In a
non-empty universe with vanishing cosmological constant, the case
first studied in detail by the Russian cosmologist Alexander Friedmann
in 1922-24, gravitational attraction ensures that the expansion rate
is always decreasing.  As a result, the Hubble radius $R_H(t)  \equiv
c H(t)^{-1}$ is increasing.  The Hubble radius of a non-empty
Friedmann universe expands even in comoving coordinates. Our backward
lightcone encompasses more of the universe as time goes on.

\subsection{Friedmann-Robertson-Walker Universes}
\label{sec:pr_cos_frw}

For a homogeneous and isotropic fluid of density $\rho$ and pressure
$p$ in a homogeneous universe with curvature $k$ and cosmological
constant $\Lambda$, Einstein's system of partial differential
equations reduces to the two ordinary differential equations labeled
in Table 1.2 FRW E(00) and E(ii), for the diagonal time and spatial
components (see, e.g., Rindler 1977, \S 9.9). Dividing E(00) by
$H_0^2$, and subtracting E(00) from E(ii) puts these equations into
more familiar forms.  Dividing the latter by 2E(00) and evaluating all
expressions at the present epoch then gives the familiar expression
for the deceleration parameter $q_0$ in terms of $\Omega_0$ and
$\Omega_\Lambda$.

Multiplying E(00) by $a^3$, differentiating with respect to $a$, and
comparing with E(ii) gives the equation of continuity. Given an
equation of state $p=p(\rho)$, this equation can be integrated to
determine $\rho(a)$; then E(00) can be integrated to determine $a(t)$.

Consider, for example, the case of vanishing pressure $p=0$, which is
presumably an excellent approximation for the present universe since
the contribution of radiation and massless neutrinos (both having
$p=\rho c^2/3$) to the mass-energy density is at the present epoch
much less than that of nonrelativistic matter (for which $p$ is
negligible). The continuity equation reduces to
$ (4\pi/3)\rho a^3 = M
= {\rm constant} $, and E(00) yields {\it Friedmann's equation}
\begin{equation}
 {\dot a}^2 = {{2 G M}\over a}  -k c^2 + {{\Lambda c^2 a^2}\over 3} .
\end{equation}
This gives an expression for the age of the universe $t_0$ which can
be integrated in general in terms of elliptic functions, and for
$\Lambda=0$ or $k=0$ in terms of elementary functions (cf. standard
textbooks, e.g. Peebles 1993, \S 13, and Felton \& Isaacman 1986).


\begin{figure}[!htbp]
\vskip6pt
\centering
\centerline{\psfig{file=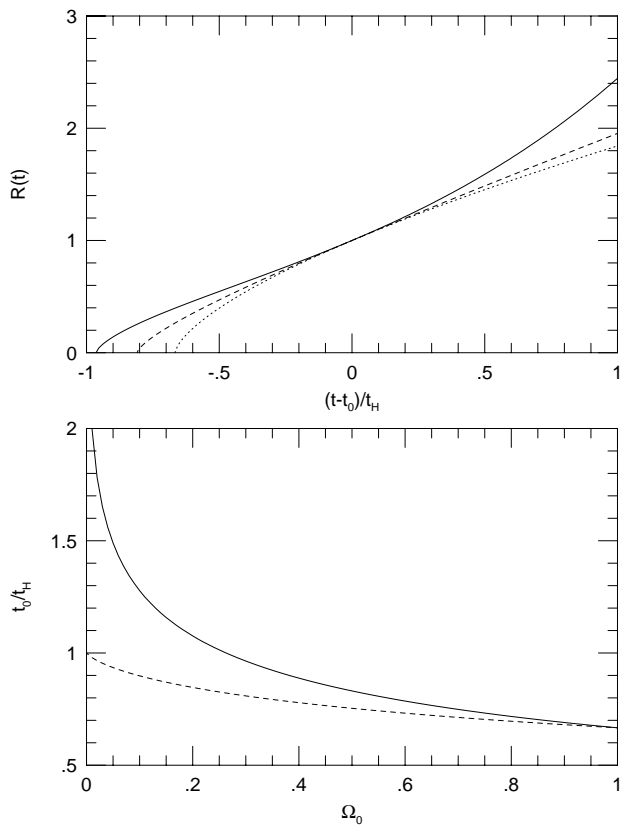,width=10cm}}
\caption{(a) Evolution of the scale factor $a(t)$ plotted vs. the time
after the present $(t-t_0)$ in units of Hubble time $t_H \equiv
H_0^{-1} = 9.78 h^{-1}$ Gyr for three different cosmologies:
Einstein-de Sitter ($\Omega_0=1$,$\Omega_\Lambda=0$ dotted curve),
negative curvature ($\Omega_0=0.3$,$\Omega_\Lambda=0$: dashed curve),
and low-$\Omega_0$ flat ($\Omega_0=0.3$, $\Omega_\Lambda=0.7$: solid
curve).  (b) Age of the universe today $t_0$ in units of Hubble time
$t_H$ as a function of $\Omega_0$ for $\Lambda=0$ (dashed curve) and
flat $\Omega_0 + \Omega_\Lambda=1$ (solid curve) cosmologies.  }
\vskip-2pc
\label{fig:pr_1}
\end{figure}


\Fig{pr_1} (a) plots the evolution of the scale factor $a$ for three
interesting examples: ($\Omega_0$,$\Omega_\Lambda$) = (1,0), (0.3,0),
and (0.3,0.7).  \Fig{pr_1} (b) shows how $t/t_H$ depends on $\Omega_0$
both for $\Lambda=0$ (dashed) and $\Omega_\Lambda=1-\Omega_0$ (solid).
Notice that for $\Lambda=0$, $t_0/t_H$ is somewhat greater for
$\Omega_0=0.3$ (0.81) than for $\Omega=1$ (2/3), while for
$\Omega_0=1-\Omega_\Lambda=0.3$ it is substantially greater: $t_0/t_H
= 0.96$.  In the latter case, the competition between the attraction
of the matter and the repulsion of space by space represented by the
cosmological constant results in a slowing of the expansion at $a \sim
0.5$; the cosmological constant subsequently dominates, resulting in
an accelerated expansion (negative deceleration $q_0=-0.55$ at the
present epoch), corresponding to an inflationary universe.  In
addition to increasing $t_0$, this
behavior has observational implications that we will explore in
\se{pr_age_lambda}.

\subsection{Is the Gravitational Force $\propto r^{-1}$ at Large $r$?}
\label{sec:pr_cos_mond}

Back to the question whether our conventional theory of gravity is
trustworthy on large scales.  The reason for raising this question is
that interpreting modern observations within the context of the
standard theory leads to the conclusion that at least 90\% of the
matter in the universe is dark.  Moreover, there is no observational
confirmation that the gravitational force falls as $r^{-2}$ on
large scales.

Tohline (1983) pointed out that a modified gravitational force law,
with the gravitational acceleration given by $a'=(G M_{lum}{/
r^2})\left(1+{r/d}\right),$ could be an alternative to dark matter
galactic halos as an explanation of the constant-velocity rotation
curves of spiral galaxies. The mass is written as $M_{lum}$ to
emphasize that there is not supposed to be any dark matter.) Indeed,
this equation implies $v^2=G M_{lum}{/d} ={\rm constant}$ for $r \gg
d$. However, with the distance scale $d$ where the force shifts from
$r^{-2}$ to $r^{-1}$ taken to be a physical constant, the same for all
galaxies, this implies that $M_{lum} \propto v^2$, whereas
observationally $M_{lum} \propto L_B \propto v^\alpha$ with $\alpha
\sim 4$ (``Tully-Fisher law''), where $L_B$ is the galaxy luminosity
in the blue band.

Milgrom (1983, 1994, 1995; cf. Mannheim \& Kazanas 1994) proposed an
alternative idea, that the separation between the classical and
modified regimes is determined by the value of the gravitational
acceleration $a'$ rather than the distance scale $r$.  Specifically,
Milgrom proposed that
\begin{equation}
 a' = G M_{lum} r^{-2}, \quad a' \gg a'_0; \qquad a'^2 = G M_{lum} r^{-2}
a'_0, \quad a \ll a'_0
\end{equation}
where the value of the critical acceleration $a'_0 \approx 8
\times 10^{-8} h^2 {\rm \, cm\, s}^{-2}$ is determined for large spiral
galaxies with $M_{lum} \sim 10^{11} M_\odot$.  (This value for $a'_0$
happens to be numerically approximately equal to $c H_0$.) This
equation implies that $v^4=a'_0 G M_{lum}$ for $a'
\ll a'_0$, which is now consistent with the Tully-Fisher law.

Although Milgrom's proposed modifications of gravity are consistent
with a large amount of data, they are entirely ad hoc. Also, not only
do they not predict the gravitational lensing by galaxies and clusters
that is observed, it has recently been shown (Bekenstein \& Sandars
1994) that in any theory that describes gravity by the usual tensor
field of GR plus one or more scalar fields,  the bending of light
cannot exceed that predicted by GR just including the actual matter.
Thus, if Milgrom's nonrelativistic theory, in which by assumption
there is no dark matter, were extended to any scalar-tensor gravity
theory, the light bending could only be due to the visible mass.
However, the evidence is becoming increasingly convincing that the
mass indicated by gravitational lensing in clusters of galaxies is at
least as large as that implied by the velocities of the galaxies and
the temperature of the gas in the clusters (see e.g. Wu \& Fang 1996).

Moreover, it is difficult to fit an $r^{-1}$ force law into the larger
framework of either cosmology or theoretical physics.  The
cosmological difficulty is that an $r^{-1}$ force never saturates:
distant masses are more important than nearby masses.  Regarding
theoretical physics, all one needs to assume in order to get the
weak-field limit of general relativity is that gravitation is carried
by a massless spin-two particle (the graviton): masslessness implies
the standard $r^{-2}$ force, and then spin two implies coupling to the
energy-momentum tensor (Weinberg 1965). In the absence of an
intrinsically attractive and plausible theory of gravity which leads
to a $r^{-1}$ force law at large distances, it seems to be preferable
by far to assume GR and take dark matter seriously, as done below. But
until the nature of the dark matter is determined --- e.g., by
discovering dark matter particles in laboratory experiments --- it is
good to remember that there may be alternative explanations for the
data.

\section{Age, Expansion Rate, and Cosmological Constant}
\label{sec:pr_age}

\subsection{Age of the Universe $t_0$}
\label{sec:pr_age_age}

The strongest lower limits for $t_0$ come from studies of the stellar
populations of globular clusters (GCs). Standard estimates of the ages
of the oldest GCs are $t_{GC} \approx 15-16$ Gyr (Bolte \& Hogan 1995;
VandenBerg, Bolte, \& Stetson 1996; Chaboyer et al. 1996).
A frequently quoted lower
limit on the age of GCs is $12$ Gyr (Chaboyer et al. 1996), which is
then an even more conservative lower limit on $t_0 = t_{GC} + \Delta
t_{GC}$, where $\Delta t_{GC} \gsim 0.5$ Gyr is the time from the Big
Bang until GC formation. The main uncertainty in the GC age estimates
comes from the uncertain distance to the GCs: a 0.25 magnitude error
in the distance modulus translates to a 22\% error in the derived
cluster age (Chaboyer 1995). (We will come back to this in the next
paragraph.) All the other obvious ways to lower the calculated
$t_{GC}$ have been considered and found to have limited effects, and
many non-obvious ideas have also been explored (VandenBerg et al.
1996).  For example, stellar mass loss is a way of lowering $t_{GC}$
(Willson, Bowen, \& Struck-Marcell 1987), but observations constrain
the reduction in $t_0$ to be less than $\sim 1$ Gyr (Shi 1995, Swenson
1995).  Helium sedimentation during the main sequence lifetime can
reduce stellar ages by $\sim1$ Gyr (Chaboyer \& Kim 1995, D'Antona et
al. 1997).  Note that the higher primordial $^4$He abundance implied
by the new Tytler et al. (1996) D/H lowers the central value of the
GC ages by perhaps 0.5 Gyr. The usual conclusion has been that $t_0
\approx 12$ Gyr is probably the lowest plausible value for $t_0$,
obtained by pushing many but not all the parameters to their limits.

However, in spring of 1997, analyses of data from the Hipparcos
astrometric satellite have indicated that the distances to GCs assumed
in obtaining the ages just discussed were systematically
underestimated.  If this is true, it follows that their stars at the
main sequence turnoff are brighter and therefore younger.  Indeed,
there are indications that this correction will be largest for the
lowest-metallicity clusters that had the oldest ages according to the
standard analysis, according to Reid (1977).  His analysis, using a
sample including 15 metal-poor stars with parallaxes determined to
better than 12\% accuracy to redefine the subdwarf main sequence,
gives distance moduli $\sim 0.3$ magnitudes ($\sim30$\%) brighter than
current standard values for his four lowest-metallicity GCs (M13, M15,
M30, and M92), and ages ({\it not} lower limits) of $\sim 12$ Gyr. The
shapes of the theoretical isochrones (Bergbusch \& Vandenberg 1992)
used in previous GC age estimates (e.g., Bolte \& Hogan 1995,
Sandquist et al. 1996) are no longer acceptable fits to the subdwarf
data with the revised distances, although the isochrones of D'Antona
et al. (1997) give better fits to the local subdwarfs and to the GCs.
Another analysis (Gratton et al. 1997) uses a sample including 11
low-metallicity non-binary subdwarf stars with Hipparcos parallaxes
better than 10\% and accurate metal abundances from high-resolution
spectroscopy to determine the absolute location of the main sequence
as a function of metallicity. They then derive ages for the old GCs
(M13, M68, M92, NGC288, NGC6752, 47 Tuc) in their GC sample of
$12.1^{+1.2}_{-3.6}$ Gyr. Their ages are lower both because of their
0.2 mag brighter distance moduli and because of their better metal
determinations of cluster and field stars.

There are systematic effects that must be taken into account in the
accurate determination of $t_{GC}$, including metallicity dependence
and reddening corrections, and various physical phenomena such as
stellar convection and helium sedimentation whose inclusion could
lower ages still further and perhaps also bring theoretical isochrones
into better agreement with the GC observations. Thus, there may be a
period during which additional data is sought and theoretical models
are revised before a new consensus emerges regarding the GC ages.  But
it does appear that the older estimates $t_{GC} \approx 15-16$ Gyr
will be revised downward substantially.  For example, in light of the
new Hipparcos data, Chaboyer et al. 1997 have redone their Monte Carlo
analysis of the effects of varying various uncertain parameters, and
obtained $t_{GC} = 11.7 \pm 1.4$ Gyr ($1\sigma$).

Stellar age estimates are also relevant to another sort of argument
for an old, low-density universe: observation of apparently old
galaxies at moderately high redshift (Dunlop et al. 1996).  In the
most extreme example presented so far (Spinrad et al. 1997), galaxy
LDBS 53W091 at redshift $z=1.55$ has a rest-frame spectrum very
similar to that of an F6 star, and the claimed minimum age of 3.5 Gyr
is based on standard stellar evolution models and assumptions about
stellar populations, reddening, etc.  The authors point out that for
3.5 Gyr to have elapsed at $z=1.55$ requires $h<0.45$ for $\Omega=1$.
(Note that the constraint on $h$ is sensitive to the claimed age of
the galaxy.   From \Fig{pr_1} (a), the age of an Einstein-de Sitter
universe at $z=1.55$ is $1.60h^{-1}$ Gyr, so for 3.0 Gyr to have
elapsed by $z=1.55$ in this cosmology imposes the less restrictive
requirement $h<0.53$, for example.) Observations of old galaxies at
high redshift will certainly constrain cosmological parameters,
especially if the assumptions that go into the analysis can be
independently verified.  However, in this case an independent analysis
(Bruzual \& Magris 1997) of the same data gives a much younger age of
1 to 2 Gyr for LDBS 53W091 (an age of 2 Gyr at $z=1.55$ poses no
problem for an $\Omega=1$ cosmology as long as $h<0.8$), which these
authors regard as more reliable since, unlike the earlier authors,
they can explain all the spectral and color data.

Stellar age estimates are of course based on standard stellar evolution
calculations.  But the solar neutrino problem reminds us that we are
not really sure that we understand how even our nearest star operates;
and the sun plays an important role in calibrating stellar evolution,
since it is the only star whose age we know independently (from
radioactive dating of early solar system material).  An important
check on stellar ages can come from observations of white dwarfs in
globular and open clusters (cf. Richer et al. 1995).  And the two
detached eclipsing binaries at the main sequence turn-off point
recently discovered in Omega Centauri can be used both to measure the
distance to this globular cluster accurately, and to determine their
ages using the mass-luminosity relation (Paczynski 1996).

What if the GC age estimates are wrong for some unknown reason? The
only other non-cosmological estimates of the age of the universe come
from nuclear cosmochronometry --- radioactive decay and chemical
evolution of the Galaxy --- and white dwarf cooling.  Cosmochronometry
age estimates are sensitive to a number of uncertain issues such as
the formation history of the disk and its stars, and possible actinide
destruction in stars (Malaney, Mathews, \& Dearborn 1989; Mathews \&
Schramm 1993). However, an independent cosmochronometry age estimate
of $13.8\pm3.7$ Gyr has been obtained for a single
ultra-low-metallicity star ([Fe/H]=-3.1), based on the measured
depletion of thorium (whose half-life is 14.2 Gyr) compared to stable
heavy r-process elements (Cowan et al. 1997; cf. Bolte 1997, Sneden et
al. 1996).  This method will become very important if it is possible
to obtain accurate measurements of r-process elements for a number of
very low metallicity stars, and the resulting age estimates are
consistent.

Independent age estimates come from the cooling of white
dwarfs in the neighborhood of the sun. The key observation is that
there is a lower limit to the luminosity, and therefore also the
temperature, of nearby white dwarfs; although dimmer ones could have
been seen, none have been found.  The only plausible explanation is
that the white dwarfs have not had sufficient time to cool to lower
temperatures, which initially led to an estimate of $9.3\pm2$ Gyr for
the age of the Galactic disk (Winget et al. 1987).  Since there was
evidence (based on the pre-Hipparcos GC distances) that the stellar
disk of our Galaxy is about 2 Gyr younger than the oldest GCs (e.g.,
Stetson, VandenBerg, \& Bolte 1996), this in turn
gave an estimate of the age of the universe of $t_0 \sim 11\pm2$ Gyr.
More recent analyses (cf. Wood 1992, Hernanz et al. 1994) conclude
that sensitivity to disk star formation history, and to effects on the
white dwarf cooling rates due to C/O separation at crystallization and
possible presence of trace elements such as $^{22}$Ne, allow a rather
wide range of ages for the disk of about $10\pm4$ Gyr. The latest
determination of the white dwarf luminosity function, using white
dwarfs in proper motion binaries, leads to a somewhat lower minimum
luminosity and therefore a somewhat higher estimate of the age of the
disk of $\sim 10.5^{+2.5}_{-1.5}$ Gyr (Oswalt et al. 1996; cf. Chabrier
1997).


\begin{figure}[htb]
\vskip-1pc
\centering
\centerline{\psfig{file=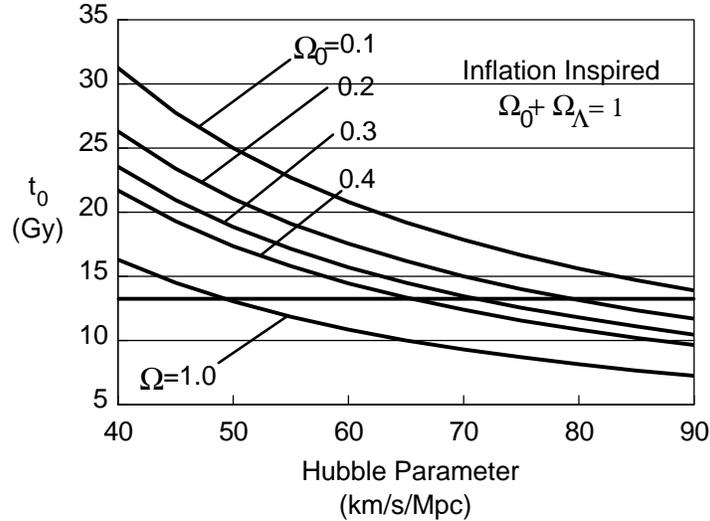,height=8cm}}
\vskip-1pc
\caption{Age of the universe $t_0$ as a function of Hubble parameter
$H_0$ in inflation inspired models with $\Omega_0 + \Omega_\Lambda = 1$,
for several values of the present-epoch cosmological density
parameter $\Omega_0$.}
\label{fig:pr_2}
\end{figure}

Suppose that the old GC stellar age estimates that $t_0 \gsim 13$ Gyr
are right, as we will assume in much of the rest of this chapter.
\Fig{pr_2} shows that $t_0 > 13$ Gyr implies that $h \leq 0.50$
for $\Omega=1$, and that $h \leq 0.73$ even for $\Omega_0 $ as small
as 0.3 in flat cosmologies (i.e., with $\Omega_0 + \Omega_\Lambda =
1$).  However, in view of the preliminary analyses using the new
Hipparcos parallaxes and other new data that give strikingly lower
age estimates for the oldest GCs, we should bear in mind that $t_0$
might actually be as low as $\sim11$ Gyr, which would allow $h$ as
high as 0.6 for $\Omega=1$.

\subsection{Hubble Parameter $H_0$}
\label{sec:pr_age_hubble}

The Hubble parameter $H_0\equiv 100 h$ km s$^{-1}$ Mpc$^{-1}$ remains
uncertain, although by less than the traditional factor of two.
de~Vaucouleurs long contended that $h \approx 1$. Sandage has long
contended that $h \approx 0.5$, and he and Tammann still conclude that
the latest data are consistent with $h=0.55\pm0.05$ (Sandage 1995;
Sandage \& Tammann 1995, 1996; Tammann \& Federspiel 1996).  A
majority of observers currently favor a value intermediate between
these two extremes, and the range of recent determinations has been
shrinking (Kennicutt, Freedman, \& Mould 1995; Tammann 1996; Freedman
1996).

The Hubble parameter has been measured in two basic ways: (1)
Measuring the distance to some nearby galaxies, typically by measuring
the periods and luminosities of Cepheid variables in them; and then
using these ``calibrator galaxies'' to set the zero point in any of
the several methods of measuring the relative distances to galaxies.
(2) Using fundamental physics to measure the distance to some distant
object(s) directly, thereby avoiding at least some of the
uncertainties of the cosmic distance ladder (Rowan-Robinson 1985). The
difficulty with method (1) was that there was only a handful of
calibrator galaxies close enough for Cepheids to be resolved in them.
However, the success of the HST Cepheid measurement of the distance to
M100 (Freedman et al. 1994, Ferrarese et al. 1996) shows that the HST
Key Project on the Extragalactic Distance Scale can significantly
increase the set of calibrator galaxies --- in fact, it already has
done so. Adaptive optics from the ground may also be able to
contribute to this effort, although the first published result of this
approach (Pierce et al. 1994) is not entirely convincing.  The
difficulty with method (2) is that in every case studied so far, some
aspect of the observed system or the underlying physics remains
somewhat uncertain. It is nevertheless remarkable that the results of
several different methods of type (2) are rather similar, and indeed
not very far from those of method (1). This gives reason to hope for
convergence.

\subsubsection{Relative Distance Methods}

One piece of good news is that the several methods of measuring the
relative distances to galaxies now mostly seem to be consistent with
each other (Jacoby et al. 1992; Fukugita, Hogan, \& Peebles 1993).
These methods use either (a) ``standard candles'' or (b) empirical
relations between two measurable properties of a galaxy, one
distance-independent and the other distance-dependent.  (a) The old
favorite standard candle is Type Ia supernovae; a new one is the
apparent maximum luminosity of planetary nebulae (Jacoby et al. 1992).
Sandage et al. (1996) and others (van den Bergh 1995, Branch et al.
1996, cf. Schaefer 1996) get low values of $h\approx0.55$ from HST
Cepheid distances to SN Ia host galaxies, including the seven SNe Ia
with what Sandage et al. characterize as well-observed maxima that lie
in six galaxies for which HST Cepheid distances are now available.
But taking account of an empirical relationship between the SN Ia light
curve shape and maximum luminosity leads to higher $h = 0.65\pm0.06$
(Riess, Press, \& Kirshner 1996) or $h = 0.63\pm0.03$ (Hamuy et al.
1996), although Tammann \& Sandage (1995) disagree that the
increase in $h$ can be so large.
(b) The old favorite empirical relation used as a relative distance
indicator is the Tully-Fisher relation between the rotation velocity
and luminosity of spiral galaxies (and the related Faber-Jackson or
$D_n - \sigma$ relation). A newer one is based on the decrease in the
fluctuations in elliptical galaxy surface brightness on a given
angular scale as comparable galaxies are seen at greater distances
(Tonry 1991); a new SBF survey gives $h = 0.81\pm0.06$ (Tonry et al.
1997).

The ``mid-term'' value of the Hubble constant from the HST key project
is $h=0.73\pm0.10$ (Freedman et al. 1997).  This is based on the
standard distance to the LMC of 50 kpc (corresponding to a distance
modulus of 18.50).  But the preliminary results from the Hipparcos
astrometric satellite suggest that the Cepheid distance scale must be
recalibrated, and that the quoted distance to the LMC is too low by
about 10\% (Feast \& Catchpole 1997, Feast \& Whitelock 1997).  An
increase in the LMC distance of about 7\% is also obtained using the
preliminary Hipparcos recalibration of the zero point and metallicity
dependence of the RR Lyrae distance scale (Gratton et al. 1997, Ried
1997; cf. Alcock et al. 1996b), thus removing a long-standing
discrepancy. The implication is that the Hubble parameter determined
by Cepheid calibrators must be decreased, by perhaps 10\%.  This
applies to the HST key project, and it also applies to the SN Ia
results for $h$, which are based on Cepheid distances; thus, for
example, the Hamuy et al. (1996) value would decrease to about $h =
0.57$, with a corresponding $t_0=11.4$ Gyr for $\Omega=1$.

\subsubsection{Fundamental Physics Approaches}

The fundamental physics approaches involve either Type Ia or Type II
supernovae, the Sunyaev-Zel'dovich (S-Z) effect, or gravitational
lensing.  All are promising, but in each case the relevant physics
remains somewhat uncertain.

The $^{56}$Ni radioactivity method for determining $H_0$ using Type Ia
SN avoids the uncertainties of the distance ladder by calculating the
absolute luminosity of Type Ia supernovae from first principles using
plausible but as yet unproved physical models.  The first result
obtained was that $h=0.61\pm0.10$ (Arnet, Branch, \& Wheeler 1985;
Branch 1992); however, another study (Leibundgut \& Pinto 1992; cf.
Vaughn et al. 1995) found that uncertainties in extinction (i.e.,
light absorption) toward each supernova increases the range of allowed
$h$. Demanding that the $^{56}$Ni radioactivity method agree with an
expanding photosphere approach leads to $h=0.60^{+0.14}_{-0.11}$
(Nugent et al. 1995). The expanding photosphere method compares the
expansion rate of the SN envelope measured by redshift with its size
increase inferred from its temperature and magnitude.  This approach
was first applied to Type II SN; the 1992 result $h=0.6\pm0.1$
(Schmidt, Kirschner, \& Eastman 1992) was subsequently revised upward
by the same authors to $h=0.73\pm0.06\pm0.07$ (1994). However, there
are various complications with the physics of the expanding envelope
(Ruiz-Lapuente et al. 1995; Eastman, Schmidt, \& Kirshner 1996).

The S-Z effect is the Compton scattering of microwave background
photons from the hot electrons in a foreground galaxy cluster.  This
can be used to measure $H_0$ since properties of the cluster gas
measured via the S-Z effect and from X-ray observations have different
dependences on $H_0$.  The result from the first cluster for which
sufficiently detailed data was available, A665 (at $z=0.182$), was
$h=(0.4-0.5)\pm0.12$ (Birkinshaw, Hughes, \& Arnoud 1991); combining
this with data on A2218 ($z=0.171$) raised this somewhat to
$h=0.55\pm0.17$ (Birkinshaw \& Hughes 1994). Early results from the
ASCA X-ray satellite gave $h=0.47\pm0.17$ for A665 and
$h=0.41^{+0.15}_{-0.12}$ for CL0016+16 ($z=0.545$) (Yamashita 1994). A
few S-Z results have been obtained using millimeter-wave observations
(Wilbanks 1994), and this method may allow more such measurements
soon.  New results for A2218 and A1413 ($z=0.14$) using the Ryle radio
telescope and ROSAT X-ray data gave $h=0.38^{+0.17}_{-0.12}$ and
$h=0.47^{+0.18}_{-0.12}$, respectively (Lasenby 1996).  New results
from the OVRO 5.5m telescope for the four X-ray brightest clusters
give $h=0.54\pm0.14$ (Myers et al. 1997).  Corrections for
the near-relativistic electron motions (Rephaeli 1995) and for lensing
by the cluster (Loeb \& Refregier 1996) may raise these estimates for
$H_0$ a little, but it seems clear that the S-Z results favor a
smaller value than many optical astronomers obtain. However, since the
S-Z measurement of $H_0$ is affected by the isothermality of the
clusters (Roettiger et al. 1997) and the unknown orientation of the
cluster ellipticity with respect to the line of sight, and the errors
in the derived values remain rather large, this lower S-Z $H_0$ can
only become convincing with more detailed observations and analyses of
a significant number of additional clusters. Perhaps this will be
possible within the next several years.

Several quasars have been observed to have multiple images separated
by a few arc seconds; this phenomenon is interpreted as arising from
gravitational lensing of the source quasar by a galaxy along the line
of sight.  In the first such system discovered, QSO 0957+561
($z=1.41$), the time delay $\Delta t$ between arrival at the earth of
variations in the quasar's luminosity in the two images has been
measured to be, e.g., $409\pm23$ days (Pelt et al. 1994), although other
authors found a value of $540\pm12$ days (Press, Rybicki, \& Hewitt
1992).  The shorter $\Delta t$ has now been confirmed by the
observation of a sharp drop in Image A of about 0.1 mag in late
December 1994 (Kundic et al. 1995) followed by a similar drop in Image
B about 405-420 days later (Kundic et al. 1997a). Since $\Delta t
\approx \theta^2 H_0^{-1}$, this observation allows an estimate of the
Hubble parameter, with the early results $h=0.50\pm0.17$ (Rhee 1991),
or $h=0.63\pm0.21$ ($h=0.42\pm0.14$) including (neglecting) dark
matter in the lensing galaxy (Roberts et al. 1991), with additional
uncertainties associated with possible microlensing and unknown matter
distribution in the lensing galaxy and the cluster in which this is
the first-ranked galaxy. Deep images allowed
mapping of the gravitational potential of the cluster (at $z=0.36$)
using weak gravitational lensing, which led to the conclusion that
$h \leq 0.70 (1.1 {\rm yr}/\Delta t)$ (Dahle, Maddox, \& Lilje 1994;
Rhee et al. 1996, Fischer et al. 1997). Detailed
study of the lensed QSO images (which include a jet) constrains the
lensing and implies $h= 0.85(1-\kappa)(1.1 {\rm yr}/\Delta t)<0.85$,
where the upper limit follows because the convergence due to the
cluster $\kappa>0$, or alternatively $h=0.85(\sigma/322\kms)^2(1.1
{\rm yr}/\Delta t)$ without uncertainty concerning the cluster if the
one-dimensional velocity dispersion $\sigma$ in the core of the giant
elliptical galaxy responsible for the lensing can be measured (Grogin
\& Narayan 1996).  The latest results for $h$ from 0957+561, using
all available data, are $h=0.64 \pm 0.13$ (95\% C.L.) (Kundic et al.
1997a), $h=0.62\pm0.07$ (Falco et al. 1997, where the error does not
include systematic errors in the assumed form of the mass distribution
in the lens; uncertainties can also be reduced with new HST images
of the system, allowing improved accuracy in the lens galaxy position).

The first quadruple-image quasar system discovered was PG1115+080.
Using a recent series of observations (Schechter et al. 1997), the
time delay between images B and C has been determined to be about
$24\pm3$ days, or $25^{+3.3}_{-3.8}$ days by an alternative analysis
(BarKana 1997).  A simple model for the lensing galaxy and the nearby
galaxies then leads to $h=0.42\pm0.06$ (Schechter et al. 1997) or
$h=0.41\pm0.12$ (95\% C.L.) (BarKana, private communication), although
higher values for $h$ are obtained by a more sophisticated analysis:
$h=0.60\pm0.17$ (Keeton \& Kochanek 1996), $h=0.52\pm0.14$ (Kundic et
al. 1997b). The results depend on how the lensing galaxy and those in
the compact group of which it is a part are modelled. Such models need
to be constrained by new HST observations, especially of the light
profile in the lensing galaxy, and spectroscopy to better determine
the velocity dispersion of the lensing galaxy and of the group.

Although the most recent time-delay results for $h$ from both lensed
quasar systems are remarkably close, the uncertainty in the $h$
determination by this method remains rather large. But it is
reassuring that this completely independent method gives results
consistent with the other determinations. The time-delay method is
promising (Blandford \& Kundic 1996), and when these systems are
better understood and/or delays are reliably measured in several other
multiple-image quasar systems, such as B1422+231 (Hammer, Rigaut, \&
Angonin-Willaime 1995, Hjorth et al. 1996), or radio Einstein-ring
systems, such as PKS 1830-211 (van Ommen et al. 1995) or B0218+357
(Corbett et al. 1996), that should lead to a more precise and reliable
value for $H_0$.

\subsubsection{Correcting for Virgocentric Infall}

What about the HST Cepheid measurement of $H_0$, giving
$h=0.80\pm0.17$ (Freedman et al. 1994), which received so much
attention in the press? This calculated value is based on neither of
the two methods (A) or (B) above, and it should not be regarded
as being very
reliable. Instead this result is obtained by assuming that M100 is at
the core of the Virgo cluster, and dividing the sum of the recession
velocity of Virgo, about 1100 km s$^{-1}$, plus the calculated
``infall velocity'' of the local group toward Virgo, about 300 km
s$^{-1}$, by the measured distance to M100 of 17.1 Mpc.  (These
recession and infall velocities are both a little on the high side,
compared to other values one finds in the literature.) Adding the
``infall velocity'' is necessary in this method in order to correct
the Virgo recession velocity to what it would be were it not for the
gravitational attraction of Virgo for the Local Group of galaxies, but
the problem with this is that the net motion of the Local Group with
respect to Virgo is undoubtedly affected by much besides the Virgo
cluster --- e.g., the ``Great Attractor.''  For example, in our CHDM
supercomputer simulations (which appear to be a rather realistic match
to observations), galaxies and
groups at about 20 Mpc from a Virgo-sized cluster often have net
outflowing rather than infalling velocities. Note that if the net
``infall'' of M100 were smaller, or if M100 were in the foreground of
the Virgo cluster (in which case the actual distance to Virgo would be
larger than 17.1 Mpc), then the indicated $H_0$ would be smaller.

Freedman et al. (1994) gave an alternative argument that avoids the
``infall velocity'' uncertainty: the relative galaxy luminosities
indicate that the Coma cluster is about six times farther away than
the Virgo cluster, and peculiar motions of the Local Group and the
Coma cluster are relatively small corrections to the much larger
recession velocity of Coma; dividing the recession velocity of the
Coma cluster by six times the distance to M100 again gives
$H_0\approx80$.  However, this approach still assumes that M100 is in
the core rather than the foreground of the Virgo cluster; and in
deducing the relative distance of the Coma and Virgo clusters it
assumes that the galaxy luminosity functions in each are comparable,
which is uncertain in view of the very different environments.  More
general arguments by the same authors (Mould et al. 1995) lead them to
conclude that $h=0.73\pm0.11$ regardless of where M100 lies in the Virgo
cluster. But Tammann et al. (1996), using all the available HST
Cepheid distances and their own complete sample of Virgo spirals,
conclude that $h \approx 0.54$.

\subsubsection{Conclusions on $H_0$}

To summarize, many observers, using mainly relative distance methods,
favor a value $h\approx 0.6-0.8$ although Sandage's group and some
others continue to get $h\approx 0.5-0.6$ and all of these values
may need to be reduced by something like 10\% if the full Hipparcos
data set bears out the preliminary reports discussed above.
Meanwhile the fundamental physics methods typically lead to $h
\approx 0.4-0.7$.  Among fundamental physics approaches, there
has been important recent progress in measuring $h$ via time delays
between different images of gravitationally lensed quasars, with
the latest analyses of both of the systems with measured time
delays giving $h\approx0.6\pm0.1$.

The fact that the fundamental physics measurements giving lower values
for $h$ (via time delays in gravitationally lensed quasars and the
Sunyaev-Zel'dovich effect) are
mostly of more distant objects has suggested to some authors (Turner,
Cen, \& Ostriker 1992; Wu et al. 1996) that the local universe may
actually be underdense and therefore be expanding faster than is
typical.  But in reasonable models where structure forms from Gaussian
fluctuations via gravitational instability, it is extremely unlikely
that a sufficiently large region has a density sufficiently smaller
than average to make more than a rather small difference in the value
of $h$ measured locally (Suto, Suginohara, \& Inagaki 1995;
Shi \& Turner 1997). Moreover,
the small dispersion in the corrected maximum luminosity of distant
Type Ia supernovae found by the LBL Supernova Cosmology Project (Kim
et al. 1997) compared to nearby SNe Ia shows directly that the local
and cosmological values of $H_0$ are approximately equal. The maximum
deviation permitted is about 10\%. Interestingly, preliminary results
using 44 nearby Type Ia supernovae as yardsticks suggest that the
actual deviation is about 5-7\%, in the sense that in our local
region of the universe, out to a radius of about $70 \hmpc$ (the
distance of the Northern Great Wall), $H_0$ is this much larger
than average (A. Dekel, private communication). The combined effect of
this and the Hipparcos correction would, for example, reduce the
``mid-term'' value $h \sim 0.73$ from the HST Key Project on the
Extragalactic Distance Scale, to $h \sim 0.63$.

There has been recent observational progress in both relative distance
and fundamental physics methods,
and it is likely that the
Hubble parameter will be known reliably to 10\% within a few years.
Most recent measurements are consistent with $h=0.6\pm0.1$, corresponding
to a range $t_0= 6.52 h^{-1} {\rm Gyr} = 9.3-13.0$ Gyr for $\Omega=1$ ---
in good agreement with the preliminary estimates of the ages of the
oldest globular clusters based on the new data from the Hipparcos
astrometric satellite.

\subsection{Cosmological Constant $\Lambda$}
\label{sec:pr_age_lambda}

Inflation is the only known solution to the horizon and flatness
problems and the avoidance of too many GUT monopoles.  And inflation
has the added bonus that at no extra charge (except the perhaps
implausibly fine-tuned adjustment of the self-coupling of the inflaton
field to be adequately small), simple inflationary models predict a
near-Zel'dovich primordial spectrum (i.e., $P_p(k) \propto k^{n_p}$
with $n_p\approx1$) of adiabatic Gaussian primordial fluctuations ---
which seems to be consistent with observations. All simple
inflationary models predict that the curvature is vanishingly small,
although inflationary models that are extremely contrived (at least,
to my mind) can be constructed with negative curvature and therefore
$\Omega_0 \lsim 1$ without a cosmological constant (see
\se{pr_origin_open} below).
Thus most authors who consider inflationary models impose the
condition $k=0$, or $\Omega_0 + \Omega_\Lambda =1$ where
$\Omega_\Lambda \equiv \Lambda/(3H_0^2)$.  This is what is assumed in
$\Lambda$CDM models, and it is what was assumed in \fig{pr_2}.
(Note that $\Omega$ is used to refer only to
the density of matter and energy, not including the cosmological
constant, whose contribution in $\Omega$ units is $\Omega_\Lambda$.)

The idea of a nonvanishing $\Lambda$ is commonly considered unattractive.
There is no known physical reason why $\Lambda$
should be so small ($\Omega_\Lambda=1$ corresponds to $\rho_\Lambda
\sim 10^{-12}$ eV$^4$, which is small
from the viewpoint of particle physics),
though there is also no known reason why it should vanish (cf.
Weinberg 1989, 1996).
A very unattractive feature of $\Lambda \ne0$ cosmologies
is the fact that $\Lambda$ must become important only at
relatively low redshift --- why not much earlier or much
later? Also $\Omega_\Lambda \gsim \Omega_0$
implies that the universe has recently entered an
inflationary epoch (with a de Sitter horizon comparable to
the present horizon).  The main motivations for $\Lambda >
0$ cosmologies are (1) reconciling inflation with
observations that seem to imply $\Omega_0 < 1$, and (2)
avoiding a contradiction between the lower limit $t_0 \gsim
13$ Gyr from globular clusters and $t_0=(2/3)H_0^{-1}=6.52
h^{-1}$ Gyr for the standard $\Omega=1$, $\Lambda=0$
Einstein-de Sitter cosmology, if it is really true that $h >
0.5$.

The cosmological effects of a cosmological constant are not difficult
to understand (Lahav et al. 1991; Carroll, Press, \& Turner 1992). In
the early universe, the density of energy and matter is far more
important than the $\Lambda$ term on the r.h.s. of the Friedmann
equation. But the average matter density decreases as the universe
expands, and at a rather low redshift ($z \sim 0.2$ for
$\Omega_0=0.3$) the $\Lambda$ term finally becomes dominant.  If it
has been adjusted just right, $\Lambda$ can almost balance the
attraction of the matter, and the expansion nearly stops: for a long
time, the scale factor $a \equiv (1+z)^{-1}$ increases very slowly,
although it ultimately starts increasing exponentially as the universe
starts inflating under the influence of the increasingly dominant
$\Lambda$ term (see \fig{pr_1}).  The existence of a period during which
expansion slows while the clock runs explains why $t_0$ can be greater
than for $\Lambda=0$, but this also shows that there is an increased
likelihood of finding galaxies at the redshift interval when the
expansion slowed, and a correspondingly increased opportunity for
lensing of quasars (which mostly lie at higher redshift $z \gsim 2$)
by these galaxies.

The frequency of such lensed quasars is about what would be expected
in a standard $\Omega=1$, $\Lambda=0$ cosmology, so this data sets
fairly stringent upper limits: $\Omega_\Lambda \leq 0.70$ at 90\% C.L.
(Maoz \& Rix 1993, Kochanek 1993), with more recent data giving even
tighter constraints: $\Omega_\Lambda < 0.66$ at 95\% confidence if
$\Omega_0 + \Omega_\Lambda =1$ (Kochanek 1996b).  This limit could
perhaps be weakened if there were (a) significant extinction by dust
in the E/S0 galaxies responsible for the lensing or (b) rapid
evolution of these galaxies, but there is much evidence that these
galaxies have little dust and have evolved only passively for
$z \lsim 1$ (Steidel, Dickinson, \& Persson 1994; Lilly et al. 1995;
Schade et al. 1996).  (An alternative analysis by Im, Griffiths, \&
Ratnatunga 1997 of some of the same optical lensing data considered by
Kochanek 1996b leads them to deduce a value $\Omega_\Lambda
=0.64_{-0.26}^{+0.15}$, which is barely consistent with Kochanek's upper
limit.  A recent paper --- Malhotra, Rhodes, \& Turner 1997 ---
presents evidence for extinction of quasars by foreground galaxies and
claims that this weakens the lensing bound to $\Omega_\Lambda<0.9$,
but there is no quantitative discussion in the paper to justify this
claim.  Maller, Flores, \& Primack 1997 shows that edge-on disk
galaxies can lens quasars very effectively, and discusses a case in
which optical extinction is significant.  But the radio observations
discussed by Falco, Kochanek, \& Munoz 1997, which give a $2\sigma$
limit $\Omega_\Lambda < 0.73$, will not be affected by extinction.)

Yet another constraint comes from number counts of bright E/S0
galaxies in HST images (Driver et al. 1996), since as was just mentioned
these galaxies appear to have evolved rather little since $z\sim 1$.
The number counts are just as expected in the $\Omega=1$, $\Lambda=0$
Einstein-de Sitter cosmology.  Even allowing for uncertainties due to
evolution and merging of these galaxies, this data would allow
$\Omega_\Lambda$ as large as 0.8 in flat cosmologies only in the
unlikely event that half the Sa galaxies in the deep HST images were
misclassified as E/S0. This number-count approach may be very
promising for the future, as the available deep HST image data and our
understanding of galaxy evolution both increase.

A model-dependent constraint comes from a detailed simulation of
$\Lambda$CDM (Klypin, Primack, \& Holtzman 1996, hereafter KPH96): a
COBE-normalized model with $\Omega_0=0.3$, $\Omega_\Lambda=0.7$, 
and $h=0.7$ has far too much
power on small scales to be consistent with observations, unless there
is unexpectedly strong scale-dependent antibiasing of galaxies with
respect to dark matter. (This is discussed in more detail in
\se{pr_models_simulations} below.) For $\Lambda$CDM models, the
simplest solution appears to be raising $\Omega_0$, lowering $H_0$,
and tilting the spectrum ($n_{p}<1$), though of course one could
alternatively modify the primordial power spectrum in other ways.

Finally, from a study of their first seven high-redshift Type Ia
supernovae, Perlmutter et al. (1996) deduced that $\Omega_\Lambda
< 0.51$ at the 95\% confidence level.  (This is discussed in more
detail in \se{pr_om_vls}, just below.)

\Fig{pr_2} shows that with $\Omega_\Lambda \leq 0.7$, the cosmological
constant does not lead to a very large increase in $t_0$ compared to
the Einstein-de Sitter case, although it may still be enough to be
significant.  For example, the constraint that $t_0 \ge 13$ Gyr
requires $h \leq 0.5$ for $\Omega=1$ and $\Lambda=0$, but this becomes
$h \leq 0.70$ for flat cosmologies with $\Omega_\Lambda \leq 0.66$.

\section{Measuring $\Omega_0$}
\label{sec:pr_om}

The present author, like many theorists, regards the Einstein-de
Sitter ($\Omega=1$, $\Lambda=0$) cosmology as the most attractive one.
For one thing, there are only three possible constant values for
$\Omega$ --- 0, 1, and $\infty$ --- of which the only one that can
describe our universe is $\Omega=1$.  Also, as will be discussed in
more detail in \se{pr_origin_inflation}, cosmic inflation is the only
known solution for several otherwise intractable problems, and all
simple inflationary models predict that the universe is flat, i.e.
that $\Omega_0 + \Omega_\Lambda=1$.  Since there is no known physical
reason for a non-zero cosmological constant, and as just discussed in
\se{pr_age_lambda} there are strong observational upper limits on it
(e.g., from gravitational lensing), it is often said that inflation
favors $\Omega=1$.

Of course, theoretical prejudice is not necessarily a reliable guide.
In recent years, many cosmologists have favored $\Omega_0 \sim 0.3$,
both because of the $H_0-t_0$ constraints and because cluster and other
relatively small-scale measurements have given low values for
$\Omega_0$. (For a recent summary of arguments favoring low $\Omega_0
\approx 0.2$ and $\Lambda=0$, see Coles \& Ellis 1997; see also the
chapters by Bahcall and Peebles in this book.  A recent review which
notes that larger scale measurements favor higher $\Omega_0$ is Dekel,
Burstein, \& White 1997.) However, in light of the new Hipparcos data,
the $H_0-t_0$ data no longer so strongly disfavor $\Omega=1$.
Moreover, as is discussed in more detail below, the small-scale
measurements are best regarded as lower limits on $\Omega_0$.  At
present, the data does not permit a clear decision whether $\Omega_0
\approx 0.3$ or 1, but there are promising techniques that may give
definitive measurements soon.

\subsection{Very Large Scale Measurements}
\label{sec:pr_om_vls}

Although it would be desirable to measure $\Omega_0$ and $\Lambda$
through their effects on the large-scale geometry of space-time, this
has proved difficult in practice since it requires comparing objects
at higher and lower redshift, and it is hard to separate selection
effects or the effects of the evolution of the objects from those of
the evolution of the universe. For example, Kellermann (1993), using the
angular-size vs. redshift relation for compact radio galaxies,
obtained evidence favoring $\Omega \approx 1$; however, selection
effects may invalidate this approach (Dabrowski, Lasenby, \& Saunders
1995).  To cite another example, in ``redshift-volume'' tests (e.g.
Loh \& Spillar 1986) involving number counts of galaxies per redshift
interval, how can we tell whether the galaxies at redshift $z\sim 1$
correspond to those at $z \sim 0$? Several galaxies at higher redshift
might have merged, and galaxies might have formed or changed
luminosity at lower redshift. Eventually, with extensive surveys of
galaxy properties as a function of redshift using the largest
telescopes such as Keck, it should be possible to perform
classical cosmological tests at least on particular classes of
galaxies --- that is one of the goals of the Keck DEEP project.
Geometric effects are also on the verge of detection in small-angle
cosmic microwave background (CMB) anisotropies (see
\se{pr_om_conclusion} and \se{pr_models_simulations}).

At present, perhaps the most promising technique involves searching
for Type Ia supernovae (SNe Ia) at high-redshift, since these are the
brightest supernovae and the spread in their intrinsic brightness
appears to be relatively small. Perlmutter et al. (1996) have recently
demonstrated the feasibility of finding significant numbers of such
supernovae, but a dedicated campaign of follow-up observations of each
one is required in order to measure $\Omega_0$ by determining how
the apparent brightness of the supernovae depends on their redshift.
This is therefore a demanding project.  It initially appeared that
$\sim 100$ high redshift SNe Ia would be required to achieve a 10\%
measurement of $q_0=\Omega_0/2 -\Omega_\Lambda$.  However, using the
correlation mentioned earlier between the absolute luminosity of a SN
Ia and the shape of its light curve (slower decline correlates
with higher peak luminosity), it now appears possible to reduce the
number of SN Ia required.  The Perlmutter group has now analyzed seven
high redshift SN Ia by this method, with the result for a flat
universe that $\Omega_0=1- \Omega_\Lambda= 0.94^{+0.34}_{-0.28}$, or
equivalently $\Omega_\Lambda= 0.06^{+0.28}_{-0.34}$ ($<0.51$ at the
95\% confidence level) (Perlmutter et al. 1996).  For a $\Lambda = 0$
cosmology, they find $\Omega_0 = 0.88^{+0.69}_{-0.60}$.
In November 1995
they discovered an additional 11 high-redshift SN Ia, and they have
subsequently discovered many more. Other groups, collaborations from ESO
and MSSSO/CfA/CTIO, are also searching successfully for high-redshift
supernovae to measure $\Omega_0$ (Garnavich et al. 1996).  There has
also been recent progress understanding the physical origin of the SN
Ia luminosity--light curve correlation, and in discovering other such
correlations.  At the present rate of progress, a reliable answer may
be available within perhaps a year or two if a consensus emerges from
these efforts.

\subsection{Large-scale Measurements}
\label{sec:pr_om_ls}

$\Omega_0$ has been measured with some precision on a scale of about
$\sim 50 \hMpc$, using the data on peculiar velocities of galaxies,
and on a somewhat larger scale using redshift surveys based on the
IRAS galaxy catalog.  Since the results of all such measurements to
date have been reviewed in detail (see
Dekel 1994, Strauss \& Willick 1995,
and Dekel's chapter in this volume), only brief comments are provided
here.  The ``POTENT''
analysis tries to recover the scalar velocity potential from the
galaxy peculiar velocities.  It looks reliable, since it reproduces
the observed large scale distribution of galaxies --- that is, many
galaxies are found where the converging velocities indicate that there
is a lot of matter, and there are voids in the galaxy distribution
where the diverging velocities indicate that the density is lower than
average. The comparison of the IRAS redshift surveys with POTENT and
related analyses typically give fairly large values for the parameter
$\beta_I \equiv \Omega_0^{0.6}/b_I$ (where $b_I$ is the biasing
parameter for IRAS galaxies), corresponding to $0.3 \lsim \Omega_0
\lsim 3$ (for an assumed $b_I=1.15$).  It is not clear
whether it will be possible to reduce the spread in these
values significantly in the near future --- probably both
additional data and a better understanding of systematic and
statistical effects will be required.

A particularly simple way to deduce a lower limit on $\Omega_0$ from
the POTENT peculiar velocity data was proposed by Dekel \& Rees
(1994), based on the fact that high-velocity outflows from voids are
not expected in low-$\Omega$ models.  Data on just one void indicates
that $\Omega_0 \ge 0.3$ at the 97\% C.L.  This argument is independent
of assumptions about $\Lambda$ or galaxy formation, but of course it
does depend on the success of POTENT in recovering the peculiar
velocities of galaxies.

However, for the particular cosmological models that are at the focus of
this review --- CHDM and $\Lambda$CDM --- stronger constraints are
available. This is because these models, in common with almost all CDM
variants, assume that the probability distribution function (PDF) of
the primordial fluctuations was Gaussian.  Evolution from a Gaussian
initial PDF to the non-Gaussian mass distribution observed today
requires considerable gravitational nonlinearity, i.e. large $\Omega$.
The PDF deduced by POTENT from observed velocities (i.e., the PDF of
the mass, if the POTENT reconstruction is reliable) is far from
Gaussian today, with a long positive-fluctuation tail.  It agrees with
a Gaussian initial PDF if and only if $\Omega$ is about unity or
larger: $\Omega_0 <1$ is rejected at the $2\sigma$ level, and
$\Omega_0 \leq 0.3$ is ruled out at $\ge  4\sigma$ (Nusser \& Dekel
1993; cf. Bernardeau et al. 1995).

\subsection{Measurements on Scales of a Few Mpc}
\label{sec:pr_om_ss}

On smaller length scales, there are many measurements that are
consistent with a smaller value of $\Omega_0$ (e.g. Peebles 1993, esp.
\S 20). For example, the cosmic virial theorem gives $\Omega(\sim 1
h^{-1}\,{\rm Mpc}) \approx 0.15 [\sigma(1 h^{-1} \, {\rm Mpc}) / (300
\, {\rm km} \, {\rm s}^{-1})]^2$, where $\sigma(1 h^{-1} \, {\rm
Mpc})$ here represents the relative velocity dispersion of galaxy
pairs at a separation of $1 h^{-1} \, {\rm Mpc}$.  Although the
classic paper (Davis \& Peebles 1983) which first measured $\sigma(1
h^{-1} \, {\rm Mpc})$ using a large redshift survey (CfA1) got a value
of 340 km s$^{-1}$, this result is now known to be in error since the
entire core of the Virgo cluster was inadvertently omitted
(Somerville, Davis, \& Primack 1996); if Virgo is included, the result
is $\sim 500-600$ km s$^{-1}$ (cf. Mo et al. 1993, Zurek et al. 1994),
corresponding to $\Omega(\sim 1 h^{-1}\,{\rm Mpc}) \approx 0.4-0.6$.
Various redshift surveys give a wide range of values for $\sigma(1
h^{-1} \, {\rm Mpc}) \sim 300-750$ km s$^{-1}$, with the most salient
feature being the presence or absence of rich clusters of galaxies;
for example, the IRAS galaxies, which are not found in clusters, have
$\sigma(1 h^{-1} \, {\rm Mpc}) \approx 320$ km s$^{-1}$ (Fisher et al.
1994), while the northern CfA2 sample, with several rich clusters, has
much larger $\sigma$ than the SSRS2 sample, with only a few relatively
poor clusters (Marzke et al. 1995; Somerville, Primack, \& Nolthenius
1996). It is evident that the $\sigma(1 h^{-1} \, {\rm Mpc})$
statistic is not a very robust one.  Moreover, the finite sizes of
the dark matter halos of galaxies and groups complicates the measurement
of $\Omega$ using the CVT, generally resulting in a significant
underestimate of the actual value (Bartlett \& Blanchard 1996, Suto \&
Jing 1996).

A standard method for estimating $\Omega$ on scales of a few Mpc is
based on applying virial estimates to groups and clusters of galaxies
to try to deduce the total mass of the galaxies including their dark
matter halos from the velocities and radii of the groups; roughly, $G
M \sim r v^2$. (What one actually does is to pretend that all galaxies
have the same mass-to-light ratio $M/L$, given by the median $M/L$ of
the groups, and integrate over the luminosity function to get the mass
density (Kirschner, Oemler, \& Schechter 1979; Huchra \& Geller 1982;
Ramella, Geller, \& Huchra 1989).  The typical result is that
$\Omega(\sim 1 \hMpc) \sim 0.1-0.2$.  However, such estimates are at
best lower limits, since they can only include the mass within the
region where the galaxies in each group can act as test particles.
It has been found in CHDM simulations (Nolthenius, Klypin, \&
Primack 1997) that the effective radius of the dark matter
distribution associated with galaxy groups is typically 2-3 times
larger than that of the galaxy distribution.  Moreover, we find a
velocity biasing (Carlberg \& Couchman 1989) factor in CHDM groups
$b_v^{grp} \equiv v_{\rm gal, rms}/v_{\rm DM, rms} \approx 0.75$,
whose inverse squared enters in the $\Omega$ estimate.  Finally, we
find that groups and clusters are typically elongated, so only part of
the mass is included in spherical estimators. These factors explain
how it can be that our $\Omega=1$ CHDM simulations produce group
velocity dispersions that are fully consistent with those of observed
groups, even with statistical tests such as the median rms internal
group velocity vs. the fraction of galaxies grouped (Nolthenius,
Klypin, \& Primack 1994, 1997).  This emphasizes the point that local
estimates of $\Omega$ are at best lower limits on its true value.

However, a new study by the Canadian Network for Observational
Cosmology (CNOC) of 16 clusters at $z\sim 0.3$ mostly chosen from the
Einstein Medium Sensitivity Survey (Henry et al. 1992) was designed to
allow a self-contained measurement of $\Omega_0$ from a field $M/L$
which in turn was deduced from their measured cluster $M/L$. The result
was $\Omega_0=0.19\pm0.06$ (Carlberg et al. 1997a,c).  These data were
mainly compared to standard CDM models, and they probably exclude
$\Omega=1$ in such models.  But it remains to be seen whether
alternatives such as a mixture of cold and hot dark matter could fit
the data.

Another approach to estimating $\Omega$ from information on relatively
small scales has been pioneered by Peebles (1989, 1990, 1994).  It is
based on using the least action principle (LAP) to reconstruct the
trajectories of the Local Group galaxies, and the assumption that the
mass is concentrated around the galaxies.  This is perhaps a
reasonable assumption in a low-$\Omega$ universe, but it is not at all
what must occur in an $\Omega=1$ universe where most of the mass must
lie between the galaxies.  Although comparison with $\Omega=1$ N-body
simulations showed that the LAP often succeeds in qualitatively
reconstructing the trajectories, the mass is systematically
underestimated by a large factor by the LAP method (Branchini \&
Carlberg 1994). Surprisingly, a different study (Dunn \& Laflamme
1995) found that the LAP method underestimates $\Omega$ by a factor of
4-5 even in an $\Omega_0=0.2$ simulation; the authors say that this
discrepancy is due to the LAP neglecting the effect of ``orphans'' ---
dark matter particles that are not members of any halo. Shaya,
Peebles, and Tully (1995) have recently attempted to apply the LAP to
galaxies in the local supercluster, again getting low $\Omega_0$. The
LAP approach should be more reliable on this larger scale, but the
method still must be calibrated on N-body simulations of both high-
and low-$\Omega_0$ models before its biases can be quantified.

\subsection{Estimates on Galaxy Halo Scales}
\label{sec:pr_om_gal}

A classic paper by Little \& Tremaine (1987) had argued that the
available data on the Milky Way satellite galaxies required that the
Galaxy's halo terminate at about 50 kpc, with a total mass of only
about $2.5 \times 10^{11} M_\odot$.  But by 1991, new data on local
satellite galaxies, especially Leo I, became available, and the
Little-Tremaine estimator increased to $1.25 \times 10^{12} M_\odot$.
A recent, detailed study finds a mass inside 50 kpc of
$(5.4\pm1.3)\times 10^{11} M_\odot$ (Kochanek 1996a).

Work by Zaritsky et al. (1993) has shown that other spiral galaxies
also have massive halos.  They collected data on satellites of
isolated spiral galaxies, and concluded that the fact that the
relative velocities do not fall off out to a separation of at least
200 kpc shows that massive halos are the norm.  The typical rotation
velocity of $\sim 200-250$ km s$^{-1}$ implies a mass within 200 kpc
of $\sim 2\times10^{12} M_\odot$.  A careful analysis taking into
account selection effects and satellite orbit uncertainties concluded
that the indicated value of $\Omega_0$ exceeds 0.13 at 90\% confidence
(Zaritsky \& White 1994), with preferred values exceeding 0.3. Newer
data suggesting that relative velocities do not fall off out to a
separation of $\sim 400$ kpc (Zaritsky et al. 1997) presumably would
raise these $\Omega_0$ estimates.

However, if galaxy dark matter halos are really so extended and
massive, that would imply that when such galaxies collide, the
resulting tidal tails of debris cannot be flung very far.  Therefore,
the observed merging galaxies with extended tidal tails such as NGC
4038/39 (the Antennae) and NGC 7252 probably have halo:(disk+bulge)
mass ratios less than 10:1 (Dubinski, Mihos, \& Hernquist 1996),
unless the stellar tails are perhaps made during the collision process
from gas that was initially far from the central galaxies (J.
Ostriker, private communication, 1996); the latter possibility can be
checked by determining the ages of the stars in these tails.

A direct way of measuring the mass and spatial extent of many galaxy
dark matter halos is to look for the small distortions of distant
galaxy images due to gravitational lensing by foreground galaxies.
This technique was pioneered by Tyson et al. (1984). Though the
results were inconclusive (Kovner \& Milgrom 1987), powerful
constraints could perhaps be obtained from deep HST images or
ground-based images with excellent seeing.  Such fields would also be
useful for measuring the correlated distortions of galaxy images from
large-scale structure by weak gravitational lensing; although a pilot
project (Mould et al. 1994) detected only a marginal signal, a
reanalysis detected a significant signal suggesting that $\Omega_0
\sigma_8 \sim 1$ (Villumsen 1995). Several groups are planning major
projects of this sort.  The first results from an analysis of the
Hubble Deep Field gave an average galaxy mass interior to $20 h^{-1}$
kpc of $5.9^{+2.5}_{-2.7}\times 10^{11} h^{-1} M_\odot$ (Dell'Antonio
\& Tyson 1996).

\subsection{Cluster Baryons vs. Big Bang Nucleosynthesis}
\label{sec:pr_om_cl-baryon}

A review (Copi, Schramm, \& Turner 1995) of Big Bang
Nucleosynthesis (BBN) and observations indicating primordial
abundances of the light isotopes concludes that $0.009h^{-2} \leq
\Omega_b \leq 0.02h^{-2}$ for concordance with all the abundances, and
$0.006h^{-2} \leq \Omega_b \leq 0.03h^{-2}$ if only deuterium is used.
For $h=0.5$, the corresponding upper limits on $\Omega_b$ are 0.08 and
0.12, respectively.  The observations (Songaila et al. 1994a,
Carswell et al. 1994) of a possible deuterium line in a hydrogen cloud
at redshift $z=3.32$ in the spectrum of quasar 0014+813,
indicating a deuterium abundance D/H$\sim 2\times
10^{-4}$ (and therefore $\Omega_b \leq 0.006h^{-2}$), are inconsistent
with D/H observations by Tytler and collaborators (Tytler et
al. 1996, Burles \& Tytler 1996) in systems at $z=3.57$ (toward
Q1937-1009) and at $z=2.504$,
but with a deuterium abundance about ten times lower.  These lower
D/H values are consistent with
solar system measurements of D and $^3$He, and they imply $\Omega_b
h^2=0.024\pm0.05$, or $\Omega_b$ in the range 0.08-0.11 for
$h=0.5$. If these represent the true D/H, then if the earlier
observations were correct they were
most probably of a Ly$\alpha$ forest line.
Rugers \& Hogan (1996) argue that the width of the $z=3.32$ absorption
features is better fit by deuterium, although they admit that only a
statistical sample of absorbers will settle the issue.  There is a new
possible detection of D at $z=4.672$ in the absorption spectrum of QSO
BR1202-0725 (Wampler et al. 1996) and at $z=3.086$ toward Q0420-388
(Carswell 1996), but they can only give upper limits on D/H.
Wampler (1996) and Songaila et al. (1997) claim that Tytler et al. (1996)
have overestimated the HI column density in their system, and therefore
underestimated D/H.  But Burles
\& Tytler (1996) argue that the two systems that they have analyzed
are much more convincing as real detections of deuterium, that their
HI column density measurement is reliable, and that the fact that they
measure the same D/H$\sim 2.4 \times 10^{-5}$ in both systems makes it
likely that this is the primordial value.  Moreover, Tytler, Burles,
\& Kirkman (1996) have recently presented a higher resolution spectrum
of Q0014+813 in which ``deuterium absorption is neither required nor
suggested,'' which would of course completely undercut the argument of
Hogan and collaborators for high D/H.  Finally, the Tytler group has
analyzed their new Keck LRIS spectra of the absorption system toward
Q1937-1009, and they say that the lower HI column density advocated
by Songaila et al. (1997) is ruled out (Burles and Tytler
1997).  Of course, one or two additional high quality
D/H measurements would be very helpful to really settle the issue.

There is an entirely different line of argument that also favors the
higher $\Omega_b$ implied by the lower D/H of Tytler et al. This is
the requirement that the high-redshift intergalactic medium contain
enough neutral hydrogen to produce the observed Lyman$\alpha$ forest
clouds given standard estimates of the ultraviolet ionizing flux from
quasars.  The minimum required $\Omega_b \gsim 0.05 h_{50}^{-2}$
(Gnedin \& Hui 1996, Weinberg et al. 1997) is
considerably higher than that advocated by higher D/H values, but
consistent with that implied by the lower D/H measurements.

Yet another argument favoring the D/H of Tytler et al. is that the D/H
in the local ISM is about $1.6 \times 10^{-5}$ (Linsky et al. 1995,
Piskunov et al. 1997), while the relatively low metallicity of the
Galaxy suggests that only a relatively modest fraction of the
primordial D could have been destroyed (Tosi et al. 1997). It thus
seems that the lower D/H and correspondingly higher $\Omega_b
\approx 0.1 h_{50}^{-2}$ are more likely to be correct, although it is
worrisome that the relatively high value $Y_p\approx 0.25$ predicted
by standard BBN for the primordial $^4$He abundance does not appear to
be favored by the data (Olive et al. 1996, but cf. Sasselov \&
Goldwirth 1995, Schramm \& Turner 1997).

White et al. (1993) have emphasized that X-ray observations of
clusters, especially Coma, show that the abundance of baryons, mostly
in the form of gas (which typically amounts to several times the
mass of the cluster galaxies), is about 20\% of the total cluster mass
if $h$ is as low as 0.5.
For the Coma cluster they find that the baryon fraction within the
Abell radius ($1.5h^{-1}$ Mpc) is
\begin{equation}
f_b \equiv {M_b \over M_{tot}} \geq 0.009+0.050h^{-3/2},
\end{equation}
where the first term comes from the galaxies and the second from gas.
If clusters are a fair sample of both baryons and dark matter, as they
are expected to be based on simulations (Evrard, Metzler, \& Navarro
1996), then this is 2-3 times the amount of baryonic mass expected on
the basis of BBN in an $\Omega=1$, $h\approx 0.5$ universe, though it
is just what one would expect in a universe with $\Omega_0 \approx
0.3$ (Steigman \& Felten 1995). The fair sample hypothesis implies
that
\begin{equation}
\Omega_0 = {\Omega_b \over f_b} = 0.3 \left({\Omega_b \over 0.06}\right)
                                  \left({0.2 \over f_b}\right).
\end{equation}
A recent review of X-ray measurements gas in a sample of clusters
(White \& Fabian 1995)
finds that the baryon mass fraction within about 1 Mpc lies between 10
and 22\% (for $h=0.5$; the limits scale as $h^{-3/2}$), and argues
that it is unlikely that {\bf (a)} the gas could be clumped enough to
lead to significant overestimates of the total gas mass --- the main
escape route considered in White et al. 1993 (cf. Gunn \& Thomas
1996). The gas mass would also be overestimated if large tangled
magnetic fields provide a significant part of the pressure in the
central regions of some clusters (Loeb \& Mao 1994, but cf. Felten
1996); this
can be checked by observation of Faraday rotation of sources behind
clusters (Kronberg 1994). If $\Omega=1$, the alternatives are then
either {\bf (b)} that clusters have more mass than virial estimates
based on the cluster galaxy velocities or estimates based on
hydrostatic equilibrium (Balland \& Blanchard 1995) of the gas at the
measured X-ray temperature (which is surprising since they agree:
Bahcall \& Lubin 1994), {\bf (c)} that the usual BBN estimate of
$\Omega_b$ is wrong, or {\bf (d)} that the fair sample hypothesis is
wrong (for which there is even some observational evidence:
Loewenstein \& Mushotzky 1996). It is interesting that there are
indications from weak lensing that at least
some clusters (e.g., for A2218 see Squires et al. 1996; for this cluster
the mass estimate from lensing becomes significantly higher than that
from X-rays when the new ASCA satellite data, indicating that the
temperature falls at large radii, is taken into account: Loewenstein
1996) may actually have extended halos of dark matter ---
something that is expected to a greater extent if the dark matter is a
mixture of cold and hot components, since the hot component clusters
less than the cold (Kofman et al. 1996).  If so, the number density of
clusters as a function of mass is higher than usually estimated, which
has interesting cosmological implications (e.g. $\sigma_8$ is higher
than usually estimated).  It is of course possible that the solution
is some combination of alternatives (a)-(d).  If none of the
alternatives is right, then the only conclusion left is that $\Omega_0
\approx 0.3$.

Notice that the rather high baryon fraction $\Omega_b \approx 0.1
(0.5/h)^2$ implied by the recent Tytler et al. measurements of low D/H
helps resolve the cluster baryon crisis for  $\Omega=1$  --- it is
escape route (c) above. With the higher $\Omega_b$ implied by the low
D/H, there is now a ``baryon cluster crisis'' for low-$\Omega_0$
models! Even with a baryon fraction at the high end of observations,
$f_b \lsim 0.2 (h/0.5)^{-3/2}$, the fair sample hypothesis with this
$\Omega_b$ implies $\Omega_0 \gsim 0.5 (h/0.5)^{-1/2}$.

Another recent development is the measurement of the cluster baryon
fraction using the Sunyaev-Zel'dovich effect (Myers et al. 1997),
giving $f_b = (0.06 \pm 0.01)h^{-1}$.  For $h \sim 0.5$, this is
considerably lower than the X-ray estimates, and consistent with the
Tytler et al. $\Omega_b$ at the $1\sigma$ level.  The S-Z decrement is
proportional to the electron density $n_e$ in the cluster, while the
X-ray luminosity is proportional to $n_e^2$; thus the S-Z measurement
is likely to be less sensitive to small-scale clumping of the gas. If
this first S-Z result for $f_b$ is confirmed by measurments on
additional clusters, the cluster baryon data will have become an
argument for $\Omega \approx 1$.

\subsection{Cluster Morphology and Evolution}
\label{sec:pr_om_cl-morphology}

{\noindent \it Cluster Morphology.}
Richstone, Loeb, and Turner (1992) showed that clusters are expected
to be evolved --- i.e. rather spherical and featureless --- in
low-$\Omega$ cosmologies, in which structures form at relatively high
redshift, and that clusters should be more irregular in $\Omega=1$
cosmologies, where they have formed relatively recently and are still
undergoing significant merger activity.  There are few known
clusters that seem to be highly evolved and relaxed, and many that
are irregular --- some of which are obviously undergoing mergers now
or have recently done so (see e.g. Burns et al. 1994).  This disfavors
low-$\Omega$ models, but it remains to be seen just how low. Recent
papers have addressed this.  In one (Mohr et al. 1995) a total of 24
CDM simulations with $\Omega=1$ or 0.2, the latter with
$\Omega_\Lambda=0$ or 0.8, were compared with data on a sample of 57
clusters.  The conclusion was that clusters with the observed range of
X-ray morphologies are very unlikely in the low-$\Omega$ cosmologies.
However, these simulations have been criticized because the
$\Omega_0=0.2$ ones included rather a large amount of ordinary matter:
$\Omega_b=0.1$.  (This is unrealistic both because $h \approx 0.8$
provides the best fit for $\Omega_0=0.2$ CDM, but then the standard
BBN upper limit is $\Omega_b<0.02h^{- 2}=0.03$; and also because
observed clusters have a gas fraction of $\sim 0.15(h/0.5)^{-3/2}$.)
Another study (Jing et al. 1995) using dissipationless simulations and
not comparing directly to observational data found that $\Lambda$CDM
with $\Omega_0=0.3$ and $h=0.75$ produced clusters with some
substructure, perhaps enough to be observationally acceptable (cf. Buote
\& Xu 1997).
Clearly, this important issue deserves study with higher resolution
hydrodynamic simulations, with a range of assumed $\Omega_b$, and
possibly including at least some of the additional physics associated
with the galaxies which must produce the metallicity observed in
clusters, and perhaps some of the heat as well.  Better statistics for
comparing simulations to data may also be useful (Buote \& Tsai 1996).

{\noindent \it Cluster Evolution.} There is evidence on the evolution
of clusters at relatively low redshift, both in their X-ray properties
(Henry et al. 1992, Castander et al. 1995, Ebeling et al. 1995) and in
the properties of their galaxies. In particular, there is a strong
increase in the fraction of blue galaxies with increasing redshift
(the ``Butcher-Oemler effect''), which may be difficult to explain in
a low-density universe (Kauffmann 1994). Field galaxies do not appear
to show such strong evolution; indeed, a recent study concludes that
over the redshift range $0.2 \leq z \leq 1.0$ there is no significant
evolution in the number density of ``normal'' galaxies (Steidel,
Dickinson, \& Persson 1994).  This is compatible with the predictions
of various models, including CHDM with two neutrinos sharing a total
mass of about 5 eV (see below), but the dependence of the number of
clusters $n_{cl}$ on redshift can be a useful constraint on theories
(Jing \& Fang 1994, Bryan et al. 1994, Walter \& Klypin 1996, Eke et
al. 1996). Some (e.g., Carlberg et al. 1997b; Bahcall, Fan, \& Cen
1997) have argued that presently available data show less fall off of
$n_{cl}(z)$ with increasing $z$ than expected in $\Omega=1$
cosmologies, and already imply that $\Omega_0 < 1$.  These arguments
are not yet entirely convincing because the cluster data at various
redshifts are difficult to compare properly since they are rather
inhomogeneous, and the data are not compared to a wide enough range of
models (Gross et al. 1997).

\subsection{Early Structure Formation}
\label{sec:pr_om_early}

In linear theory, adiabatic density fluctuations grow linearly with
the scale factor in an $\Omega=1$ universe, but more slowly if
$\Omega<1$ with or without a cosmological constant.  As a result, if
fluctuations of a certain size in an $\Omega=1$ and an $\Omega_0=0.3$
theory are equal in amplitude at the present epoch ($z=0$), then at
higher redshift the fluctuations in the low-$\Omega$ model had higher
amplitude.  Thus, structures typically form earlier in low-$\Omega$
models than in $\Omega=1$ models.

Since quasars are seen at the highest redshifts, they have been used
to try to constrain $\Omega=1$ theories, especially CHDM which because
of the hot component has additional suppression of small-scale
fluctuations that are presumably required to make early structure
(e.g., Haehnelt 1993).  The difficulty is that dissipationless
simulations predict the number density of halos of a given mass as a
function of redshift, but not enough is known about the nature of
quasars --- for example, the mass of the host galaxy --- to allow a
simple prediction of the number of quasars as a function of redshift
in any given cosmological model.  A more recent study (Katz et al.
1994) concludes that very efficient cooling of the gas in early
structures, and angular momentum transfer from it to the dark halo,
allows for formation of {\it at least} the observed number of quasars
even in models where most galaxy formation occurs late (cf.
Eisenstein \& Loeb 1995).

Observers are now beginning to see significant numbers of what may be
the central regions of galaxies in an early stage of their formation
at redshifts $z=3-3.5$ (Steidel et al. 1996; Giavalisco, Steidel, \&
Macchetto 1996) --- although, as with quasars, a danger in using
systems observed by emission is that they may not be typical.  As
additional observations (e.g., Lowenthal et al. 1996) clarify the
nature of these objects, they can perhaps be used to constrain
cosmological parameters and models. (This data is discussed in more
detail in \se{pr_models_chdm-early}.)

Another sort of high redshift object which may hold more promise for
constraining theories is damped Lyman $\alpha$ systems (DLAS).  DLAS
are high column density clouds of neutral hydrogen, generally thought
to be protogalactic disks, which are observed as wide absorption
features in quasar spectra (Wolfe 1993).  They are relatively common,
seen in roughly a third of all quasar spectra, so statistical
inferences about DLAS are possible.  At the highest redshift for which
data was published in 1995, $z=3-3.4$, the density of neutral gas in
such systems in units of critical density was reported to be
$\Omega_{gas} \approx 0.006$, comparable to the total density of visible
matter in the universe today (Lanzetta, Wolfe, \& Turnshek 1995).
Several papers (Mo \& Miralda-Escude 1994, Kauffmann \& Charlot 1994,
Ma \& Bertschinger 1994) pointed out that the CHDM model with
$\Omega_\nu=0.3$ could not produce such a high $\Omega_{gas}$.
However, it has been shown that CHDM with $\Omega_\nu=0.2$
could do so (Klypin et al. 1995, cf. Ma 1995).  The power spectrum on
small scales is a very sensitive function of the total neutrino mass
in CHDM models.  This theory makes two crucial predictions:
$\Omega_{gas}$ must fall off at higher redshifts, and the DLAS at $z
\gsim 3$ mostly correspond to systems of internal rotation velocity or
velocity dispersion less than about 100 km s$^{-1}$.  This velocity
can perhaps be inferred from the Doppler widths of the metal line
systems associated with the DLAS.  Preliminary reports regarding the
amount of neutral hydrogen in such systems deduced from the latest
data at redshifts above 3.5 appear to be consistent with the first of
these predictions (Storrie-Lombardi et al. 1996). But a possible
problem for the second (Wolfe 1996) is the large velocity widths 
and other statistical properties (Prochaska \& Wolfe 1997) of
the metal line systems associated with the highest-redshift DLAS
(e.g., Lu et al. 1996, at $z=4.4$); if these actually indicate that a
massive disk galaxy is already formed at such a high redshift, and if
discovery of other such systems shows that they are not rare, that
would certainly disfavor CHDM and other theories with relatively
little power on small scales.  However, other interpretations of such
data which would not cause such problems for theories like CHDM are
perhaps more plausible, since they are based on fairly realistic
hydrodynamic simulations (Haehnelt, Steinmetz, \& Rauch 1997; cf.
\se{pr_models_chdm-early}).  More
data will help resolve this question, along with DLAS models including
dust absorption (Pei \& Fall 1995), lensing (Bartelmann \& Loeb
1996, Maller et al. 1997), and effects of star formation (Kauffmann
1996, Somerville et al. 1997).

One of the best ways of probing early structure formation would be to
look at the main light output of the stars of the earliest galaxies,
which is redshifted by the expansion of the universe to wavelengths
beyond about 5 microns today. Unfortunately, it is not possible to
make such observations with existing telescopes; since the atmosphere
blocks almost all such infrared radiation, what is required is a large
infrared telescope in space.  The Space Infrared Telescope Facility
(SIRTF) has long been a high priority, and it will be great to have
access to the data such an instrument will produce when it is launched
sometime in the next decade.  In the meantime, the Near Infrared
Camera/Multi-Object Spectrograph (NICMOS), installed on Hubble
Space Telescope in spring 1997, will help.  Infrared spectrographs
on the largest ground-baed telescopes will also be of great value.

An alternative method is to look for the starlight from the earliest
stars as extragalactic background infrared light (EBL).  Although it
is difficult to see this background light directly because our Galaxy
is so bright in the near infrared, it may be possible to detect it
indirectly through its absorption of TeV gamma rays (via the process
$\gamma \,\gamma \to e^+ \, e^-$). Of the more than twenty active
galactic nuclei (AGNs) that have been seen at $\sim10$ GeV by the
EGRET detector on the Compton Gamma Ray Observatory, only two of the
nearest, Mk421 and Mk501, have also been clearly detected in TeV gamma
rays by the Whipple Atmospheric Cerenkov Telescope (Quinn et al. 1996,
Schubnell et al. 1996). Absorption of $\sim$ TeV gamma rays from
(AGNs) at redshifts $z\sim 0.2$ has been shown to be a sensitive probe
of the EBL and thus of the era of galaxy formation (MacMinn \& Primack
1996; MacMinn, Somerville, \& Primack 1997).

\subsection{Conclusions Regarding $\Omega$}
\label{sec:pr_om_conclusion}

The main issue that has been addressed so far is the value of the
cosmological density parameter $\Omega$.  Arguments can be made
for $\Omega_0 \approx 0.3$ (and models such as $\Lambda$CDM) or for
$\Omega=1$ (for which the best class of models
is probably CHDM), but it is too early to tell which is right.

The evidence would favor a small $\Omega_0 \approx 0.3$ if (1) the
Hubble parameter actually has the high value $H_0 \approx 75$ favored
by many observers, and the age of the universe $t_0 \geq 13$ Gyr; or
(2) the baryonic fraction $f_b=M_b/M_{tot}$ in clusters is actually
$\sim 15$\%, about 3 times larger than expected for standard Big Bang
Nucleosynthesis in an $\Omega=1$ universe. This assumes that standard
BBN is actually right in predicting that the density of ordinary
matter $\Omega_b$ lies in the range $0.009 \leq \Omega_b h^2 \leq
0.02$.  High-resolution high-redshift spectra are now providing
important new data on primordial abundances of the light isotopes that
should clarify the reliability of the BBN limits on $\Omega_b$.  If
the systematic errors in the $^4$He data are larger than currently
estimated, then it may be wiser to use the deuterium upper limit
$\Omega_b h^2 \leq 0.03$, which is also consistent with the value
$\Omega_b h^2\approx 0.024$ indicated by the only clear deuterium
detection at high redshift, with the same D/H$\approx 2.4\times
10^{-5}$ observed in two different low-metallicity quasar absorption
systems (Tytler et al. 1996); this considerably lessens the
discrepancy between $f_b$ and $\Omega_b$.  Another important
constraint on $\Omega_b$ will come from the new data on small angle
CMB anisotropies --- in particular, the location and
height of the first Doppler (or acoustic, or Sakharov)
peak (Dodelson, Gates, \& Stebbins 1996; Jungman et al. 1996; Tegmark
1996), with the latest data consistent with low $h\approx 0.5-0.6$ and
high $\Omega_b h^2 \approx 0.025$. The location of the first Doppler
peak at angular wavenumber $l\approx 250$ indicated by the presently
available data (Netterfield et al. 1997, Scott et al. 1996) is
evidence in favor of a flat universe; $\Omega_0 \lsim 0.5$ with
$\Lambda=0$ is disfavored by this data (Lineweaver \& Barbosa 1997).

The evidence would favor $\Omega=1$ if (1) the POTENT analysis of
galaxy peculiar velocity data is right, in particular regarding
outflows from voids or the inability to obtain the present-epoch
non-Gaussian density distribution from Gaussian initial fluctuations
in a low-$\Omega$ universe; or (2) the preliminary indication of high
$\Omega_0$ and low $\Omega_\Lambda$ from high-redshift Type Ia
supernovae (Perlmutter et al. 1996) is confirmed.

The statistics of gravitational lensing of quasars is incompatible
with large cosmological constant $\Lambda$ and low cosmological
density $\Omega_0$.  Discrimination between models may improve as
additional examples of lensed quasars are searched for in large
surveys such as the Sloan Digital Sky Survey. The era of structure
formation is another important discriminant between these
alternatives, low $\Omega$ favoring earlier structure formation, and
$\Omega=1$ favoring later formation with many clusters and
larger-scale structures still forming today.  A particularly critical
test for models like CHDM is the evolution as a function of redshift
of $\Omega_{gas}$ in damped Ly$\alpha$ systems. Reliable data on all
of these issues is becoming available so rapidly today that there is
reason to hope that a clear decision between these alternatives will
be possible within the next few years.

What if the data ends up supporting what appear to be contradictory
possibilities, e.g. large $\Omega_0$ {\it and} large $H_0$?  Exotic
initial conditions (e.g., ``designer'' primordial fluctuation spectra,
cf. Hodges et al. 1990) or exotic dark matter particles beyond the
simple ``cold'' vs. ``hot'' alternatives discussed in the next section
(e.g., decaying 1-10 MeV tau neutrinos, Dodelson, Gyuk, \& Turner
1994; volatile dark matter, Pierpaoli et al. 1996) could increase the
space of possible inflationary theories somewhat.  But unless new
observations, such as the new stellar parallaxes from the Hipparcos
satellite, cause the estimates of $H_0$ and $t_0$ to be lowered, it
may ultimately be necessary to go outside the framework of
inflationary cosmological models and consider models with large scale
spatial curvature, with a fairly large $\Lambda$ as well as large
$\Omega_0$. This seems particularly unattractive, since in addition to
implying that the universe is now entering a final inflationary
period, it means that inflation probably did not happen at the
beginning of the universe, when it would solve the flatness, horizon,
monopole, and structure-generation problems. Moreover, aside from the
$H_0-t_0$ problem, there is not a shred of reliable evidence in favor
of $\Lambda>0$, just increasingly stringent upper limits. Therefore,
most cosmologists are rooting for the success of inflation-inspired
cosmologies, with $\Omega_0 + \Omega_\Lambda = 1$.  With the new upper
limits on $\Lambda$ from gravitational lensing of quasars, number
counts of elliptical galaxies, and high-redshift Type Ia supernovae,
this means that the cosmological constant is probably too small to
lengthen the age of the universe significantly. So one hopes that when
the dust finally settles, $H_0$ and $t_0$ will both turn out to be low
enough to be consistent with General Relativistic cosmology. But of
course the universe is under no obligation to live up to our
expectations.

\section{Dark Matter Particles}
\label{sec:pr_dm}

\subsection{Hot, Warm, and Cold Dark Matter}
\label{sec:pr_dm_types}

The current limits on the total and baryonic
cosmological density parameters
have been summarized, and it was argued in particular
that $\Omega_0 \gsim0.3$ while $\Omega_b\lsim0.1$.  $\Omega_0 >
\Omega_b$ implies that the majority of the matter in the universe is
not made of atoms.  If the dark matter is not baryonic, what {\it is}
it?  Summarized here are the physical and astrophysical implications of
three classes of elementary particle DM candidates, which are called
hot, warm, and cold.\footnote{Dick Bond suggested this terminology to
me at the 1983 Moriond Conference, where I used it in my talk (Primack
and Blumenthal, 1983).  George Blumenthal and I had thought of this
classification independently, but we used a more complicated
terminology.} Table 1.3 gives a list of dark matter candidates,
classified into these categories.



\begin{table*}
\caption{Dark Matter Candidates}
\label{ta:DMcands}

\parindent=0pt
\def\9{\hphantom 0}

\vskip .10in
\hrule

\makeatletter
\def\Tall#1{{\hbox{$\left.\vbox to 2cm{}\right#1\n@space$}}}
\def\TALL#1{{\hbox{$\left.\vbox to 2.35cm{}\right#1\n@space$}}}
\def\huge#1{{\hbox{$\left.\vbox to 4.25cm{}\right#1\n@space$}}}
\makeatother
\vskip2pc
\begin{center}
\renewcommand{\arraystretch}{1.3}
\begin{tabular}{ccccc}
\begin{tabular}{@{}l@{}}
axion               \\
SUSY LSP neutralino \\
technibaryon        \\
pseudo Higgs         \\
\hspace*{1cm}{\vdots}  \\
shadow matter        \\
topological relics   \\
non-top. solitons   \\ [1.5pc]   
Primordial BH        \\
jupiters            \\
brown dwarfs        \\
white dwarfs        \\
neutron stars       \\
stellar BH          \\
massive BH
\end{tabular} &
\hfill
\renewcommand{\arraystretch}{2}
\begin{tabular}{@{}c@{}}
$ \TALL\}$ \\[1pc]
$ \Tall\}$
\end{tabular} &
\renewcommand{\arraystretch}{1.8}  
\begin{tabular}{@{}l@{}}
\phantom{Weakly}  \\[-6pt]
Weakly          \\
Interacting     \\
Massive         \\
Particles       \\[2.5pc]
Massive         \\
Astrophysical   \\
Compact         \\
Halo            \\
Objects
\end{tabular} &
\renewcommand{\arraystretch}{2.5}
\begin{tabular}{@{}c@{}}
$ \huge\}$
\end{tabular} &
\hfill
\renewcommand{\arraystretch}{3.2}  
\begin{tabular}{@{}p{.2\linewidth}@{}}
COLD            \\
DARK            \\
MATTER
\end{tabular}      \\
\end{tabular}
\end{center}
\vspace*{1cm}
\makeatletter
\def\tall#1{{\hbox{$\left.\vbox to .75cm{}\right#1\n@space$}}}
\makeatother
 \renewcommand{\arraystretch}{1.3}
\begin{tabular}{ccc}
\begin{tabular}{@{}p{.4\linewidth}@{}}
neutrinos $\nu_e\ \nu_\mu\ \nu_\tau\ (\nu_s?)$  \\
majorons?
\end{tabular} &
\begin{tabular}{@{}c@{}}
$ \tall\}$
\end{tabular} &
 \renewcommand{\arraystretch}{1.3}
\begin{tabular}{@{}p{.3\linewidth}@{}}
HOT     \\
DARK MATTER
\end{tabular}
\end{tabular}

\vspace*{1cm}
\begin{tabular}{ccc}
\begin{tabular}{p{.3\linewidth}l}
gravitino \\
right-handed $\nu$ \\[3pt]
decaying dark matter \\
\hspace*{1cm}\vdots
\end{tabular} &
\renewcommand{\arraystretch}{3.2} 
\begin{tabular}{@{}c@{}}
$ \tall\}$ \\
$ \tall\}$
\end{tabular}
\begin{tabular}{@{}p{.3\linewidth}@{}}
WARM DM \\
VOLATILE DM
\end{tabular} \\
\end{tabular}

\vskip .10in
\hrule

\end{table*}

Hot DM refers to low-mass neutral particles that were still in thermal
equilibrium after the most recent phase transition in the hot early
universe, the QCD confinement transition, which took place at $T_{\rm
QCD} \approx 10^2$ MeV. Neutrinos are the standard example of hot dark
matter, although other more exotic possibilities such as ``majorons''
have been discussed in the literature.  Neutrinos have the virtue that
$\nu_e$, $\nu_\mu$, and $\nu_\tau$ are known to exist, and as summarized
in \se{pr_dm_hot}
there is experimental evidence that at least some
of these neutrino species have mass, though the evidence is not yet
really convincing.  Hot DM particles have a cosmological number
density roughly comparable to that of the microwave background
photons, which implies an upper bound to their mass of a few tens of
eV: $m(\nu) = \Omega_\nu \rho_0/n_\nu = \Omega_\nu 92 h^2$ eV. Having
$\Omega_\nu \sim 1$
implies that free streaming destroys any adiabatic fluctuations
smaller than supercluster size, $\sim10^{15}M_\odot$ (Bond,
Efstathiou, \& Silk 1980).  With the COBE upper limit, HDM with
adiabatic fluctuations would lead to hardly any structure formation at
all, although Hot DM plus some sort of seeds, such as cosmic strings
(see, e.g., Zanchin et al. 1996),
might still be viable.  Another promising possibility is Cold + Hot DM
with $\Omega_\nu \sim 0.2$ (CHDM, discussed in some detail below).

Warm DM particles interact much more weakly than neutrinos. They
decouple (i.e., their mean free path first exceeds the horizon size)
at $T \gg T_{\rm QCD}$, and they are not heated by the subsequent
annihilation of hadronic species.  Consequently their number density
is expected to be roughly an order of magnitude lower, and their mass
an order of magnitude higher, than hot DM particles. Fluctuations as
small as large galaxy halos, $\gsim 10^{11}M_\odot$, could then
survive free streaming. Pagels and Primack (1982) initially suggested
that, in theories of local supersymmetry broken at $\sim 10^6$ GeV,
gravitinos could be DM of the warm variety. Other candidates have also
been proposed, for example light right-handed neutrinos (Olive \&
Turner 1982).  Warm dark matter does not lead to structure formation
in agreement with observations, since the mass of the warm particle
must be chosen rather small in order to have the power spectrum shape
appropriate to fit observations such as the cluster autocorrelation
function, but then it is too much like standard hot dark matter and
there is far too little small scale structure (Colombi, Dodelson, \&
Widrow 1996).  (This, and also the possibly promising combination of
hot and more massive warm dark matter, will be discussed in more detail
in \se{pr_models} below.)

Cold DM consists of particles for which free streaming is of no
cosmological importance. Two different sorts of cold DM consisting of
elementary particles have been proposed, a cold Bose condensate such
as axions, and heavy remnants of annihilation or decay such as
supersymmetric weakly interacting massive particles (WIMPs). As has
been summarized above, a universe dominated by cold DM looks very much
like the one astronomers actually observe, at least on galaxy to
cluster scales.

\subsection{Cold Dark Matter Candidates}
\label{sec:pr_dm_cold}

The two sorts of particle candidates for cold dark matter that are
best motivated remain supersymmetric Weakly Interacting Massive
Particles (WIMPs) and the axion. They are both well motivated because
both supersymmetry and Peccei-Quinn symmetry (associated with the
axion) are key ideas in modern particle physics that were proposed
independently of their implications for dark matter (for a review
emphasizing direct and indirect methods of detecting both of these,
see Primack, Seckel, \& Sadoulet 1988). There are many other dark
matter candidates whose motivations are more ad hoc (see Table 1.3)
from the viewpoint of particle physics.  But there is observational
evidence that Massive Compact Halo Objects (MACHOs) may comprise a
substantial part of the mass of the Milky Way's dark matter halo.

\subsubsection{Axions}
\label{sec:pr_dm_cold_axion}

Peccei-Quinn symmetry, with its associated particle the axion, remains
the best solution known to the strong CP problem.  A second-generation
experiment is currently underway at LLNL (Hagmann et al. 1996) with
sufficient sensitivity to have a chance of detecting the axions that
might make up part of the dark matter in the halo of our galaxy, if
the axion mass lies in the range 2-20 $\mu$eV.  However, it now
appears that most of the axions would have been emitted from axionic
strings (Battye \& Shellard 1994, 1997) and from the collapse of
axionic domain walls (Nagasawa \& Kawasaki 1994), rather than arising
as an axion condensate as envisioned in the original cosmological
axion scenario.  This implies that if the axion is the cold dark
matter particle, the axion mass is probably $\sim 0.1$ meV, above the
range of the LLNL experiment.  While current experiments looking for
either axion or supersymmetric WIMP cold dark matter have a chance of
making discoveries, neither type is yet sufficiently sensitive to
cover the full parameter space and thereby definitively rule out
either theory if they do not detect anything.  But in both cases this
may be feasible in principle with more advanced experiments that may
be possible in a few years.

\subsubsection{Supersymmetric WIMPs}
\label{sec:pr_dm_cold_wimp}

 From the 1930s through the early 1970s, much of the development of
quantum physics was a search for ever bigger symmetries, from spin and
isospin to the Poincar\'e group, and from electroweak symmetry to
grand unified theories (GUTs). The larger the symmetry group, the
wider the scope of the connections established between different
elementary particles or other quantum states.  The basic pattern of
progress was to find the right Lie group and understand its role ---
SU(2) as the group connecting different states in the cases of spin
and isospin; SU(3)$\times$SU(2)$\times$U(1) as the dynamical gauge
symmetry group of the ``Standard Model'' of particle physics,
connecting states without a gauge boson to states of the same
particles including a gauge boson. Supersymmetry is a generalization
of this idea of symmetry, since it mixes space-time symmetries, whose
quantum numbers include the spin of elementary particles, with
internal symmetries. It is based on a generalization of Lie algebra
called graded Lie algebra, which involves anti-commutators as well as
commutators of the operators that transform one particle state into
another. Supersymmetry underlies almost all new ideas in particle
physics since the mid-1970s, including superstrings.  If valid, it is
also bound to be relevant to cosmology.  (Some reviews: Collins,
Martin, \& Squires 1989; de Boer 1994.)

\def\half{{1/2}}

The simplest version of supersymmetry, which should be manifest at the
GUT scale ($\sim 10^{16}$ GeV) and below, has as its key prediction
that for every kind of particle that we have learned about at the
relatively low energies which even our largest particle accelerators
can reach, there should be an as-yet-undiscovered ``supersymmetric
partner particle'' with the same quantum numbers and interactions
except that the spin of this hypothetical partner particle differs
from that of the known particle by half a unit.  For example, the
partner of the photon (spin 1) is the ``photino'' (spin $\half$), and
the partner of the electron (spin $\half$) is the ``selectron'' (spin
0).  Note that if a particle is a fermion (spin $\half$ or ${3/
2}$, obeying the Pauli exclusion principle), its partner particle is a
boson (spin 0, 1, 2).  The familiar elementary particles of matter
(quarks and leptons) are all fermions, a fact that is responsible for the
stability of matter, and the force particles are all bosons.   Table 1.4
is a chart of the known families of elementary particles and their
supersymmetric partners.  It is these hypothetical partner particles
among which we can search for the cold dark matter particle.  The most
interesting candidates are underlined. (As has already been mentioned,
the gravitino is a warm dark matter particle candidate; this is
discussed further below.)


\begin{table*}
\caption{Supersymmetry}
\label{ta:Susy}

\parindent=0pt
\def\9{\hphantom 0}
\let\ul=\underline
\centerline{A hypothetical symmetry between boson and fermion fields
and interactions}

\vskip .1in
\hrule
\vskip.1in

\begin{center}
\begin{minipage}{\linewidth}
\begin{tabular}{@{}c p{2.75cm} p{2.75cm}@{\hspace{8pt}}||@{\hspace{8pt}}
                  p{3.6cm} c@{}}
{\bf Spin}& {\bf Matter} (fermions)& {\bf Forces}\break (bosons)&
                {\bf Hypothetical Superpartners}&    {\bf Spin} \\
\hline
2&  & graviton      & gravitino &  3/2 \\[3pt]
1&  & photon, $W^\pm,Z^0$ gluons & \ul{photino}, winos,
                                        \ul{zino}, gluinos&1/2\\[3pt]
1/2& quarks u,d,\dots \hfil\break leptons $e,\nu_e,\dots$&
   & \qquad squarks $\tilde u,\tilde d, \dots$ \hfill\break
    $\9\9\9\9$sleptons $\tilde e,\tilde\nu_e,\dots$ & 0\\[5pt]
0&  &Higgs bosons         & \ul{Higgsinos} &  1/2 \\
 &  & axion          & \ul{axinos}&      \\
\end{tabular}
\end{minipage}
\end{center}
\hrule
\vskip .1in

Note: Supersymmetric cold dark matter candidate particles are
underlined.

\end{table*}

Note the parallel with Dirac's linking of special relativity and
quantum mechanics in his equation for spin-$\half$ particles (Griest
1996). In modern language, the resulting CPT invariance (under the
combination of charge-conjugation C, replacing each particle with its
antiparticle; parity P, reversing the direction of each spatial
coordinate; and time-reversal T) requires a doubling of the number of
states: an anti-particle for every particle (except for particles,
like the photon, which are their own antiparticles).

There are two other key features of supersymmetry that make it
especially relevant to dark matter, $R$-parity and the connection
between supersymmetry breaking and the electroweak scale.  The
$R$-parity of any particle is $R \equiv (-1)^{L+3B+2S}$, where $L$,
$B$, and $S$ are its lepton number, baryon number, and spin.  Thus for
an electron ($L=1$, $B=0$, $S=\half$) $R=1$, and the same is true for
a quark ($L=0$, $B={1/ 3}$, $S=\half$) or a photon ($L=0$, $B=0$,
$S=1$).  Indeed $R=1$ for all the known particles.  But for a
selectron ($L=1$, $B=0$, $S=\half$) or a photino ($L=0$, $B=0$,
$S=\half$), the $R$-parity is -1, or ``odd''.  In most versions of
supersymmetry, $R$-parity is exactly conserved.  This has the powerful
consequence that the lightest $R$-odd particle --- often called the
``lightest supersymmetric partner'' (LSP) --- must be stable, for
there is no lighter $R$-odd particle for it to decay into.  The LSP is
thus a natural candidate to be the dark matter, as was first pointed
out by Pagels \& me (1982), although as mentioned above the LSP in
the early form of supersymmetry that we considered would have been a
gravitino weighing about a keV, which would now be classified as warm
dark matter.

In the now-standard version of supersymmetry, there is an answer to
the deep puzzle why there should be such a large difference in mass
between the GUT scale $M_{\rm GUT} \sim 10^{16}$ GeV and the
electroweak scale $M_W=80$ GeV.  Since both gauge symmetries are
supposed to be broken by Higgs bosons which moreover must interact
with each other, the natural expectation would be that $M_{\rm GUT}
\sim M_W$.  The supersymmetric answer to this ``gauge hierarchy''
problem is that the masses of the weak bosons $W^\pm$ and all other
light particles are zero until supersymmetry itself breaks.  Thus,
there is a close relationship between the masses of the supersymmetric
partner particles and the electroweak scale.  Since the abundance of
the LSP is determined by its annihilation in the early universe, and
the corresponding cross section involves exchanges of weak bosons or
supersymmetric partner particles --- all of which have
electromagnetic-strength couplings and masses $\sim M_W$ --- the cross
sections will be $\sigma \sim e^2 s/M_W^4$ (where $s$ is the square of
the center-of-mass energy) i.e., comparable to typical weak
interactions. This in turn has the remarkable consequence that the
resulting density of LSPs today corresponds to nearly critical
density, i.e. $\Omega_{\rm LSP} \sim 1$.
The LSP is typically a
spin-$\half$ particle called a ``neutralino'' which is its own
antiparticle --- that is, it is a linear combination of the photino
(supersymmetric partner of the photon), ``zino'' (partner of the $Z^0$
weak boson), ``Higgsinos'' (partners of the two Higgs bosons
associated with electroweak symmetry breaking in supersymmetric
theories), and  ``axinos'' (partners of the axion, if it exists).  In
much of the parameter space, the neutralino $\chi$ is a ``bino,'' a
particular linear combination of the photino and zino.  All of these
neutralino LSPs are WIMPs, weakly interacting massive particles.
Because of their large masses, several 10s to possibly 100s of GeV,
these supersymmetric WIMPs would be dark matter of the ``cold'' variety.

Having explained why supersymmetry is likely to be relevant to cold
dark matter, one should also briefly summarize why supersymmetry is so
popular with modern particle physicists.  The reasons are that it is
not only beautiful, it is even perhaps likely to be true.  The
supersymmetric pairing between bosons and fermions results in a
cancellation of the high-energy (or ``ultraviolet'') divergences due
to internal loops in Feynman diagrams.  It is this cancellation that
allows supersymmetry to solve the gauge hierarchy problem (how can
$M_{\rm GUT}/M_W$ be so big), and perhaps also unify gravity with the
other forces (``superunification,'' ``supergravity,''
``superstrings'').  The one prediction of supersymmetry (Georgi,
Quinn, \& Weinberg 1974) that has been verified so far is related to
grand unification (Amaldi, de Boer, \& Furstenau 1991).  The way this
is usually phrased today is that the three gauge couplings associated
with the three parts of the standard model --- the SU(3) ``color''
strong interactions, and the SU(2)$\times$U(1) electroweak
interactions --- do not unify at any higher energy scale unless the
effects of the supersymmetric partner particles are included in the
calculation, and they do unify with the minimal set of partners (one
partner for each of the known particles) as long as the partner
particles all have masses not much higher than the electroweak scale
$M_W$ (which, as explained above, is expected if electroweak
symmetry breaking is related to supersymmetry breaking).

The expectations for the LSP neutralino, including prospects for their
detection in laboratory experiments and via cosmic rays, have recently
been exhaustively reviewed (Jungman, Kamionkowski, \& Griest 1996).
Several ambitious laboratory search experiments for LSPs in the mass
range of tens to hundreds of GeV are now in progress (e.g., Shutt et
al. 1996), and within the next few years they will have adequate
sensitivity to probe a significant amount of the supersymmetric model
parameter space. There are also hints of supersymmetric effects from
recent experiments, which suggest that supersymmetry may be
definitively detected in the near future as collider energy is
increased --- and also hint that the LSP may be rather light (Kane \&
Wells 1996), possibly even favoring the gravitino as the LSP
(Dimopoulos et al. 1996).

\subsubsection{MACHOs}
\label{sec:pr_dm_cold_macho}

Meanwhile, the MACHO  (Alcock et al. 1996a) and EROS (Ansari et al.
1996, Renault et al. 1996)
experiments have detected microlensing of stars in the Large
Magellanic Cloud (LMC). While the number of such microlensing events
is small (six fairly convincing ones from two years of MACHO data
discussed in their latest conference presentations, and one from three
years of EROS observations), it is several times more than would be
expected just from microlensing by the known stars. The MACHO data
suggests that objects with a mass of $0.5^{+0.3}_{-0.2}M_\odot$ are
probably responsible for this microlensing, with their total density
equal to $\sim20-50$ percent of the mass of the Milky Way halo around
$\sim 20$ kpc radius (Gates, Gyuk, \& Turner 1996).  Neither the EROS
nor the MACHO groups have seen short duration microlensing events,
which implies strong upper limits on the possible contribution to the
halo of compact objects weighing less than about $0.05 M_\odot$. While
the MACHO masses are in the range expected for white dwarfs, there are
strong observational limits (Flynn, Gould, \& Bahcall 1996) and
theoretical arguments (Adams \& Laughlin 1996) against white dwarfs
being a significant fraction of the dark halo of our galaxy. Thus it
remains mysterious what objects could be responsible for the observed
microlensing toward the LMC.  But the very large number of
microlensing events observed toward the galactic bulge is probably
explained by the presence of a bar aligned almost toward our position
(Zhao, Rich, \& Spergel 1996; cf. Bissantz et al. 1996 for a
dissenting view).  Possibly the relatively small number of microlensing
events toward the LMC represent lensing by a tidal tail of
stars stretching toward us from the main body of the LMC (Zhao 1997);
there is even some data on the colors and luminosities of stars toward
the LMC suggesting that this may actually be true (D. Zaritsky,
private communication 1997).

\subsection{Hot Dark Matter: Data on Neutrino Mass}
\label{sec:pr_dm_hot}

The upper limit on the electron neutrino mass is roughly 10-15 eV; the
current Particle Data Book (Barnett et al. 1996) notes that a more
precise limit cannot be given since unexplained effects have resulted
in significantly negative measurements of $m(\nu_e)^2$ in recent
precise tritium beta decay experiments.  The (90\% C.L.) upper limit
on an effective Majorana neutrino mass 0.65 eV from the
Heidelberg-Moscow $^{76}$Ge neutrinoless double beta decay experiment
(Balysh et al. 1995).  The upper limits from accelerator experiments
on the masses of the other neutrinos are $m(\nu_\mu)< 0.17$ MeV (90\%
C.L.) and $m(\nu_\tau)< 24$ MeV (95\% C.L.).  Since stable neutrinos
with such large masses would certainly ``overclose the universe''
(i.e., prevent it from attaining its present age), the cosmological
upper limits follow from the neutrino contribution to the cosmological
density $\Omega_\nu = m(\nu)/ (92 h^2\, {\rm eV}) < \Omega_0$.  There is a
small window for an unstable $\nu_\tau$ with mass $\sim 10-24$ MeV,
which could have many astrophysical and cosmological consequences:
relaxing the Big-Bang Nucleosynthesis bound on $\Omega_b$ and $N_\nu$,
allowing BBN to accommodate a low (less than 22\%) primordial $^4$He mass
fraction or high deuterium abundance, improving significantly the
agreement between the CDM theory of structure formation and
observations, and helping to explain how type II supernovae explode
(Gyuk \& Turner 1995).

But there is mounting astrophysical and laboratory data suggesting that
neutrinos oscillate from one species to another, and therefore that
they have non-zero mass.  The implications if {\it all} these
experimental results are taken at face value are summarized in
Table 1.5.  Of these experiments, the ones that are most relevant to
neutrinos as hot dark matter are LSND and the higher energy
Kamiokande atmospheric (cosmic ray) neutrinos.  But the experimental
results that are probably most secure are those concerning solar
neutrinos, suggesting that some of the electron neutrinos undergo
MSW oscillations to another species of neutrino as they travel
through the sun (see, e.g., Hata \& Langacker 1995, Bahcall 1996).

The recent observation of events that appear to represent
$\bar\nu_\mu \to \bar\nu_e$ oscillations followed by $\bar \nu_e + p
\to n + e^+$, $n + p \to D + \gamma$, with coincident detection of
$e^+$ and the 2.2 MeV neutron-capture $\gamma$-ray in the Liquid
Scintillator Neutrino Detector (LSND) experiment at Los Alamos
suggests that $\Delta m_{e\mu}^2 \equiv |m(\nu_\mu)^2 - m(\nu_e)^2| >
0$ (Athanassopoulos et al. 1995, 1996).  The analysis of the LSND data
through 1995 strengthens the earlier LSND signal for $\bar\nu_\mu
\rightarrow \bar\nu_e$ oscillations. Comparison with exclusion plots
from other experiments implies a lower limit $\Delta m^2_{\mu e}
\equiv |m(\nu_\mu)^2-m(\nu_e)^2| \gsim 0.2$ eV$^2$, implying in turn a
lower limit $m_\nu \gsim 0.45$ eV, or $\Omega_\nu \gsim 0.02
(0.5/h)^2$. This implies that the contribution of hot dark matter to
the cosmological density is larger than that of all the visible stars
($\Omega_\ast \approx 0.004$ (Peebles 1993, eq. 5.150). More data and
analysis are needed from LSND's $\nu_\mu \rightarrow \nu_e$ channel
before the initial hint (Caldwell 1995) that $\Delta m^2_{\mu e}
\approx 6$ eV$^2$ can be confirmed.  Fortunately the KARMEN experiment
has just added shielding to decrease its background so that it can
probe the same region of $\Delta m^2_{\mu e}$ and mixing angle, with
sensitivity as great as LSND's within about two years (Kleinfeller
1996).  The Kamiokande data (Fukuda 1994) showing that the deficit of
$E > 1.3$ GeV atmospheric muon neutrinos increases with zenith angle
suggests that $\nu_\mu \rightarrow \nu_\tau$ oscillations\footnote{The
Kamiokande data is consistent with atmospheric $\nu_\mu$ oscillating
to any other neutrino species $y$ with a large mixing angle
$\theta_{\mu y}$. But as summarized in Table 1.5 (see further
discussion and references in, e.g., Primack et al. 1995, hereafter
PHKC95; Fuller, Primack, \& Qian 1995) $\nu_\mu$ oscillating to
$\nu_e$ with a large mixing angle is probably inconsistent with
reactor and other data, and $\nu_\mu$ oscillating to a sterile
neutrino $\nu_s$ (i.e., one that does not interact via the usual weak
interactions) with a large mixing angle is inconsistent with the usual
Big Bang Nucleosynthesis constraints. Thus, by a process of
elimination, if the Kamiokande data indicating atmospheric neutrino
oscillations is right, the oscillation is $\nu_\mu \rightarrow
\nu_\tau$.} occur with an oscillation length comparable to the height
of the atmosphere, implying that $\Delta m^2_{\tau \mu} \sim 10^{-2}$
eV$^2$ --- which in turn implies that if either $\nu_\mu$ or
$\nu_\tau$ have large enough mass ($\gsim 1$ eV) to be a hot dark
matter particle, then they must be nearly degenerate in mass, i.e., the
hot dark matter mass is shared between these two neutrino species. The
much larger Super-Kamiokande detector is now operating, and we should
know by about the end of 1996 whether the Kamiokande atmospheric
neutrino data that suggested $\nu_\mu \rightarrow \nu_\tau$
oscillations will be confirmed and extended. Starting in 1997 there
will be a long-baseline neutrino oscillation disappearance experiment
to look for $\nu_\mu \rightarrow \nu_\tau$ with a beam of $\nu_\mu$
from the KEK accelerator directed at the Super-Kamiokande detector,
with more powerful Fermilab-Soudan, KEK-Super-Kamiokande, and possibly
CERN-Gran Sasso long-baseline experiments later.


\begin{table}
\caption{Data Suggesting Neutrino Mass}
\label{ta:neutrmass}

\parindent=0pt
\def\9{\hphantom 0}
\let\ul=\underline

\hrule
\begin{center}
\renewcommand{\arraystretch}{2}
\begin{tabular}{@{}l}
{\bf Solar}  \qquad   $\Delta m^2_{\rm ex} = 10^{-5}$ eV$^2$,
                \quad $\sin^2 2\theta_{\rm ex}$ \quad small  \\
{\bf Atm \boldmath$\nu_\mu$ deficit \boldmath$(\theta)$} \qquad
            $\Delta m^2_{\mu y} \simeq 10^{-2}$ eV$^2$,
             \quad $\sin^2 2\theta_{\mu y}$ \quad $\sim1$\\ [-1pc]
      \9\9\9\9\9  Kamiokande $E_\nu > 1.3$ GeV        \\
{\bf Reactor \boldmath$\nu_e$} \qquad probably excludes y = e, so atm $\nu_\mu
         \to \nu_\tau$ or $\nu_s$                            \\
{\bf BBN} \qquad excludes $\nu_\mu  \to \nu_s$ with large mixing,
                                  so $y=\tau$  \\
{\bf LSND} \qquad $\Delta m^2_{\mu e} \approx 1-10$ eV$^2$,   \quad
         $\sin^2 2\theta_{\mu e}$ \quad  small            \\[-1pc]
\9 \hspace{2cm}  excludes $x=\mu$, so solar $\nu_e \to \nu_s$ \\
{\bf Cold + Hot Dark Matter} \qquad $\Sigma m_\nu \approx 5$ eV

                                      $h^2_{50}$
\end{tabular}
\end{center}

\hrule

\end{table}

Evidence for non-zero neutrino mass evidently favors CHDM, but it also
disfavors low-$\Omega$ models.  Because free streaming of the
neutrinos damps small-scale fluctuations, even a little hot dark
matter causes reduced fluctuation power on small scales and requires
substantial cold dark matter to compensate; thus evidence for even 2
eV of neutrino mass favors large $\Omega$ and would be incompatible
with a cold dark matter density $\Omega_c$ as small as 0.3
(PHKC95). Allowing $\Omega_\nu$ and the tilt to vary, CHDM can fit
observations over a somewhat wider range of values of the Hubble
parameter $h$ than standard or tilted CDM (Pogosyan \& Starobinsky
1995a, Liddle et al. 1996b).  This is especially true if the neutrino
mass is shared between two or three neutrino species (Holtzman 1989;
Holtzman \& Primack 1993; PHKC95; Pogosyan \&  Starobinsky 1995b;
Babu, Schaefer,
\& Shafi 1996), since then the lower neutrino mass results in a larger
free-streaming scale over which the power is lowered compared to CDM.
The result is that the cluster abundance predicted with $\Omega_\nu
\approx 0.2$ and $h \approx 0.5$ and COBE normalization
(corresponding to $\sigma_8 \approx 0.7$) is in reasonable agreement
with observations without the need to tilt the model (Borgani et al.
1996) and thereby reduce the small-scale power further.  (In CHDM with
a given $\Omega_\nu$ shared between $N_\nu=2$ or 3 neutrino species,
the linear power spectra are identical on large and small scales to
the $N_\nu=1$ case; the only difference is on the cluster scale, where
the power is reduced by $\sim 20\%$ (Holtzman 1989, PHKC95,
Pogosyan \& Starobinsky 1995).

\section{Origin of Fluctuations: Inflation and Topological Defects}
\label{sec:pr_origin}

\subsection{Topological defects}
\label{sec:pr_origin_defect}

A fundamental scalar field, the Higgs field, is invoked by
particle theorists to account for the generation of mass;
one of the main goals of the next generation of particle
accelerators, including the Large Hadron Collider at CERN,
will be to verify the Higgs theory for the generation
of the mass of the weak vector bosons and all the lighter
elementary particles.  Another scalar field is required to
produce the vacuum energy which may drive cosmic inflation
(discussed in the next section).  Scalar fields can also
create topological defects that might be of great importance
in cosmology.  The basic idea is that some symmetry is
broken wherever a given scalar field $\phi$ has a
non-vanishing value, so the dimensionality of the
corresponding topological defect depends on the number of
components of the scalar field: for a single-component real
scalar field, $\phi(\vec r)=0$ defines a two-dimensional
surface in three-dimensional space, a {\it domain wall}; for
a complex scalar field, the real and imaginary parts of
$\phi(\vec r)=0$ define a one-dimensional locus, a {\it
cosmic string}; for a three-component (e.g., isovector)
field, $\phi_i(\vec r)=0$ for $i=1,2,3$ is satisfied at
isolated points, {\it monopoles}; for more than three
components, one gets {\it textures} that are not
topologically stable but which can seed structure in the
universe as they unwind.

To see how this works in more detail, consider a cosmic string.  For
the underlying field theory to permit cosmic strings, we need to
couple a complex scalar field $\phi$ to a single-component (i.e.,
U(1)) gauge field $A_\alpha$, like the electromagnetic field, in the
usual way via the substitution $\partial_\alpha \rightarrow D_\alpha
\equiv (\partial_\alpha - i e A_\alpha)$, so that the scalar field
derivative term in the Lagrangian becomes ${\cal L}_{D\phi} =
|D_\alpha \phi|^2$.  Then if the scalar field $\phi$ gets a non-zero
value by the usual Higgs ``spontaneous symmetry breaking'' mechanism,
the gauge symmetry is broken because the field has a definite complex
phase.  But along a string where $\phi=0$ the symmetry is restored.
As one circles around the string at any point on it, the complex phase
of $\phi(\vec r)$ in general makes one, or possibly $n>1$, complete
circles $0 \rightarrow 2 n \pi$.  But since such a phase rotation can
be removed at large distance from the string by a gauge transformation
of $\phi$ and $A_\alpha$, the energy density associated with this
behavior of $\phi$ ${\cal L}_{D\phi} \rightarrow 0$ at large
distances, and therefore the energy $\mu$ per unit length of string is
finite.  Since it would require an infinite amount of energy to unwind
the phase of $\phi$ at infinity, however, the string is topologically
stable.  If the field theory describing the early universe includes a
U(1) gauge field and associated complex Higgs field $\phi$, a rather
high density of such cosmic strings will form when the string field
$\phi$ acquires its nonzero value and breaks the U(1) symmetry. This
happens because there is no way for the phase of $\phi$ to be aligned
in causally disconnected regions, and it is geometrically fairly
likely that the phases will actually wrap around as required for a
string to go through a given region (Kibble 1976).  The string network
will then evolve and can help cause formation of structure after the
universe becomes matter dominated, as long as the string density is
not diluted by a subsequent period of cosmic inflation (on the
difficult problem of combining cosmic defects and inflation, see,
e.g., Hodges \& Primack 1991).  A similar discussion can be given for
domain walls and local (gauge) monopoles, but these objects are
cosmologically pathological since they dominate the energy density and
``overclose'' the universe.  But cosmic strings, a sufficiently low
density of global (i.e., non-gauged monopoles), and global textures
are potentially interesting for cosmology (recent reviews include
Vilenkin \& Shellard 1994, Hindmarsh \& Kibble 1995, Shellard 1996).
Cosmic defects are the most important class of models producing
non-Gaussian fluctuations which could seed cosmic structure formation.
Since they are geometrically extended objects, they correspond to
non-local non-Gaussian fluctuations (Kofman et al. 1991).

The parameter $\mu$, usually quoted in the dimensionless
form $G\mu$ (where $G$ is Newton's constant), is the key
parameter of the theory of cosmic strings. The value
required for the COBE normalization is $G\mu_6 \equiv G\mu
\times 10^6 = 1-2$ (recent determinations include $G\mu_6 =
1.7\pm0.7$, Perivolaropoulos 1994; 2, Coulson et al. 1994;
$(1.05_{-0.20}^{+0.35})$, Allen et al. 1996; 1.7, Allen et al.
1997).
This is close enough to the value required for structure
formation, $G\mu = (2.2-2.8) b_8^{-1} \times 10^{-6}$
(Albrecht \& Stebbins 1992), with the smaller value for
cosmic strings plus cold dark matter and the higher value
for cosmic strings plus hot dark matter, so that the
necessary value of the biasing factor $b_8$ is 1.3-3, which is
high (probably leading to underproduction of clusters, and
large-scale velocities that are low compared to observations
--- cf. Perivolaropoulos \& Vachaspati 1994), but perhaps not
completely crazy. (Here $b_8$ is the factor by which galaxies
must be more clustered than dark matter, on a scale of $8
h^{-1}$ Mpc.) Since generically $G\mu \sim (M/m_{pl})^2$,
where $M$ is the energy scale at which the string field
$\phi$ acquires its nonzero value, the fact that $G\mu \sim
10^{-6}$, corresponding to $M$ at roughly the Grand
Unification scale, is usually regarded as a plus for the
cosmic string scenario.  (Even though there is no particular
necessity for cosmic strings in GUT scenarios, GUT groups
larger than the minimal SU(5) typically do contain the
needed extra U(1)s.)  Moreover, the required normalization
is well below the upper limit obtained from the requirement
that the gravitational radiation generated by the evolution
of the string network not disrupt Big Bang Nucleosynthesis,
$G\mu \lsim 6 \times 10^{-6}$.  However, there is currently
controversy whether it is also below the upper limit from
pulsar timing, which has been determined to be $G\mu \lsim 6
\times 10^{-7}$ (Thorsett \& Dewey 1996) vs. $G\mu \lsim 5
\times 10^{-6}$ (McHugh et al. 1996; cf. Caldwell, Battye, \&
Shellard 1996).

As for cosmic strings, the COBE normalization for global texture
models also implies a high bias $b_8 \approx 3.4$ for $h=0.7$ (Bennett
\& Rhie 1993), although the needed bias is somewhat lower for $\Omega
\approx 0.3$ (Pen \& Spergel 1995).  The latest global defect
simulations (Pen, Seljak, \& Turok, 1997) show that the matter
power spectrum in all such models also has a shape very different
than that suggested by the available data on galaxies and clusters.

But both cosmic string and global defect models have a problem which
may be even more serious: they predict a small-angle CMB fluctuation
spectrum in which the first peak is at rather high angular wavenumber
$\ell \sim 400$ (Crittenden \& Turok 1995, Durrer et al. 1996,
Magueijo et al. 1996) and in any case is rather low in amplitude,
partly because of incoherent addition of scalar, vector, and tensor
modes, according to the latest simulations (strings: Allen et al.
1997; global defects: Pen, Seljak, \& Turok 1997; cf. Albrecht, Battye,
\& Robinson 1997).  This is in conflict with the currently available
small-angle CMB data (Netterfield et al. 1997, Scott et al. 1996),
which shows a peak at $\ell \sim 250$ and a drop at $\ell \gsim 400$,
as predicted by flat ($\Omega_0 + \Omega_\Lambda=1$) CDM-type models.
Since the small-angle CMB data is still rather preliminary, it is
premature to regard the cosmic defect models as being definitively
ruled out. It will be interesting to see the nature of the predicted
galaxy distribution and CMB anisotropies when more complete
simulations of cosmic defect models are run.  This is more difficult
than simulating models with the usual inflationary fluctuations, both
because it is necessary to evolve the defects, and also because the
fact that these defects represent rare but high amplitude fluctuations
necessitates a careful treatment of their local effects on the
ordinary and dark matter.  It may be difficult to sustain the effort
such calculations require, because the poor agreement between the
latest defect simulations and current small-angle CMB data does not
bode well for defect theories.  Fortunately, there have been
significant technical breakthroughs in calculational techniques (cf.
Allen et al. 1997, Pen et al. 1997).

\subsection{Cosmic Inflation: Introduction}
\label{sec:pr_origin_inflation}

The basic idea of inflation is that before the universe
entered the present adiabatically expanding Friedmann era,
it underwent a period of de Sitter exponential expansion of
the scale factor, termed {\it inflation} (Guth 1981).
Actually, inflation is never precisely de Sitter, and
any superluminal (faster-than-light) expansion is now
called inflation.  Inflation was originally invented to
solve the problem of too many GUT monopoles, which, as
mentioned in the previous section, would otherwise be
disastrous for cosmology.

The de Sitter cosmology corresponds to the solution of
Friedmann's equation in an empty universe (i.e.,
with $\rho=0$) with vanishing
curvature ($k=0$) and positive cosmological constant
($\Lambda>0$).  The solution is $a=a_o e^{Ht}$,
with constant Hubble parameter $ H=(\Lambda/3)^{1/2} $.
There are analogous solutions for $k=+1$ and $k=-1$ with
$a\propto \cosh Ht$ and $a\propto \sinh Ht$ respectively.
The scale factor expands exponentially because the
positive cosmological constant corresponds effectively to a
negative pressure.  de~Sitter space is discussed in textbooks
on general relativity (for example,
Rindler 1977, Hawking \& Ellis 1973) mainly for its
geometrical interest. Until cosmological inflation was considered,
the chief significance of
the de Sitter solution in cosmology was that it is a
limit to which all indefinitely expanding models with $\Lambda>0$
must tend, since as $a\rightarrow \infty$, the cosmological
constant term ultimately dominates the right hand side of the
Friedmann equation.

As Guth (1981) emphasized, the de Sitter solution might also have been
important in the very early universe because the vacuum energy that
plays such an important role in spontaneously broken gauge theories
also acts as an effective cosmological constant.  A period of de
Sitter inflation preceding ordinary radiation-dominated Friedmann
expansion could explain several features of the observed universe that
otherwise appear to require very special initial conditions: the
horizon, flatness/age, monopole, and structure formation problems.
(See Table 1.6.)


\begin{table}
\caption{Inflation Summary}
\label{ta:Inflsummary}

\parindent=0pt
\def\9{\hphantom 0}
\font\caps=cmcsc10

\vskip .1in
\hrule
\vskip .1in

{\caps Problem Solved}

\begin{center}
\renewcommand{\arraystretch}{1.2}
\setlength{\tabcolsep}{1.5pc}
\begin{tabular}{ll}
Horizon&        Homogeneity, Isotropy, Uniform T  \\
Flatness/Age&   Expansion and gravity balance  \\
``Dragons"&     Monopoles, doman walls,\dots banished \\
Structure&  Small fluctuations to evolve into galaxies, \\[-3pt]
&           \hskip1cm   clusters, voids \\
\end{tabular}
\end{center}

\vskip1pc
\raggedright
Cosmological constant $\Lambda > 0 \Rightarrow$ space repels space, so
the more space the more repulsion, $\Rightarrow$ de Sitter exponential
expansion $a \propto e^{\sqrt\Lambda t}$.

\vskip1pc
Inflation is exponentially accelerating expansion caused by effective
cosmological constant (``false vacuum" energy)
associated with hypothetical scalar field (``inflaton").

\vskip1pc
\begin{center}
\renewcommand{\arraystretch}{1.2}
\setlength{\tabcolsep}{9pt}
\begin{tabular}{llc}
&\multicolumn{1}{c}{\caps Forces of Nature}&   {\bf  Spin}  \\ [3pt]
             &      Gravity&                            2 \\ [-1.3pc]
\hfill Known\quad $\Bigg\{$ & &\\[-1.3pc]
             &     Strong, weak, and electromagnetic&  1 \\ [2pt]
 Goal of LHC \qquad\quad&  Mass (Higgs Boson)&                 0 \\
Early universe        & Inflation (Inflaton)&               0
\end{tabular}
\end{center}

\vskip1pc
Inflation lasting only $\sim$10$^{-32}$s suffices to solve all the
problems listed above.  Universe must then convert to ordinary
expansion through conversion of false to true vacuum (``re-''heating).

\vskip .1in
\hrule

\end{table}

Let us illustrate how inflation can help with the horizon
problem.  At recombination ($p^+ + e^- \rightarrow H$),
which occurs at $a/a_o \approx 10^{-3}$, the mass encompassed
by the horizon was $M_H \approx 10^{18} M_\odot$, compared
to $M_{H,o} \approx 10^{22} M_\odot$ today.  Equivalently,
the angular size today of the causally connected regions at recombination
is only $\Delta \theta \sim 3^\circ$.  Yet the fluctuation
in temperature of the cosmic background radiation from
different regions is very small: $\Delta T/T \sim
10^{-5}$.  How could
regions far out of causal contact have come to temperatures
that are so precisely equal?  This is the ``horizon
problem''.  With inflation, it is no problem because the
entire observable universe initially lay inside a single
causally connected region that subsequently inflated to a
gigantic scale.  Similarly, inflation exponentially dilutes
any preceeding density of monopoles or other unwanted
relics (a modern version of the ``dragons'' that decorated
the unexplored borders of old maps).

In the first inflationary models, the dynamics of the very early
universe was typically controlled by the self-energy of the
Higgs field associated with the breaking of a Grand Unified
Theory (GUT) into the standard 3-2-1 model: GUT$\rightarrow
SU(3)_{color}\otimes [SU(2)\otimes U(1)]_{electroweak}$.
This occurs when the cosmological temperature drops to the
unification scale $T_{GUT} \sim 10^{14}$ GeV at about
$10^{-35}$ s after the Big Bang.
Guth (1981) initially considered a scheme in
which inflation occurs while the universe is trapped in an
unstable state (with the GUT unbroken) on the wrong side of
a maximum in the Higgs potential.  This turns out not to
work: the transition from a de Sitter to a Friedmann
universe never finishes (Guth \& Weinberg 1981).
The solution in the ``new inflation'' scheme
(Linde 1982; Albrecht and Steinhardt 1982)
is for inflation to occur {\it after} barrier penetration
(if any).  It is necessary that the potential of the
scalar field controlling inflation (``{\it inflaton}'') be nearly
flat (i.e., decrease very slowly with increasing inflaton field)
for the inflationary period to last long enough.  This nearly
flat part of the potential must then be followed by a
very steep minimum, in order that the energy contained in
the Higgs potential be rapidly shared with the other degrees
of freedom (``reheating'').  A more general approach,
``chaotic'' inflation, has been worked out by Linde (1983, 1990)
and others; this works for a wide range of inflationary
potentials, including simple power laws such as $\lambda
\phi^4$.  However, for the amplitude of the fluctuations to
be small enough for consistency with observations, it is necessary
that the inflaton self-coupling be very small, for example $\lambda
\sim 10^{-14}$ for the $\phi^4$ model. This requirement prevents a
Higgs field from being the inflaton, since Higgs fields by definition
have gauge couplings to the gauge field (which are expected to be of
order unity), and these would generate self-couplings of similar
magnitude even if none were present.  Both the Higgs and inflaton are
hypothetical fundamental (or possibly composite) scalar fields (see
Table 1.6).

It turns out to be necessary to inflate by a factor $\gsim
e^{66}$ in order to solve the flatness problem, i.e., that
$\Omega_0 \sim 1$.  (With $H^{-1}\sim 10^{-34}$ s during the
de Sitter phase, this implies that the inflationary period
needs to last for only a relatively small time $\tau \gsim
10^{-32}$ s.)  The ``flatness problem'' is essentially the
question why the universe did not become curvature dominated
long ago.  Neglecting the cosmological constant on the
assumption that it is unimportant after the inflationary
epoch, the Friedmann equation can be written
\begin{equation}
 \left({\dot a \over a}\right)^2={{8\pi G}\over 3} {\pi^2
 \over 30} g(T) T^4 - {{k T^2}\over {(aT)^2}}
\end{equation}
where the first term on the right hand side is the
contribution of the energy density in relativistic particles
and $g(T)$ is the effective number of degrees of freedom.
The second term on the
right hand side is the curvature term.  Since $aT \approx $
constant for adiabatic expansion, it is clear that as the
temperature $T$ drops, the curvature term becomes
increasingly important.  The quantity $K\equiv k/(aT)^2$ is
a dimensionless measure of the curvature.
Today, $\left |K\right |=\left | \Omega-1\right |
H_o^2/T_o^2 \leq 2\times 10^{-58}$.  Unless
the curvature exactly vanishes, the most ``natural'' value
for $K$ is perhaps $K \sim 1$.
Since inflation increases $a$ by a tremendous factor
$e^{H\tau}$ at essentially constant $T$ (after reheating),
it increases $aT$ by the same tremendous factor and thereby
decreases the curvature by that factor squared.  Setting
$e^{-2 H \tau} \lsim 2 \times 10^{-58}$ gives the needed
amount of inflation: $H\tau \gsim 66$.  This much
inflation turns out to be enough to take care of the other
cosmological problems mentioned above as well.

Of course, this is only the minimum amount of inflation needed; the
actual inflation might have been much greater. Indeed it is frequently
argued that since the amount of inflation is a tremendously sensitive
function of (e.g.) the initial value of the inflaton field, it is
extremely likely that there was much more inflation than the minimum
necessary to account for the fact that the universe is observed to be
nearly flat today. It then follows (in the absence of a cosmological
constant today) that $\Omega_0=1$ to a very high degree of accuracy.
A way of evading this that has recently been worked out is discussed
in \se{pr_origin_open}.

\subsection{Inflation and the Origin of Fluctuations}
\label{sec:pr_origin_fluct}

Thus far, it has been sketched how inflation stretches, flattens,
and smooths out the universe, thus greatly increasing the
domain of initial conditions that could correspond to the
universe that we observe today.  But inflation also can
explain the origin of the fluctuations necessary in the
gravitional instability picture of galaxy and cluster
formation.  Recall that the very existence of these
fluctuations is a problem in the standard Big Bang picture,
since these fluctuations are much larger than the horizon at
early times.  How could they have arisen?

The answer in the inflationary universe scenario is that
they arise from quantum fluctuations in the inflaton field
$\phi$ whose vacuum energy drives inflation.  The scalar
fluctuations $\delta \phi$ during the de Sitter phase are of
the order of the Hawking temperature $H/2\pi$.  Because of these
fluctuations, there is a time spread $\Delta t \approx
\delta \phi /\dot \phi$ during which different regions of
the same size complete the transition to the Friedmann
phase.  The result is that the density fluctuations when a
region of a particular size re-enters the horizon are equal
to (Guth \& Pi 1982; see Linde 1990 for alternative
approaches)
$\delta_H \equiv ({{\delta\rho}/\rho})_H
 \sim \Delta t/t_H = H \Delta t$.
The time spread $\Delta t$ can be estimated from the
equation of motion of $\phi$ (the free Klein-Gordon equation
in an expanding universe):
$\ddot\phi + 3H\dot\phi = -(\partial V/\partial \phi)$.
Neglecting the $\ddot\phi$ term, since the scalar potential
$V$ must be very flat in order for enough inflation to
occur (this is called the ``slow roll'' approximation),
$\dot\phi \approx -V'/(3H)$, so $\delta_H \sim H^3/V' \sim
V^{3/2}/V'$.  Unless there is a special feature in the potential
$V(\phi)$ as $\phi$ rolls through the scales of importance in
cosmology (producing such ``designer inflation'' features generally
requires fine tuning --- see e.g. Hodges et al. 1990), $V$ and $V'$
will hardly vary there and hence $\delta_H$ will be essentially
constant. These are fluctuations of all the contents of the universe,
so they are adiabatic fluctuations.

Thus {\it inflationary models typically predict a nearly constant
curvature spectrum} $\delta_H={\rm constant}$
{\it of adiabatic fluctuations}.
Some time ago Harrison (1970), Zel'dovich (1972),
and others had emphasized that this is the only scale-invariant
(i.e., power-law) fluctuation spectrum that avoids trouble
at both large and small scales.  If
$\delta_H \propto M_H^{-\alpha}$, where $M_H$ is the mass inside
the horizon, then if $-\alpha$ is too large the universe will be less
homogeneous on large than small scales, contrary to
observation; and if $\alpha$ is too large, fluctuations
on sufficiently small scales will enter the horizon with
$\delta_H \gg 1$ and collapse to black holes
(see e.g. Carr, Gilbert, \& Lidsey 1995, Bullock \& Primack 1996);
thus $\alpha \approx 0$.  The $\alpha=0$ case has come to be
known as the Zel'dovich spectrum.

Inflation predicts more: it allows the calculation of the value of the
constant $\delta_H$ in terms of the properties of the scalar potential
$V(\phi)$.  Indeed, this proved to be embarrassing, at least
initially, since the Coleman-Weinberg potential, the first potential
studied in the context of the new inflation scenario, results in
$\delta_H \sim 10^2$ (Guth \& Pi 1982) some six orders of magnitude
too large. But this does not seem to be an insurmountable difficulty;
as was mentioned above, chaotic inflation works, with a sufficiently
small self-coupling. Thus inflation at present appears to be a
plausible solution to the problem of providing reasonable cosmological
initial conditions (although it sheds no light at all on the
fundamental question why the cosmological constant is so small now).
Many variations of the basic idea of inflation have been worked out,
and the following sections will discuss two recent developments
in a little more detail. Linde (1995) recently classified these
inflationary models in an interesting and useful way: see Table 1.7.


\begin{table*}
\caption{Linde's Classification of Inflation Models}
\label{ta:LindeInfl}

\font\caps=cmcsc10
\makeatletter
\def\lsim{\mathrel{\mathpalette\@versim<}}
\def\gsim{\mathrel{\mathpalette\@versim>}}
\def\@versim#1#2{\vcenter{\offinterlineskip
        \ialign{$\m@th#1\hfil##\hfil$\crcr#2\crcr\sim\crcr } }}
\makeatother

\vskip .1in
\hrule

\parindent=0pt
\def\9{\hphantom 0}
\let\ul=\underline
\def\bigfrac#1#2{{\displaystyle{#1\over #2}}}

\vskip1pc
{\caps How Inflation Begins}
\vskip6pt

\begin{tabular}{@{\hspace{9pt}}ll}
Old Inflation& $T_{\rm initial}$ high, $\phi_{\rm in}\approx0$ is
                         false vacuum until phase transition \\
     &     Ends by bubble creation; Reheat by bubble collisions
\end{tabular}
\vskip6pt

\quad New Inflation \quad Slow roll down $V(\phi)$, no phase transition
\vskip6pt

\quad Chaotic Inflation \quad  Similar to New Inflation, but $\phi_{\rm in}$ essentially arbitrary:

\quad \hphantom{Chaotic Inflation}
   \qquad any region with ${1\over2}\dot\phi^2+{1\over2} (\partial_i\phi)^2
                         \lsim V(\phi)$ inflates
\vskip6pt

\quad Extended Inflation \quad Like Old Inflation, but slower (e.g., power $a\propto t^p$),
\quad \hphantom{Extended Inflation} \qquad\quad so phase transition can finish

\vskip1pc
{\caps Potential $V(\phi)$ During Inflation}

\vskip6pt

\quad Chaotic typically $V(\phi) = \Lambda \phi^n$, can also use $V = V_0
    e^{\alpha \phi}$, etc.

\hskip7.3cm $\Rightarrow a \propto t^p,\ p = 16\pi/\alpha^2 \gg1$

\vskip1pc
{\caps How Inflation Ends}
\vskip3pt

\quad  First-order phase transition --- e.g., Old or Extended inflation

\quad   Faster rolling $\rightarrow$ oscillation --- e.g., Chaotic
   $V(\phi)^2 \Lambda \phi^n$

\quad  ``Waterfall" --- rapid roll of $\sigma$ triggered by slow roll
      of $\phi$

\vskip1pc
{\caps (Re)heating}
\vskip3pt

\quad  Decay of inflatons

\quad  ``Preheating" by parametric resonance, then decay

\vskip1pc
{\caps Before Inflation?}
\vskip3pt

\quad Eternal Inflation?  Can be caused by

\qquad$\bullet$\enskip  Quantum $\delta \phi \sim H/2\pi >$ rolling $\Delta\phi =
        \phi \Delta t = \phi H^{-1} \approx V'/V$

\qquad$\bullet$\enskip Monopoles or other topological defects

\vskip .1in
\hrule

\end{table*}

\subsection{Eternal Inflation}
\label{sec:pr_origin_eternal}

Vilenkin (1983) and
Linde (1986, 1990) pointed out that if one extrapolates
inflation backward to try to imagine what might have
preceeded it, in many versions of inflation the answer is
``eternal inflation'': in most of the volume of the universe
inflation is still happening, and our part of the expanding
universe (a region encompassing far more than our entire
cosmic horizon) arose from a tiny part of such a region.  To see how
eternal inflation works, consider the simple chaotic model
with $V(\phi)=(m^2/2)\phi^2$.  During the de Sitter Hubble
time $H^{-1}$, where as usual $H^2=(8 \pi G/3)V$, the slow
rolling of $\phi$ down the potential will reduce it by
\begin{equation}
\Delta \phi = \dot \phi \Delta t = -{V' \over {3H}} \Delta t
= {{m_{pl}^2}\over {4 \pi \phi}}.
\end{equation}
Here $m_{Pl}$ is the Planck mass (see Table 1.1).  But there will
also be quantum fluctuations that will change $\phi$ up or
down by
\begin{equation}
\delta \phi = {H\over {2\pi}} = {m \phi \over {{\sqrt{3 \pi}
m_{Pl}}}}
\end{equation}
These will be equal for $\phi_\ast = m_{pl}^{3/2}/2m^{1/2}$,
$V(\phi_\ast) = (m/8m_{Pl})m_{Pl}^4$.  If $\phi \gsim
\phi_\ast$, {\it positive quantum fluctuations dominate} the
evolution: after $\Delta t \sim H^{-1}$, an initial region
becomes $\sim e^3$ regions of size $\sim H^{-1}$, in half of
which $\phi$ increases to $\phi+\delta \phi$.  Since $H
\propto \phi$, this drives inflation faster in these
regions.  Various mechanisms probably cut this off as $\phi
\rightarrow m_{Pl}^2/m$ and $V \rightarrow m_{Pl}^4$ ---
for further discussion and references, see Linde
(1995).  Thus, although $\phi$ at any given point is
likely eventually to roll down the potential and end
inflation, in most of the volume of the metauniverse
$\phi>\phi_\ast$ and inflation is proceeding at
a very fast rate.

Eternal Inflation is eternal in the sense that, once started, it never
ends.  But it remains uncertain whether or not it could have begun an
infinite length of time ago.  Assuming the ``weak energy condition''
$T_{\mu \nu} V^\mu V^\nu \geq 0$ for all timelike vectors $V^\mu$,
i.e. that any observer will measure a positive energy density, Borde
\& Vilenkin (1994) proved that a future-eternal inflationary model
cannot be extended into the infinite past.  However, Borde \& Vilenkin
(1997) have recently shown that the weak energy condition is quite
likely to be violated in inflating spacetimes (except the open
universe inflation models discussed below, \se{pr_origin_open}), so a
``steady-state'' eternally inflating universe may be possible after
all, with no beginning as well as no end.

\subsection{A Supersymmetric Inflation Model}
\label{sec:pr_origin_susy}

We have already considered, in connection with cold dark matter
candidates, why supersymmetry is likely to be a feature of the
fundamental theory of the particle interactions, of which the present
``Standard Model'' is presumably just a low-energy approximation.  If
the higher-energy regime within which cosmological inflation occurs is
described by a supersymmetric theory, there are new cosmological
problems that initially seemed insuperable.  But recent work has
suggested that these problems can plausibly be overcome, and that
supersymmetric inflation might also avoid the fine-tuning otherwise
required to explain the small inflaton coupling corresponding to the
COBE fluctuation amplitude.  Here the problems
will be briefly summarized, and an explanation will be given of
how one such model, due to Ross \& Sarkar (1996;
hereafter RS96) overcomes them.  (An interesting alternative
supersymmetric approach to inflation is sketched in Dine et
al. 1996.)

\def\Mp{M_{\rm Pl}}
\def\mgr{m_{3/2}}

When Pagels and I (1982) first suggested that the lightest supersymmetric
partner particle (LSP), stable because of R-parity, might be the dark
matter particle, that particle was the gravitino in the early version of
supersymmetry then in fashion.  Weinberg (1982) immediately pointed out
that if the gravitino were not the LSP, it could be a source or real
trouble because of its long lifetime $\sim \Mp^2/\mgr^3 \sim (\mgr/{\rm
TeV})^{-3} 10^3$ s, a consequence of its gravitational-strength
coupling to other fields.  Subsequently, it was realized that
supersymmetric theories can naturally solve the gauge hierarchy problem,
explaining why the electroweak scale $M_{\rm EW} \sim 10^2$ GeV is so much
smaller than the GUT or Planck scales.  In this version of supersymmetry,
which has now become the standard one, the gravitino mass will typically
be $\mgr \sim$ TeV; and the late decay of even a relatively small number
of such massive particles can wreck BBN and/or the thermal spectrum of
the CBR. The only way to prevent this is to make sure that the reheating
temperature after inflation is sufficiently low: $T_{\rm RH} \lsim 2
\times 10^9$ GeV (for $\mgr =$ TeV) (Ellis, Kim, \& Nanopoulos 1984,
Ellis et al. 1992).

This can be realized in supergravity theories rather naturally
(RS96).  Define $M \equiv \Mp/(8 \pi)^{1/2} = 2.4\times10^{18}$
Gev.  Break GUT by the Higgs field $\chi$ with vacuum expectation
value (vev) $<\chi> \sim 10^{16}$ GeV.  Break supersymmetry by a
gaugino condensate $<\lambda \lambda> \sim (10^{13} {\rm GeV})^3$;
then the gravitino mass is $\mgr \sim <\lambda\lambda>/M^2 \sim $
TeV. Inflation is expected to inhibit such breaking, so it must occur
afterward.  The inflaton superpotential has the form $I=\Delta^2 M
f(\phi/M)$, with the corresponding potential
\begin{equation}
V(\phi)=e^{|\phi|^2/M^2} \left[ \left( {{\partial I} \over {\partial \phi}} +
{{\phi I} \over {M^2}} \right)^2 - {{3 I^2} \over {M^2}} \right],
\end{equation}
with minimum at $\phi_0$.  Demanding that at this minimum
the potential actually vanishes $V(\phi_0)=0$, i.e., that the
cosmological constant vanishes, implies that
$I(\phi_0)=(\partial I/\partial \phi)_{\phi_0} = 0$.  The
simplest possibility is $I=\Delta^2 (\phi - \phi_0)^2 /M$.
Requiring that $\partial V / \partial \phi|_0 = 0$ for a
sufficiently flat potential, implies that $\phi_0=M$ and
that the second derivative also vanishes at the origin; thus
\begin{equation}
V(\phi) = \Delta^4 \left[ 1 - 4({\phi \over M})^3 + {13\over 2} ({\phi \over M})^4 -
8({\phi \over M})^5 + ... \right]
\end{equation}
(Holman, Raymond, \& Ross 1984).  This particular inflaton potential is
of the ``new inflation'' type, and corresponds to tilt $n_p=0.92$ and
a number of $e$-folds during inflation
\begin{equation}
N =  \int_{\phi_{\rm in}}^{\phi_{\rm end}} \left({{-V'}\over V}\right)
\, d\phi =   {M\over {12\Delta}},
\end{equation}
assuming that the starting value of the inflaton field $\phi_{\rm in}$
is sufficiently close to the origin (which has relatively small but
nonvanishing probability --- the $\phi$ field presumably has a broad
initial distribution).  Matching The COBE fluctuation amplitude
requires that $\Delta/M=1.4\times 10^{-4}$, which in turn implies that
$N\sim 10^3$, $m_\phi \sim \Delta^2 /M \sim 10^{11}$ GeV, $T_{\rm RH}
\sim 10^5$ GeV (parametric resonance reheating does not occur).  Such
a low reheat temperature insures that there will be no gravitino
problem, and requires that the baryon asymmetry be generated by
electroweak baryogenesis --- which appears to be viable as long as the
theory contains adequately large CP violation.

Note the following features of the above scenario: inflation occurs at an
energy scale far below the GUT scale, so there is essentially no gravity
wave contribution to the large-angle CMB fluctuations (i.e., $T/S
\approx 0$) even through there is significant tilt ($n_p=0.92$ for the
particular potential above); there is a low reheat temperature, so
electroweak baryogenesis is required; and the universe is predicted to be
very flat since there are many more $e$-folds than required to solve the
flatness problem.

\subsection{Inflation with $\Omega_0<1$}
\label{sec:pr_origin_open}

Can inflation produce a region of negative curvature larger than our present
horizon --- for example, a region with $\Omega_0 < 1$ and $\Lambda=0$?
The old approach to this problem was to imagine that there might be
just enough inflation to solve the horizon problem, but not quite enough
to oversolve the flatness problem, e.g. $N \sim 60$ (Steinhardt 1990).
This requires fine tuning, but the real problem with this approach is
that the resulting region will not be smooth enough to agree with the
small size of the quadrupole anisotropy $Q$ measured by COBE.  According
to the Grischuk-Zel'dovich (1978) theorem (cf. Garcia-Bellido et al. 1995),
$\delta \sim 1$ fluctuations
on a super-horizon scale $L > H_0^{-1}$ imply $Q \sim (L H_0)^{-2}$.
COBE measured $Q_{\rm rms} < 2 \times 10^{-5}$, which implies in turn
that the region containing our horizon must be homogeneous on a scale $L
\gsim 500 H_0^{-1}$, i.e. $N \gsim 70$, $|1-\Omega_0| \lsim 10^{-4}$.

Recently a new approach was discovered, based on the fact that a bubble
created from de Sitter space by quantum tunneling tends to be spherical
and homogeneous if the tunneling is sufficiently improbable.  The
interior of such bubbles are quite empty, i.e., they are a region of
negative curvature with $\Omega \rightarrow 0$.  That was why, in ``old
inflation,'' the bubbles must collide to fill the universe with energy;
and the fact that this does not happen (because the bubbles grow only at
the speed of light while the space between them grows superluminally) was
fatal for that approach to inflation (Guth \& Weinberg
1983).\footnote{Although there have been attempts to revive Old
Inflation within scenarios in which the inflation is slower so that the
bubbles can collide, it remains to be seen whether any such
Extended Inflation model can be
sufficiently homogeneous to be entirely satisfactory.}  But now this
defect is turned into a virtue by arranging to have a second burst of
inflation inside the bubble, to drive the curvature back toward zero,
i.e., $\Omega_0 \rightarrow 1$.  By tuning the amount of this second
period of inflation, it is possible to produce any desired value of
$\Omega_0$ (Sasaki, Tanaka, \& Yamamoto 1995; Bucher,
Goldhaber, \& Turok 1995; Yamamoto, Sasaki, \& Tanaka 1995).
The old problem of too much inhomogeneity beyond the horizon producing
too large a value of the quadrupole anisotropy is presumably solved
because the interior of the bubble produced in the first inflation is
very homogeneous.

I personally regard this as an existence proof that inflationary models
producing $\Omega_0 \sim 0.3$ (say) can be constructed which are not
obviously wrong.  But I do not regard such contrived  models as being as
theoretically attractive as the simpler models in which the universe after
inflation is predicted to be flat.  (Somewhat simpler two-inflaton models
giving $\Omega_0<1$ have been constructed by Linde \& Mezhlumian 1995.)
Note also that if varying amounts of inflation are possible, much greater
volume is occupied by the regions in which more inflation has occurred,
i.e., where $\Omega_0 \approx 1$.  But the significance of such arguments
is uncertain, since no one knows whether volume is the appropriate
measure to apply in calculating the probability of our horizon having any
particular property.

The spectra of density fluctuations produced in inflationary models with
$\Omega_0 < 1$ tend to have a lot of power on very large scales.
However, when such spectra are normalized to the COBE CMB anisotropy
observations, the spherical harmonics with angular wavenumber $\ell
\approx 8$ have the most weight statistically, and all such models have
similar normalization (Liddle et al. 1996a).

\subsection{Inflation Summary}
\label{sec:pr_origin_summary}

The key features of all inflation scenarios are a period of superluminal
expansion, followed by (``re-'')heating which converts the energy stored
in the inflaton field (for example) into the thermal energy of the hot
big bang.

Inflation is {\it generic}: it fits into many versions of particle
physics, and it can even be made rather natural in modern supersymmetric
theories as we have seen.  The simplest models have inflated away all
relics of any pre-inflationary era and result in a flat universe after
inflation, i.e., $\Omega=1$ (or more generally $\Omega_0 + \Omega_\Lambda
= 1$).  Inflation also produces scalar (density) fluctuations that
have a primordial spectrum
\begin{equation}
 \left({\delta \rho \over \rho}\right)^2
 \sim \left(V^{3/2} \over {m_{Pl}^3 V'}\right)^2  \propto k^{n_p},
\end{equation}
where $V$ is the inflaton potential and $n_p$ is the primordial spectral
index, which is expected to be near unity (near-Zel'dovich spectrum).
Inflation also produces tensor (gravity wave) fluctuations, with spectrum
\begin{equation}
 P_t(k) \sim \left({V \over m_{Pl}}\right)^2 \propto k^{n_t},
\end{equation}
where the tensor spectral index $n_t \approx (1-n_p)$ in many models.

The quantity $(1-n_p)$ is often called the ``tilt'' of the spectrum;
the larger the tilt, the more fluctuations on small spatial scales
(corresponding to large $k$) are suppressed compared to those on
larger scales. The scalar and tensor waves are generated by
independent quantum fluctuations during inflation, and so their
contributions to the CMB temperature fluctuations add in quadrature.
The ratio of these contributions to the quadrupole anisotropy
amplitude $Q$ is often called $T/S \equiv Q_t^2/Q_s^2$; thus the
primordial scalar fluctuation power is decreased by the ratio
$1/(1+T/S)$ for the same COBE normalization, compared to the situation
with no gravity waves ($T=0$).  In power-law inflation,
$T/S = 7(1-n_p)$.  This is an approximate equality in other popular
inflation models such as chaotic inflation with $V(\phi) = m^2 \phi^2$
or $\lambda \phi^4$.  But note that the tensor wave amplitude is just
the inflaton potential during inflation divided by the Planck mass, so
the gravity wave contribution is negligible in theories like the
supersymmetric model discussed above in which inflation occurs at an
energy scale far below $m_{Pl}$.  Because gravity waves just redshift
after they come inside the horizon, the tensor contributions to CMB
anisotropies corresponding to angular wavenumbers $\ell \gg 20$, which
came inside the horizon long ago, are strongly suppressed compared to
those of scalar fluctuations.  The indications from presently
available data (Netterfield et al. 1996; cf. Tegmark 1996, and Silk's
article in this volume) are that the CMB amplitude is rather high for
$\ell \approx 200$, approximately in agreement with the predictions of
standard CDM with $h \approx 0.5$, $\Omega_b \approx 0.1$, and scalar
spectral index $n_p=1$.  This suggests that there is little room for
gravity-wave contributions to the low-$\ell$ CMB anisotropies, i.e.,
that $T/S \ll 1$.  Thus tests of inflation involving the gravity-wave
spectrum will be very difficult.  Fortunately, inflation can be tested
with the data expected soon from the next generation of CMB
experiments, since it makes very specific and discriminatory
predictions regarding the relative locations of the acoustic peaks in
the spectrum, for example the ratio of the first peak location
to the spacing between the peaks $\ell_1/\Delta \ell \approx 0.7-0.9$
(Hu \& White 1996, Hu et al. 1997).

On the other hand, inflation is also {\it Alice's restaurant} where,
according to the Arlo Guthrie song, ``...you can get anything you
want ... excepting Alice''.  It's not even clear what ``Alice'' you can't
get from inflation.  It was initially believed that inflation predicts a
flat universe.  But now we know that you
\begin{itemize}
\item can get $\Omega_0 < 1$ (with $\Lambda=0$), as discussed in the
previous section.
\item can make models consistent with supergravity and the sort of
four-dimen\-sional physics expected from superstrings, in which case one
may expect that inflation occurs at a relatively low energy scale, which
implies $T/S \approx 0$, a low reheat temperature implying no production
of topological defects and presumably requiring that baryosynthesis occur
at the electroweak phase transition, and plenty of inflation implying
that $\Omega_0 \approx 1$.
\item can alternatively get strings or other topological defects such as
textures during or at the end of inflation (e.g. Hodges \& Primack 1991)
--- which however probably requires tuning of the inflation and/or string
model, for example to avoid a fractal pattern of structure-forming
defects, which would conflict with the observed homogeneity of structure
on very large scales.
\end{itemize}

And in many versions of inflation, the most reasonable answer to the
question ``what happened before inflation'' appears to be eternal
inflation, which implies that in most of the meta--universe,
exponentially far beyond our horizon, inflation never stopped.

\def\lcdm{$\Lambda$CDM}

\section{Comparing DM Models to Observations: \lcdm\ vs. CHDM}
\label{sec:pr_models}

\subsection{Building a Cosmology: Overview}
\label{sec:pr_models_overview}

\begin{figure}[htb]   
\vskip5.5pc
\centering
\centerline{\psfig{file=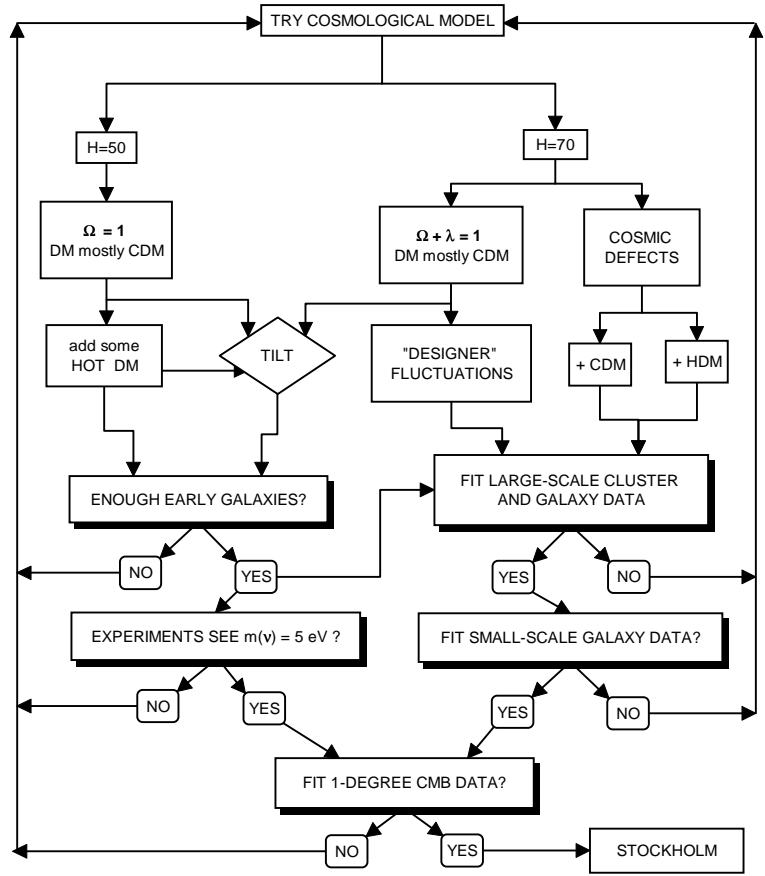,width=10cm}} 
\caption{Building a Cosmological Model. (This figure was inspired by
similar flow-charts on inventing dark matter
candidates, by David Weinberg and friends, and by Rocky Kolb.)}
\label{fig:pr_3}
\end{figure}

An effort has been made to summarize the main issues in cosmological
model-building in \fig{pr_3}.  Here
the choices of cosmological parameters, dark matter composition, and
initial fluctuations that specify the model are shown at the top of
the chart, and the types of data that each cosmological model must
properly predict are shown in the boxes with shaded borders in the
lower part of the chart.  Of course, the chart only shows a few of the
possibilities.  Models in which structure arises from gravitational
collapse of adiabatic inflationary fluctuations and in which most of
the dark matter is cold are very predictive.  Since such models have
also been studied in greatest detail, this class of models
will be the center of attention here.

Perhaps the most decisive issue in model building is the value of the
cosmological expansion rate, the Hubble parameter $h$.  If $h \approx
0.7$ as some observers still advocate, and the age of the universe $t_0
\gsim 13$ Gyr, then only low-$\Omega_0$ models can be consistent with
general relativity.\footnote{It is important to appreciate that the
possible $t_0-H_0$ (age-expansion rate) conflict goes to the heart of
GR and does not depend on cosmological-model-dependent issues like the
growth rate of fluctuations.  As explained in \se{pr_cos} besides GR
itself the only other theoretical input needed is the cosmological
principle: we do not live in the center of a spherical universe; any
observer would see the same isotropy of the distant universe, as
reflected in particular in the COBE observations. That is enough to
imply the Friedmann-Robertson-Walker equations, which give the
$t_0-H_0$ connection. GR is not just a theory whose intrinsic beauty
and great success in describing data on relatively small scales
encourage us to extrapolate it to the scale of the entire observable
universe.  It is the only decent theory of gravity and cosmology that
we have.} Depending on just how large $h$ and $t_0$ are, a positive
cosmological constant may also be necessary for consistency with GR,
since even in a universe with $\Omega \rightarrow 0$ the age $t_0
\rightarrow H_0^{-1} = 9.78 h^{-1}$ Gyr (see \fig{pr_1}).  Thus, with
$\Lambda=0$ and $\Omega_0 \rightarrow 0$, $h < 0.75 (13 {\rm
Gyr}/t_0)$.  The upper limit on $h$ is stronger, the larger $\Omega_0$
is: with $\Lambda=0$ and $\Omega_0 \geq 0.3$, $h < 0.61 (13 {\rm
Gyr}/t_0)$; with $\Lambda=0$ and $\Omega_0 \geq 0.5$, $h < 0.57 (13
{\rm Gyr}/t_0)$.  It has been argued above that the evidence strongly
suggests that $\Omega_0 \geq 0.3$, especially if the initial
fluctuations were Gaussian; thus, if we assume values of $h=0.7$ and
$t_0=13$ Gyr, we must include a positive cosmological constant.  For
definiteness, the specific choice shown is $\Omega_0 + \Omega_\Lambda
= 1$, corresponding to the flat cosmology inspired by standard
inflation.

In such a $\Lambda$CDM model, one might initially try the Zel'dovich
primordial fluctuation spectrum, i.e. $P_s(k)=A_s k^{n_p}$ with $n_p=1$.
However, this might not predict the observed abundance of clusters when
the amplitude of the spectrum is adjusted to agree with the COBE data on
large scales. If $\Omega_0 > 0.3$, then COBE-normalized $\Lambda$CDM
predicts more rich clusters than are actually observed.  In that case, it
will be necessary to change the spectrum.  The simplest way to do that is
to add some ``tilt'' --- i.e., assume that $n_p < 1$.  This adds one
additional parameter.  One can also consider more complicated
``designer'' primordial spectra with two or more parameters, which as I
have mentioned can also be produced by inflation.  In any case, it is then
also necessary to check that the large-scale cluster and galaxy
correlations on large scales are also in agreement with experiment.  As
we will see, this is indeed the case for typical $\Lambda$CDM models.  In
order to see whether such a model also correctly predicts the galaxy
distribution on smaller spatial scales $\lsim 10 \hMpc$, on which the
fluctuations in the number counts of galaxies $N_g$ are nonlinear ---
i.e., $(\delta N_g/N_g)_{\rm rms} \geq 1$ --- it is necessary to do N--body
simulations.  As we will discuss, these are probably rather accurate in
showing the distribution of dark matter on intermediate scales.  But
simulations are not entirely reliable on the scales of clusters, groups,
and individual galaxies since even the best available simulations include
only part of the complicated physics of galaxy formation, and omit or
treat superficially crucial aspects such as the feedback from supernovae.
Thus one of the main limitations of simulations is ``galaxy
identification'' --- locating the likely sites of galaxies in the
simulations, and assigning them appropriate morphologies and
luminosities.

If $h \approx 0.5$ and $t_0 \lsim 13$ Gyr, or if $h \approx 0.6$ and
$t_0 \lsim 11$ Gyr, then models with critical density, $\Omega=1$, are
allowed.  Since the COBE-normalized CDM model greatly overproduces
clusters, it will be necessary to make some modification to decrease
the fluctuation power on cluster scales --- for example, tilt the
spectrum or change the assumed dark matter composition. As we have
discussed, hot dark matter cannot preserve fluctuations on small
scales, so adding a little hot dark matter to the mix of cold dark
matter and baryons will indeed decrease the amount of cluster-scale
power.  A possible problem is that tilting or adding hot dark matter
will also decrease the amount of power on small scales, which means
that protogalaxies will form at lower redshift. So such models must be
checked against data indicating the amount of small-scale structure at
redshifts $z \geq 3$ --- for example, against the abundance of neutral
hydrogen in damped Lyman $\alpha$ absorption systems in quasar
spectra, or the protogalaxies seen in emission at high redshift.
Acceptable models must of course also fit the data on large and
small-scale galaxy distributions.  As we will see, $\Omega=1$
COBE-normalized models with a mixture of Cold and Hot Dark Matter
(CHDM) can do this if the hot fraction $\Omega_\nu
\approx 0.2$.

The ultimate test for all such cosmological models is whether they
will agree with the CMB anisotropies on scales of a degree and below.
Such data is just beginning to become available from ground-based and
balloon-borne experiments, and continuing improvements in the
techniques and instruments insure that the CMB data will become
steadily more abundant and accurate. CMB maps of the whole sky must
come from satellites, and it is great news for cosmology that NASA has
approved the MAP satellite which is expected to be ready for launch by
2001, and that the European Space Agency is planning the even more
ambitious COBRAS/SAMBA satellite, recently renamed Planck, to be
launched a few years later.

Both sorts of models that have been discussed --- $\Omega=1$ tilted CDM
(tCDM) or CHDM, and $\Omega_0+\Omega_\Lambda=1$ \lcdm --- are simple,
one-parameter modifications of the original standard CDM model.
The astrophysics community has been encouraged by the great initial
success of this theory in explaining the existence of galaxies and
fitting galaxy and cluster data (BFPR, DEFW), and the fact that biased CDM
only missed predicting the COBE observations by a factor of about 2.
The other reason why the CDM-variant models have been studied in much
more detail than other cosmological models is that they are so
predictive: they predict the entire dark matter distribution in terms
of only one or two model parameters (in addition to the usual
cosmological parameters), unlike non-Gaussian models based on randomly
located seeds, for example.  Of course, despite the relatively good
agreement between observations and the predictions of the best CDM
variants, there is no guarantee that such models will ultimately be
successful.

Although the cosmic defect models (cosmic strings, textures) are in
principle specified in terms of only a small number of parameters (in
the case of cosmic strings, the string tension parameter $G\mu$ plus
perhaps a couple of parameters specifying aspects of the evolution of
the string network), in practice it has not yet been possible for any
group to work out the predicted galaxy distribution in such models.
Most proponents of cosmic defect models have assumed an $\Omega=1$,
$H_0 \approx 50$ cosmology, but the chart refers instead to a cosmic
defects option under $H_0=70$.  This is done because it would be
worthwhile to work out a low-$\Omega$ case as well, since in defect
models there is less motivation to assume the inflation-inspired flat
($\Omega_0 + \Omega_\Lambda=1$) cosmology.

\subsection{Lessons from Warm Dark Matter}
\label{sec:pr_models_warm}

As has been said, the chart in \fig{pr_3} only includes a few
of the possibilities.  But many possibilities that have been examined
are not very promising.  The problems with a pure Hot Dark Matter
(HDM) adiabatic cosmology have already been mentioned.  It will be
instructive to look briefly at Warm Dark Matter, to see that some
variants of CDM have less success than others in fitting cosmological
observations, and also because there is renewed interest in WDM.
Although CHDM and WDM are similar in the sense that both are
intermediate models between CDM and HDM, CHDM and WDM are quite
different in their implications.  The success of some but not other
modifications of the original CDM scenario shows that more is required
than merely adding another parameter.

As explained above, WDM is a simple modification of HDM, obtained by
changing the assumed average number density $n$ of the particles. In
the usual HDM, the dark matter particles are neutrinos, each species
of which has $n_\nu=113$ cm$^{-3}$, with a corresponding mass of
$m(\nu) = \Omega_\nu \rho_0/n_\nu = \Omega_\nu 92 h^2$ eV.  In WDM,
there is another parameter, $m/m_0$, the ratio of the mass of the warm
particle to the above neutrino mass; correspondingly, the number
density of the warm particles is reduced by the inverse of this
factor, so that their total contribution to the cosmological density
is unchanged. It is true of both of the first WDM particle candidates,
light gravitino and right-handed neutrino, that these particles
interact much more weakly than neutrinos, decouple earlier from the
hot big bang, and thus have diluted number density compared to
neutrinos since they do not share in the entropy released by the
subsequent annihilation of species such as quarks. This is analogous
to the neutrinos themselves, which have lower number density today
than photons because the neutrinos decouple before e$^+$e$^-$
annihilation (and also because they are fermions).

\begin{figure}[htb]   
\centering
\centerline{\psfig{file=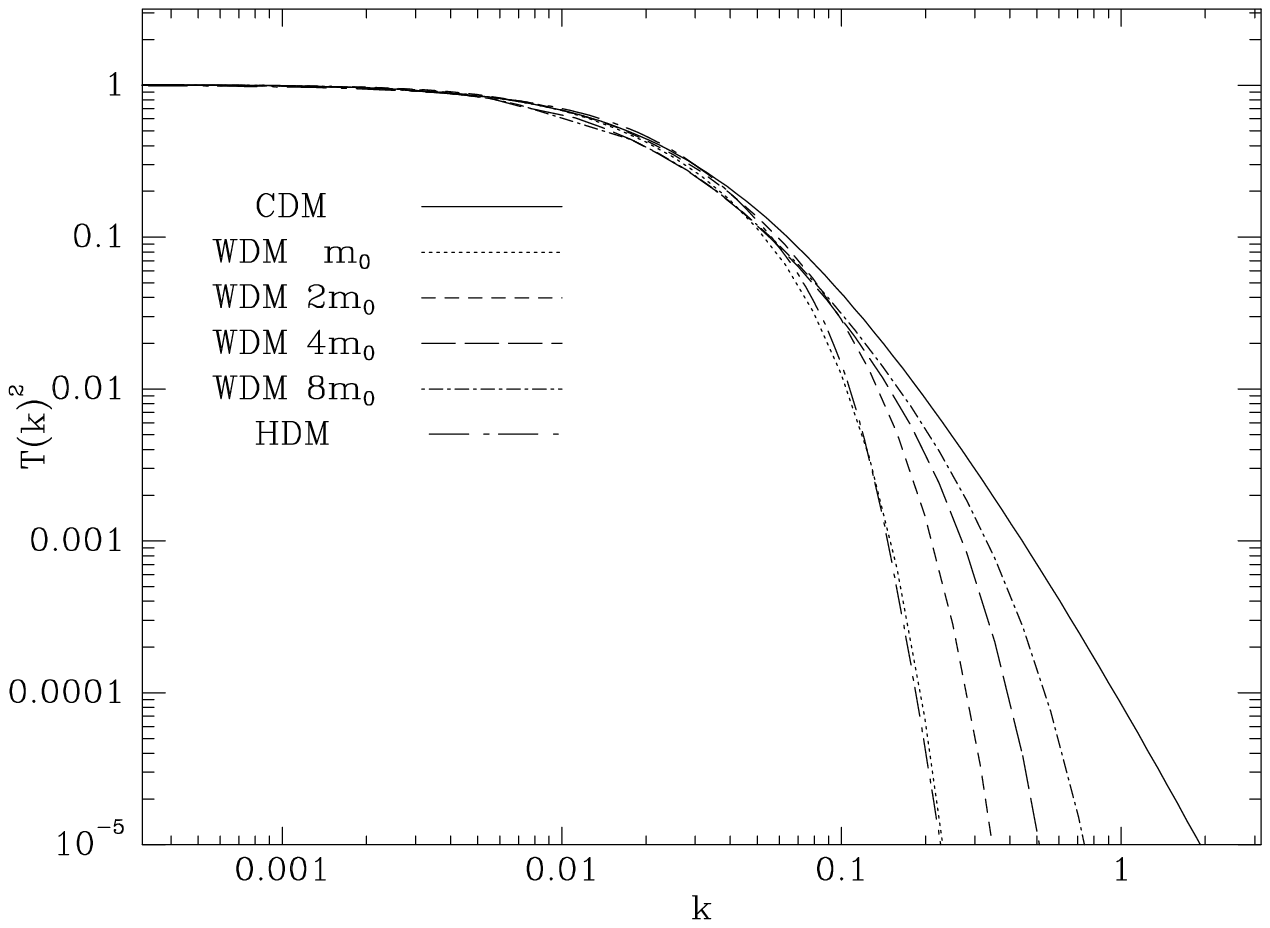,width=10cm}}
\centerline{\psfig{file=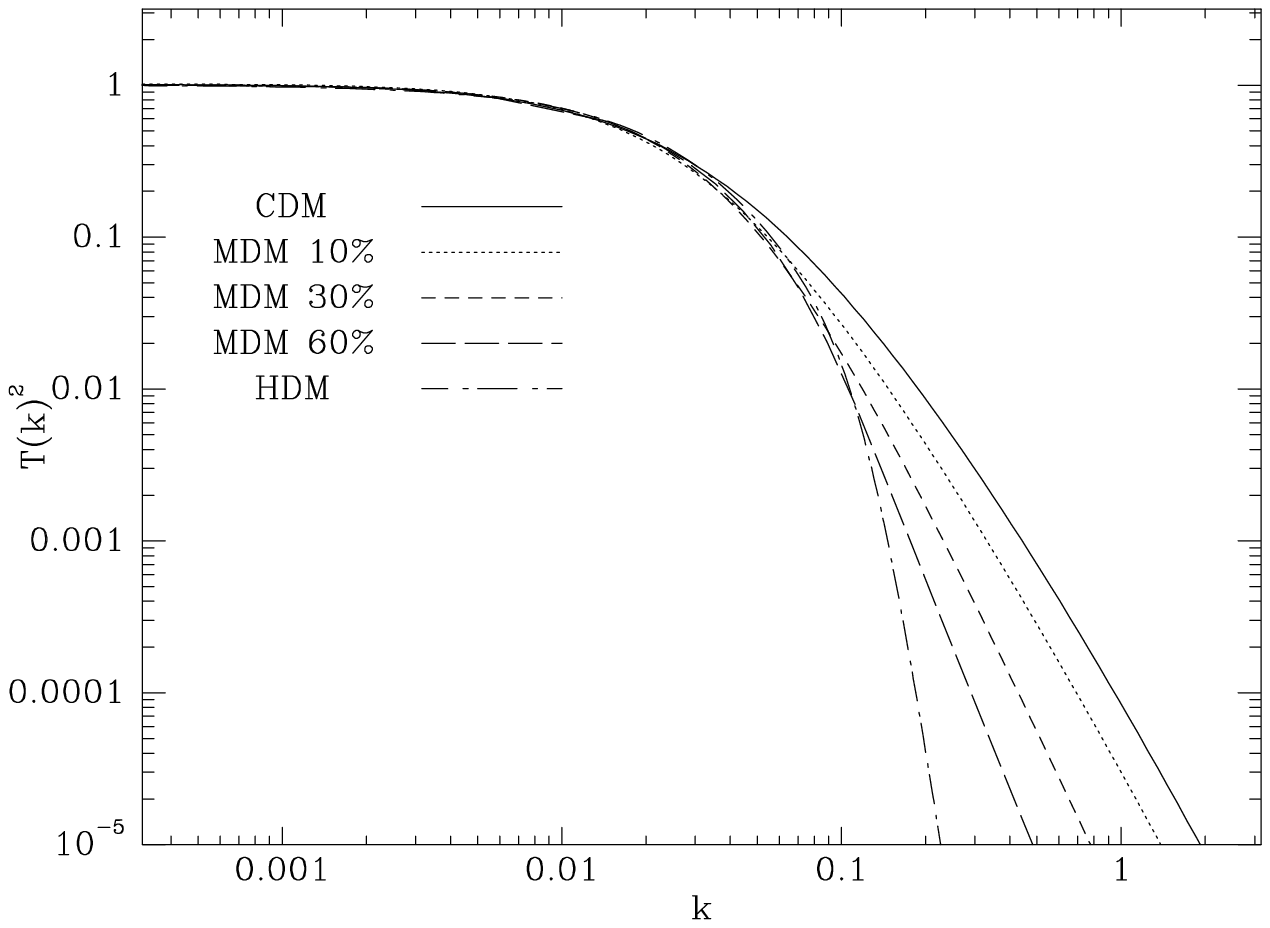,width=10cm}}
\vskip-1pc
\caption{%
The square of the linear transfer function $T(k)$
vs. wavenumber $k=(2\pi)/\lambda$ (in units of $h$ Mpc$^{-1}$) for (a)
Warm Dark Matter (WDM), and (b) Mixed Dark Matter (MDM --- CHDM with
$N_\nu=1$ neutrino species).  (From Colombi, Dodelson, \& Widrow 1996,
used by permission.)}
\label{fig:pr_4}
\end{figure}

In order to investigate the cosmological implications of any dark
matter candidate, it is necessary to work out the gravitational
clustering of these particles, first in linear theory, and then after
the amplitude of the fluctuations grows into the nonlinear regime.
Colombi, Dodelson, \& Widrow (1996) did this for WDM, and \fig{pr_4}
from their paper compares the square of the linear transfer functions
$T(k)$ for WDM and CHDM.  The power spectrum $P(k)$ of fluctuations is
given by the quantity plotted times the assumed primordial power
spectrum $P_p(k)$, $P(k) = P_p(k) T(k)^2$.  The usual assumption
regarding the primordial power spectrum is $P_p(k)=Ak^{n_p}$, where
the ``tilt'' equals $1 - n_p$, and the untilted, or Zel'dovich,
spectrum corresponds to $n_p=1$.

\begin{figure}[htb]   
\centering
\centerline{\psfig{file=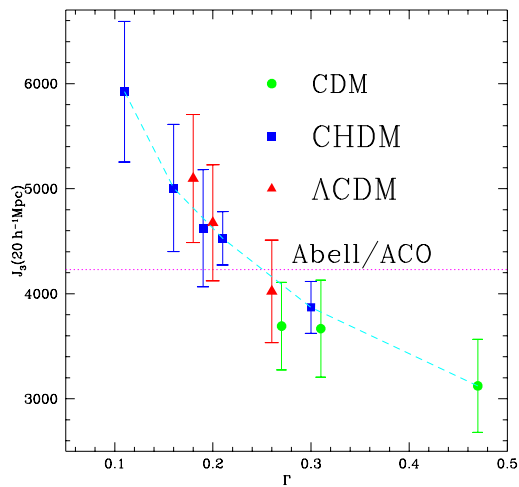,width=7cm}}
\caption{%
The value of the $J_3$ integral for sCDM and a number of \lcdm\ and
CHDM models evaluated at $R=20\hmpc$ is plotted against the value of
the shape parameter $\Gamma$ defined in the text.  As usual, $J_3(R) =
\int_0^R \xi_{cc}(r)r^2 dr$, where $\xi_{cc}$ is the cluster
correlation function.  The horizontal dotted line is the $J_3$ value
for the Abell/ACO sample.  The squares connected by the dashed line
correspond from left to right to CHDM with $n_\nu=1$ neutrino species
and $\Omega_\nu=0.5,\, 0.3,\, 0.2,\, 0.1,$ and 0 (sCDM); the square
slightly below the dashed line corresponds to CHDM with $N_\nu=2$
and $\Omega_\nu=0.2$; all these models have $\Omega=1$, $h=0.5$. and
no tilt.  The triangles correspond (l-to-r) to \lcdm\ with
($\Omega_0$,$h$) = (0.3,0.7), (0.4,0.6) and (0.5,0.6).  The two
circles on the left correspond to CDM with $h=0.4$ and (l-to-r) tilt
$(1-n_p)=0.1$ and 0.  These points and error bars are from a suite of
truncated Zel'dovich approximation (TZA) simulations, checked by
N-body simulations. (From Borgani et al. 1996.)}
\label{fig:pr_5}
\end{figure}

One often can study large scale structure just on the basis of such
linear calculations, without the need to do computationally expensive
simulations of the non-linear gravitational clustering.  Such studies
have shown that matching the observed cluster and galaxy correlations
on scales of about 20-30 $h^{-1}$ Mpc in CDM-type theories requires
that the ``Excess Power'' $EP\approx 1.3$, where
\begin{equation}
EP \equiv {{\sigma(25 \hMpc)/\sigma(8 \hMpc)} \over
              {(\sigma(25 \hMpc)/\sigma(8 \hMpc))_{sCDM}}},
\end{equation}
and as usual $\sigma(r)=(\delta \rho/\rho)(r)$ is the rms fluctuation
amplitude in randomly placed spheres of radius $r$.
The $EP$ parameter was introduced by Wright et al. (1992), and Borgani et
al. (1996) has shown that $EP$ is related to the spectrum shape parameter
$\Gamma$ introduced by Efstathiou, Bond, and White (1992) (cf. Bardeen et
al. 1986) by $\Gamma \approx 0.5 (EP)^{-3.3}$. For CDM and the \lcdm\
family of models, $\Gamma=\Omega h$; for CHDM and other models, the
formula just quoted is a useful generalization of the spectrum shape
parameter since the cluster correlations do seem to be a function of this
generalized $\Gamma$, as shown in \fig{pr_5}.  As this
figure shows, $\Gamma \approx 0.25$ to match cluster correlation data.
Peacock \& Dodds (1994) have shown that $\Gamma \approx 0.25$ also is
required to match large scale galaxy clustering data.  This
corresponds to $EP \approx 1.25$.

\begin{figure}[htb]   
\centering
\centerline{\psfig{file=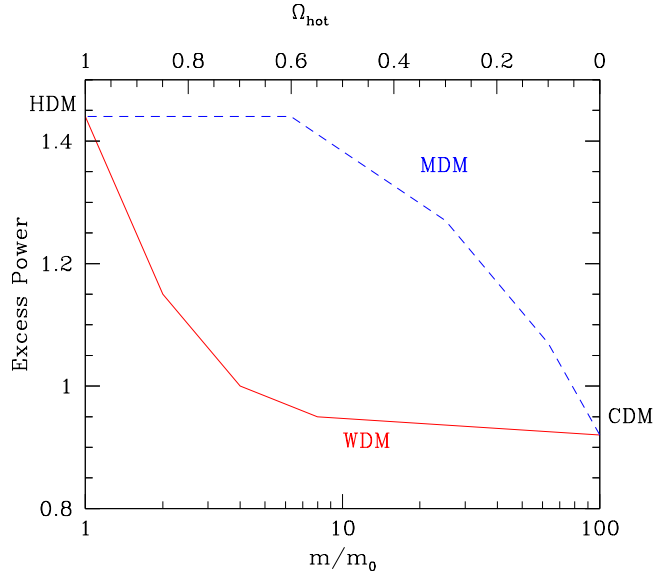,width=9cm}}
\vskip-1pc
\caption{%
Excess power $EP$ in the two models discussed that
interpolate between CDM and HDM.  Solid curve shows $EP$ as a function
of the WDM parameter $m/m_0$; note how quickly it becomes similar to
CDM.  Dashed curve shows how $EP$ for Mixed Dark Matter (MDM --- CHDM
with $N_\nu=1$ neutrino species) depends on $\Omega_\nu$.  The
observationally preferred value is $EP \approx 0.25$.  (From Colombi,
Dodelson, \& Widrow 1996, used by permission.)}
\label{fig:pr_6}
\end{figure}

Since calculating $\sigma(r)$ is a simple matter of integrating the power
spectrum times the top-hat window function,
\begin{equation}
\sigma^2(r)=\int_0^\infty P(k) W(kr) k^2 dk
\end{equation}
the linear calculations shown in \fig{pr_4} immediately
allow determination of $EP$ for WDM and CHDM.  The results are shown in
\fig{pr_6}, in which the lower horizontal axis represents
the values of the WDM parameter $m/m_0$ (with $m/m_0=1$ representing the
HDM limit), and the upper horizontal axis represents the values of the
CHDM parameter $\Omega_\nu$.  This figure shows that for WDM to give the
required $EP$, the parameter value $m/m_0 \approx 1.5-2$, while for CHDM
the required value of the CHDM parameter is $\Omega_\nu \approx 0.3$.
But one can see from \fig{pr_4} that in WDM with $m/m_0
\gsim 2$, the spectrum lies a lot lower than the CDM spectrum at $k\gsim
0.3 h^{-1}$ Mpc (length scales $\lambda \lsim 20 \hMpc$).  This in
turn implies that formation of galaxies, corresponding to the
gravitational collapse of material in a region of size $\sim 1$ Mpc,
will be strongly suppressed compared to CDM.  Thus WDM will not be
able to accommodate simultaneously the distribution of clusters and
galaxies.  But CHDM will do much better --- note how much lower
$T(k)^2$ is at $k \gsim 0.3 h$ Mpc$^{-1}$ for WDM with $m/m_0 = 2$
than for CHDM with $\Omega_\nu = 0.3$. Actually, as we will discuss
in more detail shortly, CHDM with
$\Omega_\nu=0.3$ turns out, on more careful examination, to have
several defects --- too many intermediate-size voids, too few early
protogalaxies. Lowering $\Omega_\nu$ to about 0.2, corresponding to a
total neutrino mass of about $4.6(h/0.5)^{-2}$ eV, in a model in which
$N_\nu=2$ neutrino species share this mass, fits all this data
(PHKC95).

Probably the only way to accommodate WDM in a viable cosmological
model is as part of a mixture with hot dark matter (Malaney, Starkman,
\& Widrow 1995), which might even arise naturally in a supersymmetric model
(Borgani, Masiero, \& Yamaguchi 1996) of the sort in which the
gravitino is the LSP (Dimopoulos et al. 1996).  Cold plus ``volatile''
dark matter is a related possibility (Pierpaoli et al. 1996); in
these models, the hot component arises from decay of a heavy unstable
particle rather than decoupling of relativistic particles.

There are many more parameters needed to describe the presently
available data on the distribution of galaxies and clusters and their
formation history than the few parameters needed to specify a CDM-type
model. Thus it should not be surprising that at most a few CDM variant
theories can fit all this data.  Once it began to become clear that
standard CDM was likely to have problems accounting for all the data,
after the discovery of large-scale flows of galaxies was announced in
early 1986 (Burstein et al. 1986),
Jon Holtzman in his dissertation research worked
out the linear theory for a wide
variety of CDM variants (Holtzman 1989; cf. also Blumenthal, Dekel, \&
Primack 1988) so that we could see which ones would best fit the data
(Primack \& Holtzman 1992, Holtzman \& Primack 1993; cf. Schaefer
\& Shafi 1993).  The clear
winners were CHDM with $\Omega_\nu\approx 0.3$ if $h\approx 0.5$, and
\lcdm\ with $\Omega_0 \approx 0.2$ if $h\approx 1$.  CHDM had first
been advocated several years earlier (Bonometto \& Valdarnini 1984,
Dekel \& Aarseth 1984, Fang et al. 1984, Shafi \& Stecker 1984) but
was not studied in detail until more recently (starting with Davis
et al. 1992, Klypin et al. 1993).

\subsection{$\Lambda$CDM vs. CHDM --- Linear Theory}
\label{sec:pr_models_lcdm}

\begin{figure}[hp]   
\centering
\centerline{\psfig{file=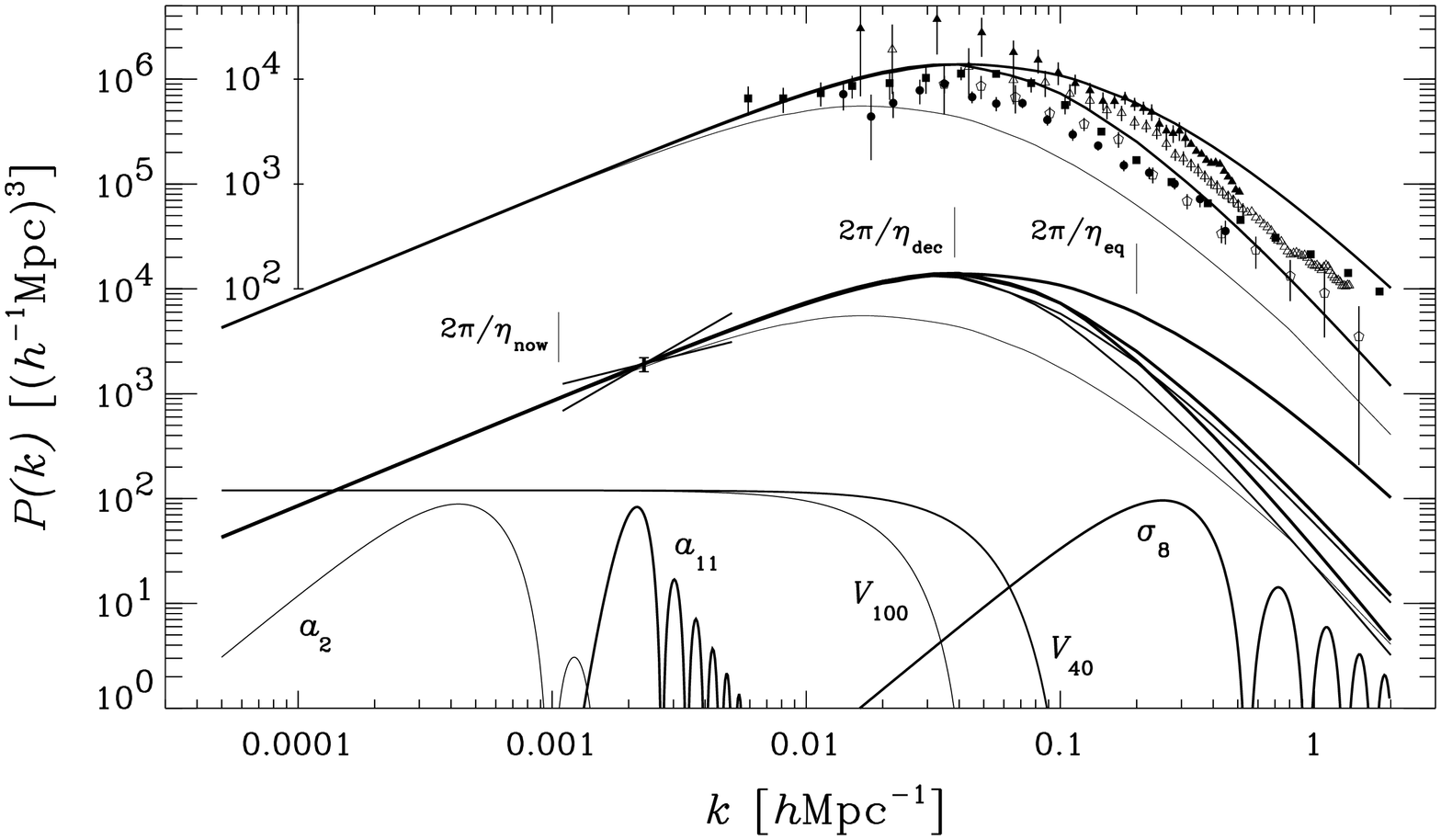,width=11cm,angle=90}}
\vskip1pc
\centerline{\psfig{file=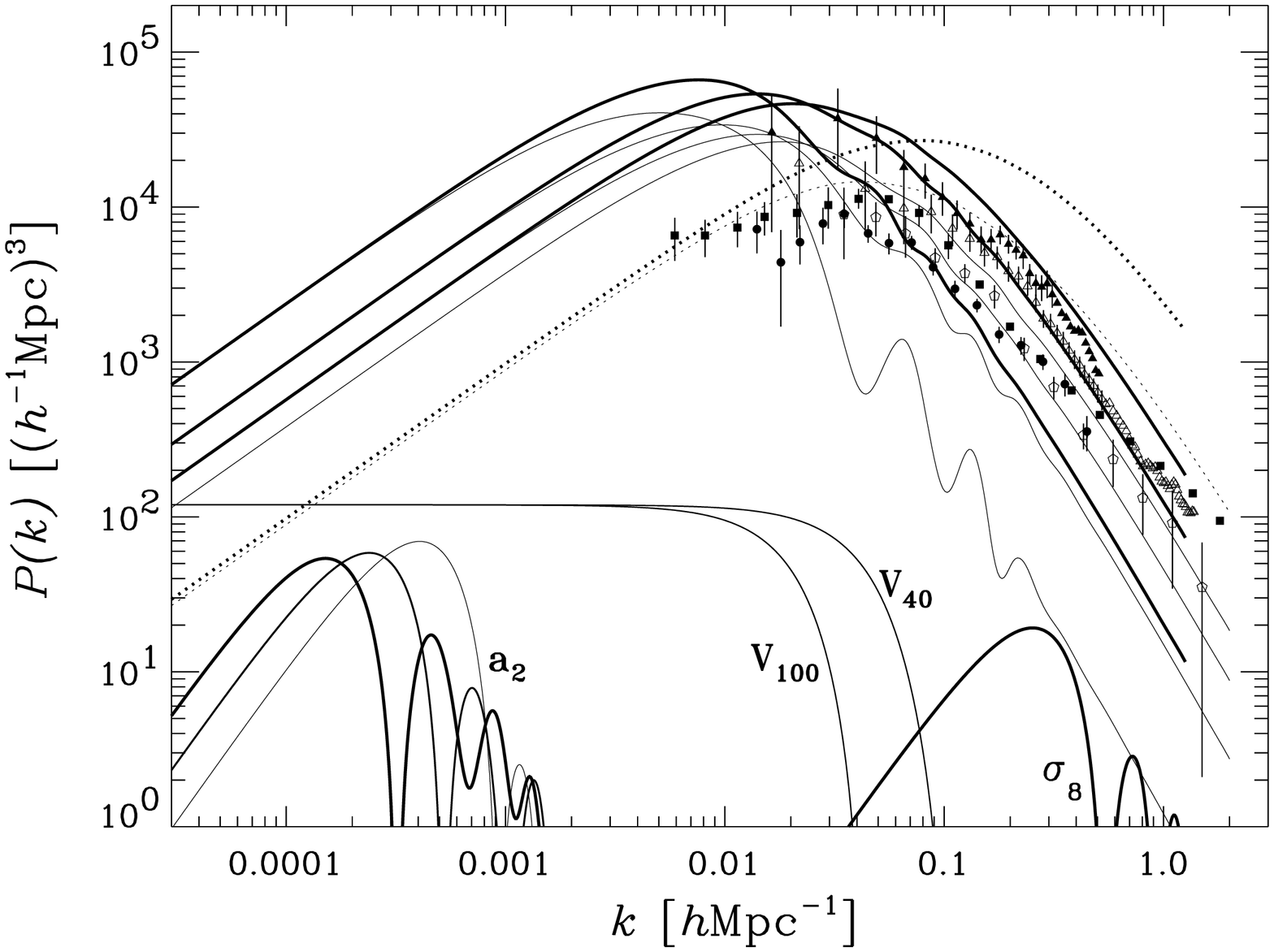,width=12.75cm,angle=90}}
\vskip1pc
\caption{%
Fluctuation power
spectra for COBE-DMR-normalized models: Panel (a) $\Omega=1$ CDM and
CHDM models, Panel (b) $\Omega_0 + \Omega_\Lambda=1$ \lcdm\ models.
The theoretical spectra are discussed in the text.  The data plotted for
comparison is squares -- real space $P(k)$ from angular APM data (Baugh
\& Efstathiou 1993), filled circles -- estimate of real space $P(k)$
from redshift galaxy and cluster data (Peacock \& Dodds 1994), pentagons
-- IRAS 1.2 Jy redshift space $P(k)$ (Fisher et al. 1993), open and
filled triangles -- CfA2 and SSRS2 redshift space $P(k)$ (da Costa et al.
1994).  At the bottom of each panel are plotted window functions for CMB
anisotropy expansion coefficients $a_\ell$ (Panel (a): quadrupole $a_2$,
and $a_{11}$; Panel (b) left to right: $a_2$ for $\Omega_0=0.1,$ 0.3, and
1), bulk flows $V_R$, and the rms mass fluctuation in a sphere of
$8h^{-1}$ Mpc $\sigma_8$. (From Stompor, Gorski, \& Banday 1995, used by
permission.)}
\label{fig:pr_7}
\end{figure}

These two CDM variants were identified as the best bets in the COBE
interpretation paper (Wright et al. 1992, largely based on Holtzman
1989).  In order to discuss them in more detail, it will be best
to start by considering the rather complicated but very illuminating
\fig{pr_7}, showing
COBE-normalized linear CHDM and \lcdm\ power spectra $P(k)$ compared with
four observational estimates of $P(k)$.\footnote{The normalization is
actually according to the two-year COBE data, which is about 10\% higher
in amplitude than the final four-year COBE data
(Gorski et al. 1996), but this relatively small difference will not be
important for our present purposes.}  Panel (a) shows the $\Omega=1$ CHDM
models, and Panel (b) shows the \lcdm\ models.  The heavy solid curves in
Panel (a) are for $h=0.5$ and $\Omega_b=0.05$.  In the middle section of
the figure, the highest of these curves represents the standard CDM
model, and the lower ones standard CHDM ($N_\nu=1$) with $\Omega_\nu=0.2$
(higher) and 0.3; the medium-weight solid curves represent the
corresponding CHDM models with two neutrinos equally sharing the same
total neutrino mass ($N_\nu=2$).  Note that the $N_\nu=2$ CHDM power
spectra are significantly smaller than those for $N_\nu=1$ for $k\approx
0.04-0.4 h$ Mpc$^{-1}$; this arises because for $N_\nu=2$ the neutrinos
weigh half as much and correspondingly free stream over a longer
distance.  The result is that $N_\nu=2$ COBE-normalized CHDM with
$\Omega_\nu\approx 0.2$ can simultaneously fit the abundance and
correlations of clusters (PHKC95, cf. Borgani et al. 1996).
The light solid curve is CDM with $\Gamma=\Omega h=0.2$.

The ``bow'' superimposed on these curves represents the approximate
``pivot point'' (cf. Gorski et al. 1994) for COBE-normalized ``tilted''
models (i.e., with $n_p \ne 1$), and the error bar there represents the
$1\sigma$ COBE normalization uncertainty. The window functions for
various spherical harmonic coefficients $a_\ell$, bulk velocities $V_R$,
and $\sigma_8$ are shown in the bottom part of this figure (see caption).
The bow lies above the $a_{11}$ window because the statistical weight of
the COBE data is greatest for angular wavenumber $\ell \approx 11$
(cosmic variance is greater for lower $\ell$, and the $\sim 7^\circ$
resolution of the COBE DMR makes the uncertainty increase for higher
$\ell$).

The upper section of Panel (a) reproduces the curves for sCDM (top),
$\Omega_\nu = 0.2$ $N_\nu=1$ CHDM, and $\Gamma=0.2$ (light) $P(k)$,
compared with several observational $P(k)$ (see caption). Beware of
comparing apples to oranges to bananas!  Note that the only one of
these observational data sets, that of Baugh \& Efstathiou
(1993, 1994)
(squares) is the real-space $P(k)$ reconstructed from the angular APM
data; that of Peacock \& Dodds (1994) (filled circles) is based on the
redshift-space data with a bias-dependent and $\Omega$-dependent
correction for redshift
distortions and a model-dependent (Peacock \& Dodds 1996, Smith
et al. 1997) correction for nonlinear evolution;
the others are in redshift space. Also, the observations are of
galaxies, which are likely to be a biased tracer of the dark matter,
while the theoretical spectra are for the dark matter itself.
Moreover, as will be discussed in more detail shortly, the real-space
linear $P(k)$ are only a good approximation to the true real-space
$P(k)$ for $k \lsim 0.2 h$ Mpc$^{-1}$; nonlinear gravitational
clustering makes the actual $P(k)$ rise about an order of magnitude
above the linear power spectrum for $k \gsim 1 h$ Mpc$^{-1}$.  Thus
one can see that COBE-normalized sCDM predicts a considerably higher
$P(k)$ than observations indicate. COBE-normalized $\Gamma=0.2$ CDM
predicts a power spectrum shape in better agreement with the data, but
with a normalization that is too low. But the $P(k)$ for
$\Omega_\nu=0.2$ CHDM, especially with $N_\nu=2$, is a pretty good fit
both in shape and amplitude.  The fact that the linear spectrum lies
lower than the data for large $k$ is good news for this model, since,
as was just mentioned, nonlinear effects will increase the power there.

The three heavy solid curves in Panel (b) represent the $P(k)$ for \lcdm\
with $h=0.8$, $\Omega_b=0.02$ for $\Omega_0=0.1$ (top, for $k=0.001h$
Mpc$^{-1}$), 0.2, and 0.3 (bottom).  The lighter curves are for the same
three values of $\Omega_0$ plus 0.4 (bottom) with $h=0.5$,
$\Omega_b=0.05$ (the large wiggles in the latter reflect the effect of
the acoustic oscillations with a relatively large fraction of baryons).
Dotted curves are for sCDM models with the same pair of $h$ values.  The
observational $P(k)$ are as in Panel (a).

Note that the power increases at small $k$ as $\Omega_0$ decreases, with
opposite behavior at large $k$.  Also, the COBE-normalized power spectra
are unaffected by the value of $h$ for small $k$, but increase with $h$
for larger $k$ (the fact that the light $h=0.2$ curve in Panel (a) is
lower than sCDM reflects the same trend).  The fact that the data points
lie lower than any of the \lcdm\ models for $k\lsim 0.02$ is worrisome
for the success of \lcdm, but it is too early to rule out these models on
this basis since various effects such as sparse sampling can lead the
current observational estimates of $P(k)$ to be too low on large scales
(Efstathiou 1996).  A better measurement of $P(k)$ on such large scales
$k \lsim 10^{-2} h$ Mpc$^{-1}$ will be one of the most important early
outputs of the next-generation very large redshift surveys: the $2^\circ$
field (2DF) survey at the Anglo-Australian Telescope, and the Sloan
Digital Sky Survey (SDSS) using a dedicated 2.5 m telescope at the Apache
Point Observatory in New Mexico.  $P(k)$ is much better determined for
larger $k$ by the presently available data, and the fact that the linear
$\Omega_0=0.2$ and 0.3 curves lie higher than many of the data points for
larger $k$ means that these $h=0.8$ models will lie far above the data
when nonlinear effects are taken into account.  This means that, unless
some physical process causes the galaxies to be much less clustered than
the dark matter (``anti-biasing''), such models could be acceptable only
with a considerable amount of tilt --- but that can make the shape of the
spectrum fit more poorly.

\subsection{Numerical Simulations to Probe Smaller Scales}
\label{sec:pr_models_simulations}

\begin{figure}[!htb]   
\centering
\centerline{\psfig{file=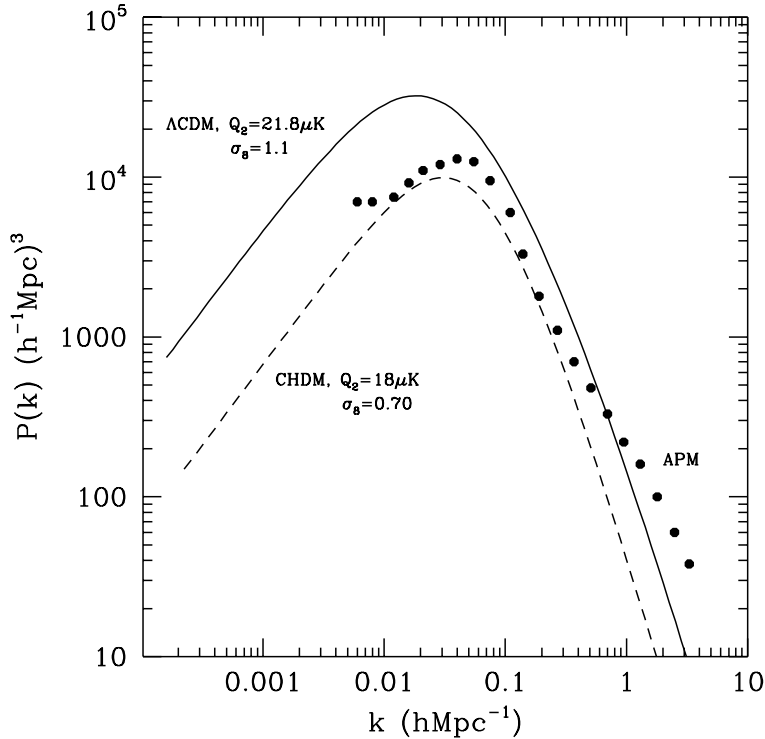,width=10cm}}
\caption{%
Power spectrum of dark matter for \lcdm\ and CHDM models considered
here, both normalized to COBE, compared to the APM galaxy
real-space power spectrum. (\lcdm: $\Omega_0=0.3$,
$\Omega_\Lambda=0.7$, $h=0.7$, thus $t_0=13.4$ Gy; CHDM: $\Omega=1$,
$\Omega_\nu=0.2$ in $N_\nu=2$ $\nu$ species, $h=0.5$, thus $t_0=13$
Gy; both models fit cluster abundance with no tilt, i.e., $n_p=1$.
(From Primack \& Klypin 1996.) }
\label{fig:pr_8}
\end{figure}

``Standard'' $\Omega=1$ Cold Dark Matter (sCDM) with $h \approx 0.5$
and a near-Zel'dovich spectrum of primordial fluctuations (Blumenthal
et al. 1984) until a few years ago seemed to many theorists to be the
most attractive of all modern cosmological models. But although sCDM
normalized to COBE nicely fits the amplitude of the large-scale flows
of galaxies measured with galaxy peculiar velocity data (Dekel 1994),
it does not fit the data on smaller scales: it predicts far too many
clusters (White, Efstathiou, \& Frenk 1993) and does not account for
their large-scale correlations (e.g. Olivier et al. 1993, Borgani et
al. 1996), and the shape of the power spectrum $P(k)$ is wrong (Baugh
\& Efstathiou 1994, Zaroubi et al. 1996). But as discussed above,
variants of sCDM can do better.  Here the focus is on CHDM and \lcdm.  The
linear {\it matter} power spectra for these two models are compared in
\fig{pr_8} to the real-space {\it galaxy} power spectrum obtained from
the two-dimensional APM galaxy power spectrum (Baugh \& Efstathiou
1994), which in view of the uncertainties is not in serious
disagreement with either model for $10^{-2} \lsim k \lsim 1 h$
Mpc$^{-1}$. The \lcdm\ and CHDM models essentially bracket the range
of power spectra in currently popular cosmological models that are
variants of CDM.


\begin{figure}[!htb]   
\centering
\centerline{\psfig{file=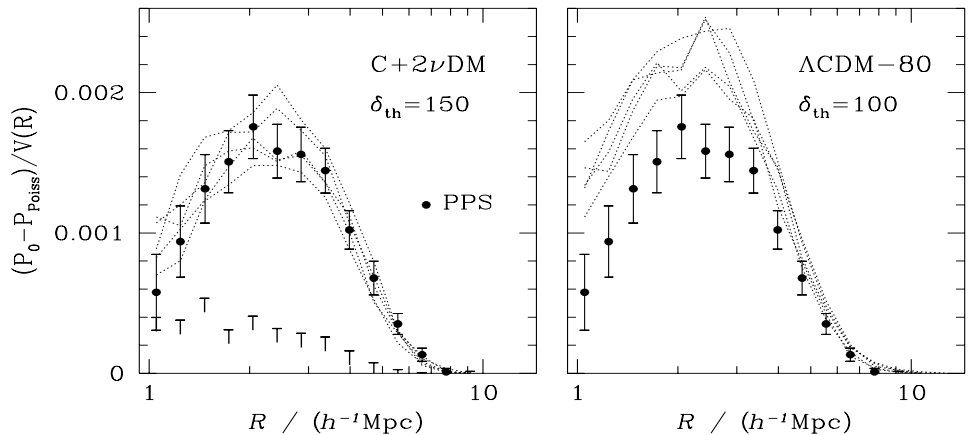,width=\textwidth}}
\caption{%
Void Probability Function $P_0(R)$ for (left panel) CHDM with $h=0.5$
and $\Omega_\nu=0.2$ in $N_\nu=2$ species of neutrinos and (right
panel) \lcdm\ with $h=0.7$ and $\Omega_0=0.3$.  What is plotted here
is difference between the actual VPF and that for a Poisson
distribution, divided by $V(R) = 4 \pi R^3/3$.  Each plot shows also
$P_0 (R)$ for five typical different locations in the simulations
(dotted lines) to give an indication of the sky variance. Data points
are the VPF from the Perseus-Pisces Survey, with $3\sigma$ error bars;
the VPF from the CfA2 survey is very similar.  We have chosen the
$\delta_{th}$ for which the $P_0$ of each model best approaches the
observational data. In the top--left panel, the heavy ``T'' symbols at
the bottom sets the boundary of the region where the signal is
indistinguishable from Poissonian. They are obtained from the
3$\sigma$ scatters among measures for 50 different realizations of the
Poissonian distribution in the same volume as our samples. (From
Ghigna et al. 1996.) }
\label{fig:pr_9}
\end{figure}

The Void Probability Function (VPF) is the probability $P_0(r)$ of
finding no bright galaxy in a randomly placed sphere of radius $r$.
It has been shown that CHDM with $\Omega_\nu=0.3$ predicts a VPF
larger than observations indicate (Ghigna et al. 1994), but newer
results based on our $\Omega_\nu=0.2$ simulations in which the
neutrino mass is shared equally between $N_\nu =2$ neutrino species
(PHKC95) show that the VPF for this model is in excellent
agreement with observations (Ghigna et al. 1996), as shown
in \fig{pr_9}. However, our simulations (Klypin, Primack, \& Holtzman
1996, hereafter KPH96) of COBE-normalized \lcdm\ with $h=0.7$ and
$\Omega_0=0.3$ lead to a VPF that is too large to be compatible with a
straightforward interpretation of the data. Acceptable \lcdm\  models
probably need to have $\Omega_0 >0.3$ and $h<0.7$, as discussed
further below.


Another consequence of the reduced power in CHDM on small scales is
that structure formation is more recent in CHDM than in \lcdm.  As
discussed above (in \se{pr_om_early}), this
may conflict with observations of damped Lyman $\alpha$ systems in
quasar spectra, and other observations of protogalaxies at high
redshift, although the available evidence does not yet permit a clear
decision on this (see below). While the original $\Omega_\nu=0.3$ CHDM
model (Davis, Summers, \& Schlegel 1992, Klypin et al. 1993) certainly
predicts far less neutral hydrogen in damped Lyman $\alpha$ systems
(identified as protogalaxies with circular velocities $V_c \geq
50\kms$) than is observed, as discussed already, lowering the hot
fraction to $\Omega_\nu \approx 0.2$ dramatically improves this
(Klypin et al. 1995). Also, the evidence from preliminary data of a
fall-off of the amount of neutral hydrogen in damped Lyman $\alpha$
systems for $z \gsim 3 $ (Storrie-Lombardi et al. 1996) is in accord
with predictions of CHDM (Klypin et al. 1995).

However, as for all $\Omega=1$ models, $h \gsim 0.55$ implies $t_0
\lsim 12$ Gyr, which conflicts with the pre-Hipparcos
age estimates from globular clusters. The only way to accommodate both
large $h$ and large $t_0$ within the standard FRW framework of General
Relativity, if in fact both $h \gsim 0.65$ and $t_0\gsim 13$ Gyr,
is to introduce a positive cosmological constant
($\Lambda>0)$.  Low-$\Omega_0$ models with $\Lambda=0$ don't help much
with $t_0$, and anyway are disfavored by the latest small-angle cosmic
microwave anisotropy data (Netterfield et al. 1997, Scott et al.
1996, Lineweaver \& Barbosa 1997; cf. Ganga, Ratra, \& Sugiyama 1996
for a contrary view).

\lcdm\  flat cosmological models with $\Omega_0 = 1 - \Omega_\Lambda
\approx 0.3$, where $\Omega_\Lambda \equiv \Lambda/(3H_0^2)$,
were discussed as an alternative to $\Omega=1$ CDM since the beginning
of CDM (Blumenthal et al. 1984, Peebles 1984, Davis et al. 1985). They
have been advocated more recently (e.g., Efstathiou, Sutherland, \&
Maddox 1990; Kofman, Gnedin, \& Bahcall 1993; Ostriker \& Steinhardt
1995; Krauss \& Turner 1995) both because they can solve the $H_0-t_0$
problem and because they predict a larger fraction of baryons in
galaxy clusters than $\Omega=1$ models (this is discussed in
\se{pr_om_cl-baryon} above).

\begin{figure}[!htb]
\vskip2pc
\centering
\centerline{\psfig{file=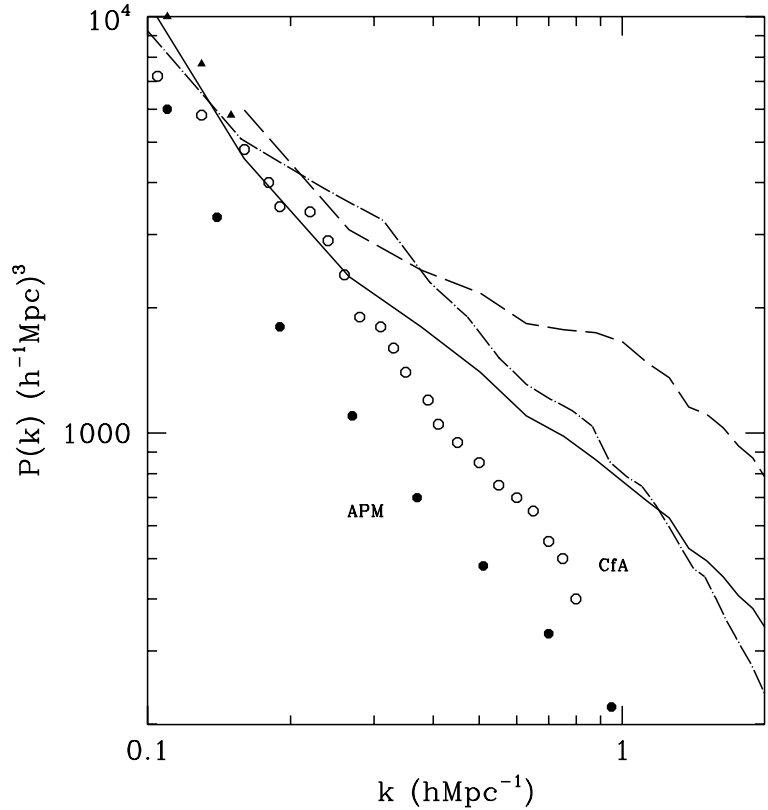,width=10cm}\hspace*{3pt}}
\caption{%
Comparison of the nonlinear power spectrum in the $\Omega_0=0.3$,
$h=0.7$ \lcdm\  model with observational results. Dots are results for the
APM galaxy survey.  Results for the real-space power spectrum for the
CfA survey are shown as open circle ($101 h^{-1}$ Mpc sample) and triangles
($130 h^{-1}$ Mpc sample). Formal error bars for each of the surveys are
smaller than the difference between the open and filled points, which
should probably be regarded as a more realistic estimate of the range of
uncertainty.  The full curve
represents the power spectrum of the dark matter. Lower limits on the
power spectrum of galaxies predicted by the \lcdm\  model are shown as
the dashed curve (higher resolution $\Lambda$CDM$_f$ simulation in
KPH96) and the dot-dashed curve (lower resolution $\Lambda$CDM$_c$
simulation). }
\label{fig:pr_10}
\end{figure}

\begin{figure}[!htb]
\centering
\centerline{\psfig{file=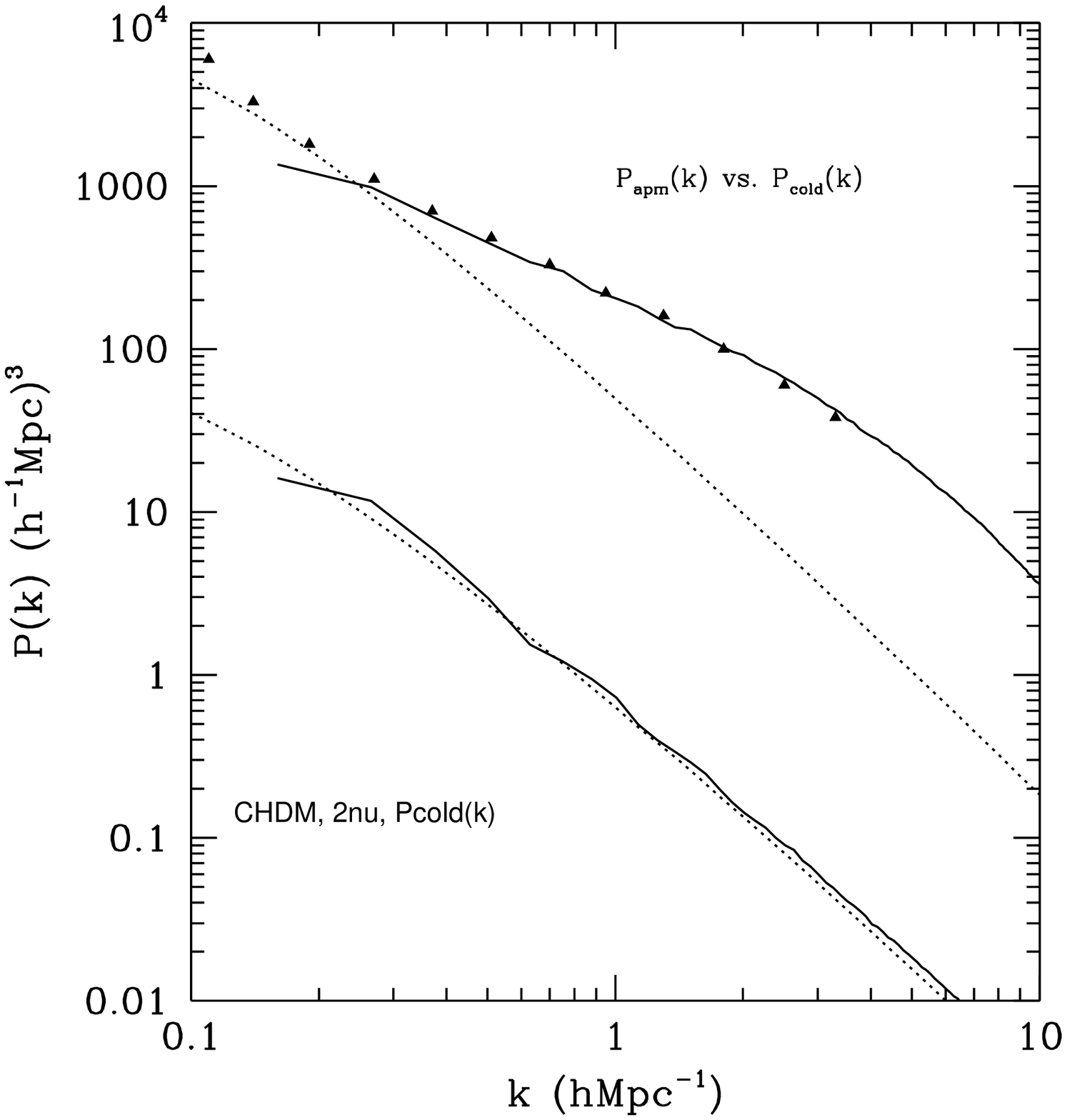,width=10cm}}
\caption{%
Comparison of APM galaxy power spectrum (triangles) with nonlinear
cold particle power spectrum from CHDM model considered in this paper
(upper solid curve).  The dotted curves are linear theory; upper
curves are for $z=0$, lower curves correspond to the higher redshift
$z=9.9$. (From Primack \& Klypin 1996.)}
\label{fig:pr_11}
\end{figure}

Early galaxy formation also is
often considered to be a desirable feature of these models. But early
galaxy formation implies that fluctuations on scales of a few Mpc
spent more time in the nonlinear regime, as compared with CHDM models.
As has been known for a long time, this results in excessive
clustering on small scales.  It has been found that a
typical $\Lambda$CDM model with $h=0.7$ and $\Omega_0=0.3$, normalized
to COBE on large scales (this fixes $\sigma_8\approx 1.1$ for this
model), is compatible with the number-density of galaxy clusters
(Borgani et al. 1997), but predicts a power spectrum of galaxy
clustering in real space that is much too high for wavenumbers
$k=(0.4-1)h/{\rm Mpc}$ (KPH96). This conclusion holds if we assume
either that galaxies trace the dark matter, or just that a region with
higher density produces more galaxies than a region with lower
density. One can see immediately from \fig{pr_7} and \fig{pr_8} that
there will
be a problem with this \lcdm\ model, since the APM power spectrum is
approximately equal to the linear power spectrum at wavenumber $k
\approx 0.6 h$ Mpc$^{-1}$, so there is no room for the extra power
that nonlinear evolution certainly produces on this scale ---
illustrated in \fig{pr_10} for \lcdm\ and in \fig{pr_11} for CHDM.  The
only way to reconcile the $\Omega_0=0.3$ \lcdm\ model considered here
with the observed power spectrum is to assume that some mechanism
causes strong anti-biasing --- i.e., that regions with high dark
matter density produce fewer galaxies than regions with low density.
While theoretically possible, this seems very unlikely; biasing rather
than anti-biasing is expected, especially on small scales (e.g.,
Kauffmann, Nusser, \& Steinmetz 1997). Numerical hydro+N-body
simulations that incorporate effects of UV radiation, star formation,
and supernovae explosions (Yepes et al. 1997) do not show any antibias
of luminous matter relative to the dark matter.


Our motivation to investigate this particular \lcdm\  model was to
have $H_0$ as large as might possibly be allowed in the \lcdm\ class
of models, which in turn forces $\Omega_0$ to be rather small in order
to have $t_0 \gsim 13$ Gyr. There is little room to lower the
normalization of this \lcdm\  model by tilting the primordial power
spectrum $P_p(k)=A k^{n_p}$ (i.e., assuming $n_p$ significantly
smaller than the ``Zel'dovich'' value $n_p=1$), since then the fit to
data on intermediate scales will be unacceptable --- e.g., the number
density of clusters will be too small (KPH96).  Tilted \lcdm\ models
with higher $\Omega_0$, and therefore lower $H_0$ for $t_0
\gsim 13$ Gyr, appear to have a better hope of fitting the available
data, based on comparing quasi-linear calculations to the data (KPH96,
Liddle et al. 1996c). But all models with a cosmological constant
$\Lambda$ large enough to help significantly with the $H_0-t_0$
problem are in trouble with the observations summarized above
providing strong upper limits on $\Lambda$: gravitational lensing, HST
number counts of elliptical galaxies, and especially the preliminary
results from measurements using high-redshift Type Ia supernovae.

It is instructive to compare the $\Omega_0=0.3$, $h=0.7$ \lcdm\  model
that we have been considering with standard CDM and with CHDM.  At
$k=0.5 h$ Mpc$^{-1}$, Figures~5 and 6 of Klypin, Nolthenius, \&
Primack (1997) show that the $\Omega_\nu=0.3$ CHDM spectrum and that
of a biased CDM model with the same $\sigma_8=0.67$ are both in good
agreement with the values indicated for the power spectrum $P(k)$ by
the APM and CfA data, while the CDM spectrum with $\sigma_8=1$ is
higher by about a factor of two. As \fig{pr_11} shows, CHDM with
$\Omega_\nu=0.2$ in two neutrino species (PHKC95) also gives nonlinear
$P(k)$ consistent with the APM data.


\subsection{CHDM: Early Structure Troubles?}
\label{sec:pr_models_chdm-early}

Aside from the possibility mentioned at the outset that the Hubble
constant is too large and the universe too old for any $\Omega=1$
model to be viable, the main potential problem for CHDM appears to be
forming enough structure at high redshift. Although, as mentioned
above, the prediction of CHDM that the amount of gas in damped Lyman
$\alpha$ systems is starting to decrease at high redshift $z \gsim 3$
seems to be in accord with the available data, the large velocity
spread of the associated metal-line systems {\it may} indicate that
these systems are more massive than CHDM would predict (see e.g., (Lu
et al. 1996, Wolfe 1996).  Also, results from a recent CDM
hydrodynamic simulation (Katz et al. 1996) in which the amount of
neutral hydrogen in protogalaxies seemed consistent with that observed
in damped Lyman $\alpha$ systems (DLAS) led the authors to speculate
that CHDM models would produce less than enough DLAS (cf. Ma et al.
1997, Gardner et al. 1997); however, since
the regions identified as DLAS in these simulations were not actually
resolved gravitationally, this will need to be addressed by higher
resolution simulations for all the models considered before
their arguments can be considered completely convincing.  As mentioned
above, fairly realistic simulations by Haehnelt, Steinmetz, \& Rauch
(1997) are able to account in rather impressive detail for the
statistics characterizing the DLAS metal line systems measured by
Prochaska \& Wolfe (1997), but Haehnelt et al. find that it is not
large disks but rather clouds of gas in dark matter halos which
account for most of these metal lines.

Finally, Steidel et al. (1996) have found objects by their emitted
light at redshifts $z=3-3.5$ apparently with relatively high velocity
dispersions (indicated by the equivalent widths of absorption lines),
which they tentatively identify as the progenitors of giant elliptical
galaxies. {\it Assuming} that the indicated velocity dispersions are
indeed gravitational velocities, Mo \& Fukugita (1996, hereafter MF96)
have argued that the abundance of these objects is higher than
expected for the COBE-normalized $\Omega=1$ CDM-type models that can
fit the low-redshift data, including CHDM, but in accord with
predictions of the \lcdm\ model considered here. (In more detail, the
MF96 analysis disfavors CHDM with $h=0.5$ and $\Omega_\nu
\gsim 0.2$ in a single species of neutrinos. They apparently would
argue that this model is then in difficulty since it overproduces rich
clusters --- and if that problem were solved with a little tilt $n_p
\approx 0.9$, the resulting decrease in fluctuation power on small
scales would not lead to formation of enough early objects. However,
if $\Omega_\nu \approx 0.2$ is shared between two species of
neutrinos, the resulting model appears to be at least marginally
consistent with both clusters and the Steidel objects even with the
assumptions of MF96.  The \lcdm\ model with $h=0.7$ consistent with
the most restrictive MF96 assumptions has $\Omega_0 \gsim 0.5$, hence
$t_0 \lsim 12$ Gyr.  \lcdm\ models having tilt and lower $h$, and
therefore more consistent with the small-scale power constraint
discussed above, may also be in trouble with the MF96 analysis.) But
in addition to uncertainties about the actual velocity dispersion and
physical size of the Steidel et al. objects, the conclusions of the
MF96 analysis can also be significantly weakened if the gravitational
velocities of the observed baryons are systematically higher than the
gravitational velocities in the surrounding dark matter halos, as is
perhaps the case at low redshift for large spiral galaxies (Navarro,
Frenk, \& White 1996), and even more so for elliptical galaxies which
are largely self-gravitating stellar systems in their central regions.

Given the irregular morphologies of the high-redshift objects seen in
the Hubble Deep Field (van den Bergh et al. 1996) and other deep HST
images, it seems more likely that they are mostly relatively low mass
objects undergoing starbursts, possibly triggered by mergers, rather than
galactic protospheroids (Lowenthal et al. 1996).
Since the number density of the brightest of
such objects may be more a function of the probability and duration of
such starbursts rather than the nature of the underlying cosmological
model, it may be more useful to use the star formation or metal
injection rates (Madau et al. 1996) indicated by the total observed
rest-frame ultraviolet light to constrain models (Somerville et al.
1997). The available data on the history of star formation (Gallego et
al. 1996, Lilly et al. 1996, Madau et al. 1996, Connolly et al. 1997) 
suggests that most of
the stars and most of the metals observed formed relatively recently,
after about redshift $z\sim 1$; and that the total star formation rate
at $z\sim 3$ is perhaps a factor of 3 lower than at $z \sim 1$, with
yet another factor  of $\sim 3$ falloff to $z \sim 4$ (although the
rates at $z\gsim 3$ could be higher if most of the star formation is
in objects too faint to see). This is in accord with indications from
damped Lyman $\alpha$ systems (Fall, Charlot, \& Pei 1996) and
expectations for $\Omega=1$ models such as CHDM, but perhaps not with
the expectations for low-$\Omega_0$ models which have less growth of
fluctuations at recent epochs, and therefore must form structure
earlier.  But this must be investigated using more detailed modeling,
including gas cooling and feedback from stars and supernovae (e.g.,
Kauffmann 1996, Somerville et al. 1997), before strong conclusions can
be drawn.

There is another sort of constraint from observed numbers of
high-redshift protogalaxies that would appear to disfavor \lcdm. The
upper limit on the number of $z\gsim 4$ objects in the Hubble Deep
Field (which presumably correspond to smaller-mass galaxies than most
of the Steidel objects) is far lower than the expectations in
low-$\Omega_0$ models, especially with a positive cosmological
constant, because of the large volume at high redshift in such
cosmologies (Lanzetta et al. 1996). Thus evidence from high-redshift
objects cuts both ways, and it is too early to tell whether high- or
low-$\Omega_0$ models will ultimately be favored by such data.

\subsection{Advantages of Mixed CHDM Over Pure CDM Models}
\label{sec:pr_models_chdm-cdm}

There are three basic reasons why a mixture of cold plus hot dark
matter works better than pure CDM without any hot particles: {\bf (1)}
the power spectrum shape $P(k)$ is a better fit to observations, {\bf
(2)} there are indications from observations for a more weakly
clustering component of dark matter, and {\bf (3)} a hot component may
help avoid the too-dense central dark matter density in pure CDM dark
matter halos.  Each will be discussed in turn.

{\bf (1) Spectrum shape.}  As explained in discussing WDM vs. CHDM
above, the pure CDM spectrum $P(k)$ does not fall fast enough on the
large-$k$ side of its peak in order to fit indications from galaxy and
cluster correlations and power spectra.  The discussion there of
``Excess Power'' is a way of quantifying this.  This is also related
to the overproduction of clusters in pure CDM. The obvious way to
prevent $\Omega=1$ sCDM normalized to COBE from overproducing clusters
is to tilt it a lot (the precise amount depending on how much of the
COBE fluctuations are attributed to gravity waves, which can be
increasingly important as the tilt is increased).  But a constraint on
CDM-type models that is likely to follow both from the high-$z$ data
just discussed and from the preliminary indications on cosmic
microwave anisotropies at and beyond the first acoustic peak from the
Saskatoon experiment (Netterfield et al. 1997) is that viable models
cannot have much tilt, since that would reduce too much both their
small-scale power and the amount of small-angle CMB anisotropy. As
already explained, by reducing the fluctuation power on cluster
scales and below, COBE-normalized CHDM naturally fits both the CMB
data and the cluster abundance without requiring much tilt.  The need
for tilt is further reduced if a high baryon fraction $\Omega_b
\gsim 0.1$ is assumed (M. White et al. 1996), and this also boosts the
predicted height of the first acoustic peak. No tilt is necessary for
$\Omega_\nu=0.2$ shared between $N_\nu=2$ neutrino species with
$h=0.5$ and $\Omega_b=0.1$. Increasing the Hubble parameter in
COBE-normalized models increases the amount of small-scale power, so
that if we raise the Hubble parameter to $h=0.6$ keeping
$\Omega_\nu=0.2$ and $\Omega_b=0.1(0.5/h)^2=0.069$, then fitting the
cluster abundance in this $N_\nu=2$ model requires tilt $1-n_p \approx
0.1$ with no gravity waves (i.e., $T/S=0$; alternatively if
$T/S=7(1-n_p)$ is assumed, about half as much tilt is needed, but the
observational consequences are mostly very similar, with a little more
small scale power). The fit to the small-angle CMB data is still good,
and the predicted $\Omega_{\rm gas}$ in damped Lyman $\alpha$ systems
is a little higher than for the $h=0.5$ case. The only obvious problem
with $h=0.6$ applies to any $\Omega=1$ model --- the universe is
rather young: $t_0=10.8$ Gyr.  But the revision of the globular
cluster ages with the new Hipparcos data may permit this.

{\bf (2) Need for a less-clustered component of dark matter.}  The
fact that group and cluster mass estimates on scales of $\sim 1
\hMpc$ typically give values for $\Omega$ around 0.1-0.2
while larger-scale estimates give larger values around 0.3-1 (Dekel
1994) suggests that there is a component of dark matter that does not
cluster on small scales as efficiently as cold dark matter is expected
to do.  In order to  quantify this,
the usual group $M/L$ measurement of $\Omega_0$ on small scales
has been performed
in ``observed'' $\Omega=1$ simulations of both CDM and CHDM (Nolthenius,
Klypin, \& Primack 1997). We found that COBE-normalized
$\Omega_\nu=0.3$ CHDM gives $\Omega_{M/L} = 0.12-0.18$ compared to
$\Omega_{M/L} = 0.15$ for the CfA1 catalog analyzed exactly the same
way, while for CDM $\Omega_{M/L} = 0.34-0.37$, with the lower value
corresponding to bias $b=1.5$ and the higher value to $b=1$ (still
below the COBE normalization).  Thus local measurements of the density
in $\Omega=1$ simulations can give low values, but it helps to have a
hot component to get values as low as observations indicate.  We found
that there are three reasons why this virial estimate of the mass in
groups misses so much of the matter in the simulations: (1) only the
mass within the mean harmonic radius $r_h$ is measured by the virial
estimate, but the dark matter halos of groups continue their roughly
isothermal falloff to at least $2r_h$, increasing the total mass by
about a factor of 3 in the CHDM simulations; (2) the velocities of the
galaxies are biased by about $70\%$ compared to the dark matter
particles, which means that the true mass is higher by about another
factor of 2; and (3) the groups typically lie along filaments and are
significantly elongated, so the spherical virial estimator misses
perhaps 30\% of the mass for this reason.  Visualizations of these
simulations (Brodbeck et al. 1997) show clearly how extended the hot
dark matter halos are.  An analysis of clusters in CHDM found similar
effects, and suggested that observations of the velocity distributions
of galaxies around clusters might be able to discriminate between pure
cold and mixed cold + hot models (Kofman et al. 1996). This is an area
where more work needs to be done --- but it will not be easy since it
will probably be necessary to include stellar and supernova feedback
in identifying galaxies in simulations, and to account properly for
foreground and background galaxies in observations.

{\bf (3) Preventing too dense centers of dark matter halos.} Flores
and Primack (1994) pointed out that dark matter density profiles with
$\rho(r) \propto r^{-1}$ near the origin from high-resolution
dissipationless CDM simulations (Dubinski \& Carlberg 1991; Warren et
al. 1992; Crone, Evrard, \& Richstone 1994) are in serious conflict
with data on dwarf spiral galaxies (cf. Moore 1994), and in possible
conflict with data on larger spirals (Flores et al. 1993) and on
clusters (cf. Miralda-Escud\'e 1995, Flores \& Primack 1996). Navarro,
Frenk, \& White (1996; cf. Cole \& Lacey 1996) agree that rotation
curves of small spiral galaxies such as DDO154 and DDO170 are strongly
inconsistent with their universal dark matter profile $\rho_{NFW}(r)
\propto 1/[r(r+a)^2]$.  Navarro, Eke, \& Frenk (1996) proposed a possible
explanation for the discrepancy regarding dwarf spiral galaxies
involving slow accretion followed by explosive ejection of baryonic
matter from their cores, but it is implausible that such a process
could be consistent with the observed regularities in dwarf spirals
(Burkert 1995); in any case it will not work for
low-surface-brightness galaxies.  Work is in progress with Stephane
Courteau, Sandra Faber, Ricardo Flores, and others to see whether the
$\rho_{NFW}$ universal profile is consistent with data from high- and
low-surface-brightness galaxies with moderate to large circular
velocities, and with Klypin, Kravtsov, and Bullock to see whether
higher resolution simulations for a wider variety of models continue
to give $\rho_{NFW}$.  The failure of earlier simulations to form
cores as observed in dwarf spiral galaxies either is a clue to a
property of dark matter that is not understood, or is telling us that
the simulations were inadequate. It is important to discover whether
this is a serious problem, and whether inclusion of hot dark matter or
of dissipation in the baryonic component of galaxies can resolve it.
It is clear that including hot dark matter will decrease the central
density of dark matter halos, both because the lower fluctuation power
on small scales in such models will prevent the early collapse that
produces the highest dark matter densities, and also because the hot
particles cannot reach high densities because of the phase space
constraint (Tremaine \& Gunn 1979, Kofman et al. 1996).  But this may
not be necessary, since even our simulations without any hot dark
matter appear to be consistent with the rotation curves observed in
dwarf irregular and low surface brightness galaxies (Kravtsov et al.
1997).

\subsection{Best Bet CDM-Type Models}
\label{sec:pr_models_best}

As said at the outset, the fact that the original CDM model
did so well at predicting both the CMB anisotropies discovered by COBE
and the distribution of galaxies makes it likely that a large fraction
of the dark matter is cold --- i.e., that one of the variants of the
sCDM model might turn out to be right.  Of these, CHDM
is the best bet if $\Omega_0$ turns out to be near unity and the
Hubble parameter is not too large, while \lcdm\ is the best bet if the
Hubble parameter is too large to permit the universe to be older than
its stars with $\Omega=1$.

Both theories do seem less ``natural'' than sCDM, in that they are
both hybrid theories.  But although sCDM won the beauty contest, it
doesn't fit the data.

CHDM is just sCDM with some light neutrinos.  After all, we know that
neutrinos exist, and there is experimental evidence --- admittedly not
yet entirely convincing --- that at least some of these neutrinos have
mass, possibly in the few-eV range necessary for CHDM.
Isn't it an unnatural coincidence to have three different sorts of
matter --- cold, hot, and baryonic --- with contributions to the
cosmological density that are within an order of magnitude of each
other?  Not necessarily.  All of these varieties of matter may have
acquired their mass from (super?)symmetry breaking associated with the
electroweak phase transition, and when we understand the nature of the
physics that determines the masses and charges that are just
adjustable parameters in the Standard Model of particle physics, we
may also understand why $\Omega_c$, $\Omega_\nu$, and $\Omega_b$ are so
close.  In any case, CHDM is certainly not uglier than \lcdm.

In the \lcdm\ class of models, the problem of too much power on small
scales that has been discussed here at some length for $\Omega_0=0.3$
and $h=0.7$ \lcdm\ implies either that there must be some physical
mechanism that produces strong, scale-dependent anti-biasing of the
galaxies with respect to the dark matter, or else that higher
$\Omega_0$ and lower $h$ are preferred, with a significant amount of
tilt to get the cluster abundance right and avoid too much small-scale
power (KPH96). Higher $\Omega_0 \gsim 0.5$ also is more consistent
with the evidence summarized above against large $\Omega_\Lambda$ and
in favor of larger $\Omega_0$, especially in models such as \lcdm\
with Gaussian primordial fluctuations.

Among CHDM models, having $N_\nu=2$ species share the neutrino mass
gives a better fit to COBE, clusters, and small-scale data than
$N_\nu=1$, and moreover it appears to be favored by the available
experimental data (PHKC95).  But it remains to be seen whether CHDM
models can fit the data on structure formation at high redshifts,
and whether any models of the CDM type can fit all the data
--- the data on the values of the cosmological parameters,
the data on the distribution and structure of galaxies at
low and high redshifts, and the increasingly precise CMB
anisotropy data.  Reliable data is becoming available so
rapidly now, thanks to the wonderful new ground and
space-based instruments, that the next few years will be
decisive.

The fact that NASA and the European Space Agency plan to launch the
COBE follow-up satellites MAP and COBRAS/SAMBA in the early years of
the next decade, with ground and balloon-based detectors promising to
provide precise data on CMB anisotropies even earlier, means that we
are bound to know much more soon about the two key questions of modern
cosmology: the nature of the dark matter and of the initial
fluctuations.  Meanwhile, many astrophysicists, including my
colleagues and I, will be trying to answer these questions using data
on galaxy distribution, evolution, and structure, in addition to the
CMB data.  And there is a good chance that in the next few years
important inputs will come from particle physics experiments on dark
matter candidate particles or the theories that lead to them, such as
supersymmetry.

\bigskip

\noindent {\bf Acknowledgments.}  This work was partially supported by
NASA and NSF grants at UCSC. JRP thanks his Santa Cruz colleagues and
all his collaborators, especially Anatoly Klypin, for many helpful
discussions of the material presented here.  Special thanks to James
Bullock, Avishai Dekel, Patrik Jonsson, and Tsafrir Kolatt for reading
an earlier draft and for helpful suggestions for its improvement, and
to Nora Rogers for TeX help.


\end{document}